\documentclass[journal]{IEEEtran}

\usepackage{eurosym,makecell,comment}
\usepackage{multirow}
\usepackage{threeparttable}
\usepackage{eurosym} 
\usepackage{framed}
\usepackage{hyperref}
\usepackage[pdftex]{graphicx}
\usepackage{url}
\usepackage{verbatim}
\usepackage{graphicx}
\usepackage{subcaption}
\usepackage{float}
\usepackage{xcolor}
\usepackage{colortbl}
\usepackage{soul}
\usepackage{amsmath}
\usepackage{mathtools}
\usepackage{ragged2e, microtype}
\usepackage{xcolor,pifont}
\usepackage{stfloats}

\newcommand\sbullet[1][.5]{\mathbin{\vcenter{\hbox{\scalebox{#1}{$\bullet$}}}}}
\newcommand{\yestick}{{\color{olive}\ding{51}}}
\newcommand{\notick}{{\color{red}\ding{55}}}

\usepackage{array} 
\newcolumntype{L}[1]{>{\raggedright\let\newline\\\arraybackslash\hspace{0pt}}m{#1}}

\usepackage{etoolbox}
\usepackage{soul} 

\newtoggle{finalPaper}

\setstcolor{red}

\toggletrue{finalPaper} 

\iftoggle{finalPaper} {
	\newcommand{\addtxtRev}[1]{#1}
	\newcommand{\changeRev}[2]{#2}
	\newcommand{\rmvtxtRev}[1]{}
	\newcommand{\addtxt}[1]{#1}
	\newcommand{\change}[2]{#2}
	\newcommand{\rmvtxt}[1]{}
	}{
	\newcommand{\addtxtRev}[1]{\textcolor{red}{#1}}
	\newcommand{\changeRev}[2]{\st{#1}\textcolor{red}{#2}}
	\newcommand{\rmvtxtRev}[1]{\st{#1}}
	\newcommand{\addtxt}[1]{#1}
	\newcommand{\change}[2]{#2}
	\newcommand{\rmvtxt}[1]{}
}

\begin{document}

\title{A Survey on Device Behavior Fingerprinting: Data Sources, Techniques, Application Scenarios, and Datasets}





\author{Pedro Miguel S\'anchez S\'anchez, Jos\'e Mar\'ia Jorquera Valero,
Alberto Huertas Celdr\'an, G\'er\^ome Bovet,\\Manuel Gil P\'erez, and Gregorio Mart\'inez P\'erez,~\IEEEmembership{Member,~IEEE}

\thanks{Pedro Miguel S\'anchez S\'anchez, Jos\'e Mar\'ia Jorquera Valero, Manuel Gil P\'erez, and Gregorio Mart\'inez P\'erez are with the Department of Information and Communications Engineering, University of Murcia, 30100 Murcia, Spain (e-mail: pedromiguel.sanchez@um.es; josemaria.jorquera@um.es; mgilperez@um.es; gregorio@um.es) \textit{(Corresponding author: Pedro Miguel S\'anchez S\'anchez)}}
\thanks{Alberto Huertas Celdr\'an is with the Communication Systems Group (CSG) at the Department of Informatics (IfI), University of Zurich UZH, 8050 Zürich, Switzerland (e-mail: huertas@ifi.uzh.ch).}
\thanks{G\'{e}r\^{o}me Bovet is with the Cyber-Defence Campus within armasuisse Science \& Technology, 3602 Thun, Switzerland (e-mail: gerome.bovet@armasuisse.ch)}

\thanks{© 2021 IEEE.  Personal use of this material is permitted.  Permission from IEEE must be obtained for all other uses, in any current or future media, including reprinting/republishing this material for advertising or promotional purposes, creating new collective works, for resale or redistribution to servers or lists, or reuse of any copyrighted component of this work in other works.}}


\maketitle

\begin{abstract}
In the current network-based computing world, where the number of interconnected devices grows exponentially, their diversity, malfunctions, and cybersecurity threats are increasing at the same rate. To guarantee the correct functioning and performance of novel environments such as Smart Cities, Industry 4.0, or crowdsensing, it is crucial to identify the capabilities of their devices (e.g., sensors, actuators) and detect potential misbehavior that may arise due to cyberattacks, system faults, or misconfigurations. With this goal in mind, a promising research field emerged focusing on creating and managing fingerprints that model the behavior of both the device actions and its components. The article at hand studies the recent growth of the device behavior fingerprinting field in terms of application scenarios, behavioral sources, and processing and evaluation techniques. First, it performs a comprehensive review of the device types, behavioral data, and processing and evaluation techniques used by the most recent and representative research works dealing with two major scenarios: device identification and device misbehavior detection. After that, each work is deeply analyzed and compared, emphasizing its characteristics, advantages, and limitations. This article also provides researchers with a review of the most relevant characteristics of existing datasets as most of the novel processing techniques are based on Machine Learning and Deep Learning. Finally, it studies the evolution of these two scenarios in recent years, providing lessons learned, current trends, and future research challenges to guide new solutions in the area.

\end{abstract}


%
\IEEEpeerreviewmaketitle

\section{Introduction}
\label{sec:introduction}

\begin{figure*}
    \centering
    \includegraphics[width=\textwidth]{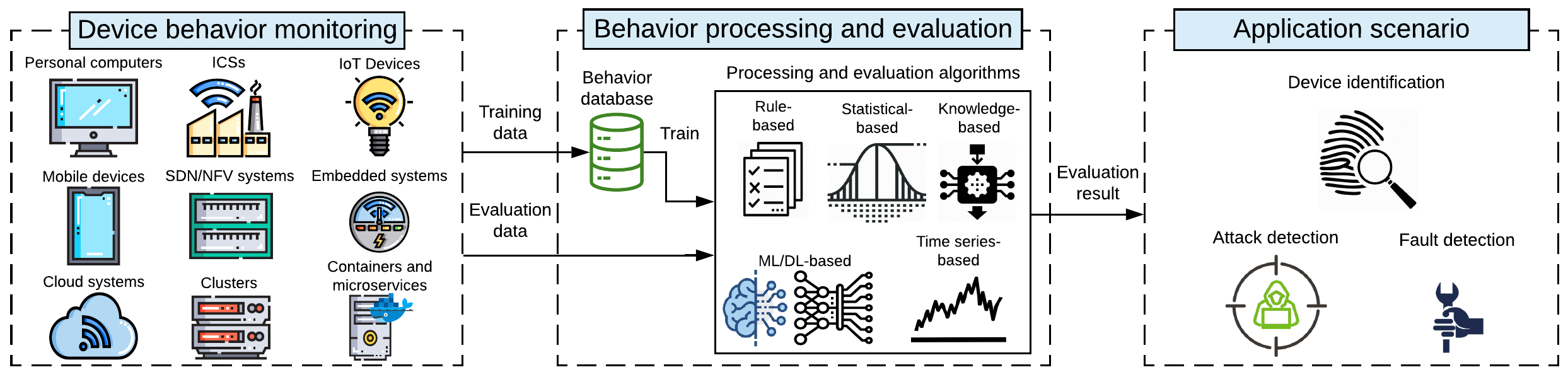}
    \caption{Common life cycle implemented by device behavior fingerprinting solutions.}
    \label{fig:life_cycle}
\end{figure*}

Previsions for 2025 estimate nearly 64 billion IoT devices connected to each other into diverse cutting-edge environments such as Smart Cities, Industry 4.0, or crowdsensing (e.g., Flightradar24, OpenSky, ElectroSense), among others \cite{riad2020dynamic}. These environments have their own particularities in terms of devices, data, communications, and purposes, which increase the complexity of achieving one of their common challenges: to optimize the performance of devices and provide accurate services. To meet this challenge, the advancement of communication networks and computing paradigms has influenced that behavioral data science evolved from studying theoretical and empirical issues related to human behaviors \cite{fuentes2017human} --its initial scope-- to conquer the cyberworld and offer a promising alternative to model device behaviors \cite{shone2013misbehaviour}. Nowadays, a thriving research field within behavior data science focuses on creating device behavior patterns (\textit{fingerprints}) able to optimize their performance and detect potential issues in the early stages \cite{Sivanathan1,Sivanathan2}. 
In this context, this article studies the recent growth of the device behavior research field in terms of application scenarios, behavioral sources, and processing and evaluation techniques. \figurename~\ref{fig:life_cycle} shows an overview of the typical life cycle implemented by the literature, where different devices, techniques, and application scenarios are considered.

The first step to build a device fingerprint is to identify the application scenario where it will be needed. By keeping in mind the goal of optimizing devices and systems performance, the literature has recognized two critical application scenarios. The first one consists in identifying devices with different granularity levels --to differentiate them and fully exploit their capabilities \cite{Marchal2019Audi}-- while the second focuses on detecting cyberattacks \cite{Haefner2019CompleIoT}, malfunction \cite{manco2017fault}, or misbehavior \cite{Ferrando2017Network} --to mitigate them. The nature of each scenario influences the selection of behavioral sources, data, and techniques employed to create fingerprints since the detection of misbehavior produced by a given cyberattack is different from identifying several IoT devices of the same family. Even in the same application scenario, the behavioral data might be different as well; this is the case of some cyberattacks affecting network communications \cite{hamza2019detecting}, while others impact the CPU usage \cite{ravichandiran2018anomaly}.


In both application scenarios, the literature contains an extensive number of works where device fingerprinting has been applied \cite{Miettinen2017IoTSentinel,selis2018classification,Sivanathan1,Jafari_Fingerprinting_Deep_Learning_2018,Nguyen2019DioT,shone2013misbehaviour,samir2020detecting}. On the one hand and in terms of device identification, behavioral data science has dramatically improved the limitations of traditional solutions, mainly focused on using names, identifiers, labels, or tags to identify devices \cite{liu2014unknown}. The main limitation of these approaches is that they can be modified or even duplicated in an environment where the number of devices grows exponentially. Another relevant drawback appears when device identification is performed at different granularity levels, requiring multiple labels and increasing management complexity. Nowadays, the literature categorizes the following identification granularity levels: \textit{type}, with the main goal of creating fingerprints able to detect different types of devices \cite{Marchal2019Audi}; 
\textit{model}, focused on identifying different models of devices based on common hardware and software \cite{Ortiz2019DeviceMien}; and 
\textit{individual}, probably the most challenging level because it tries to identify identical physical devices according to minor differences occurred during manufacturing processes \cite{Jafari_Fingerprinting_Deep_Learning_2018}. 

On the other hand and with the goal of detecting misbehavior or malfunction caused by cybersecurity issues, novel and sophisticated cyberattacks are influencing the replacement of traditional cybersecurity techniques. Existing mechanisms based on signatures are no longer effective against unseen, encrypted, or large-scale cyberattacks, and device fingerprinting has been identified as one of the most promising solutions to tackle this challenge \cite{mishra2017intrusion}. A relevant number of works found in the literature rely on creating ``normal'' behavioral fingerprints to spot changes caused by some previous issues \cite{Haefner2019CompleIoT,Nguyen2019DioT,li2019mad}. In this case, fingerprint evaluation is usually tackled from an anomaly detection perspective \cite{Haefner2019CompleIoT,barbhuiya2018rads}.

In this context, the article at hand performs a comprehensive analysis of the main characteristics --devices, behavioral sources, data, and techniques-- considered by the most representative and recent works of device identification and malfunctioning detection scenarios. Besides, it studies how characteristics of device identification, and misbehavior and malfunction detection scenarios are evolving since last years.

Once having the fingerprints, there is another exciting research area focused on applying the most suitable techniques to process and evaluate \change{the behavior profiles}{them}. Statistical approaches have been dominating the field for the last decades. However, the incursion of Artificial Intelligence (AI), \addtxt{and more concretely Machine and Deep Learning (ML and DL) as the dominating trend}, shifted the field and generated an open discussion concerning the most suitable methods per scenario. This manuscript seeks to help readers understand the trend concerning behavior processing and evaluation techniques, as well as the most appropriate techniques for each application scenario. 

Influenced by the rise of AI techniques, there is also a crescent necessity of exhaustive datasets with which algorithms can train models able to learn and infer valuable information aligned with the target scenarios. Datasets are also critical to have standard benchmarks enabling fair comparisons of existing techniques and solutions. In this direction, this article also pretends to support researchers working on the device behavior research field with a review of the most relevant characteristics of existing datasets.

\begin{table*}[htpb]
    \centering
    \scriptsize
    \begin{tabular}{ >{\Centering}m{0.6cm} >{\Centering}m{0.6cm}  >{\Centering}m{1.5cm} >{\Centering}m{1.5cm} >{\Centering}m{1.2cm}
    >{\Centering}m{1.5cm}
    >{\Centering}m{1.0 cm} >{\RaggedRight\arraybackslash}m{8cm} } 
    \hline
    \textbf{Work} & \textbf{Year} & \textbf{Device Types / Area}& \textbf{Device Identification} & \textbf{Intrusion Detection} & \textbf{Malfunction Detection }& \textbf{Dataset review} & \makecell[c]{\textbf{Focus and solution categorization}} \\
    \hline
    \cite{xu2015device} & 2015 & Wireless devices & \yestick & \notick & \notick & \notick & $\sbullet[0.75]$ Survey on device fingerprinting in wireless networks. \newline $\sbullet[0.75]$ Authors differentiate between white list-based and unsupervised algorithms.\\
    \hline
    \cite{Baldini2017Survey} & 2017 & Mobile phones & \yestick & \notick & \notick & \notick & $\sbullet[0.75]$ Survey on mobile device identification based on physical components. \newline $\sbullet[0.75]$ Fingerprinting techniques are classified in two different categories, emitted signal-based and electronic component-based.  \\
    \hline
    \cite{mishra2017intrusion} & 2017 & Cloud environments & \notick & \yestick & \notick & \notick & $\sbullet[0.75]$ Survey on IDSs applications focused on cloud computing environments. \newline $\sbullet[0.75]$ Intrusion detection techniques are divided into misuse detection (rule-based) and anomaly detection (behavior-based).\\
    \hline
    \cite{Liu2018HIDS} & 2018 & Any, focus on embedded devices & \notick & \yestick & \notick & \yestick & $\sbullet[0.75]$ Survey on IDSs deployed in hosts and based on system calls. \newline $\sbullet[0.75]$ IDSs solutions are divided into anomaly and detection-based and misuse detection-based. \\
    \hline
    \cite{elrawy2018intrusion} & 2018 & IoT Environments & \notick & \yestick & \notick & \notick & $\sbullet[0.75]$ Survey on IDSs focused on IoT-based smart environments. \newline $\sbullet[0.75]$ IDS types are divided into anomaly, specification and misuse-based.\\
    \hline
    \cite{khraisat2019survey} & 2019 & Any & \notick & \yestick & \notick & \yestick & $\sbullet[0.75]$ IDS survey, groups the solutions in signature-based and anomaly-based. \newline $\sbullet[0.75]$ Data sources divided into network and system logs and audits.\\
    \hline
    This work & 2020 & Any\addtxt{, focus on IoT} & \yestick & \yestick & \yestick & \yestick & $\sbullet[0.75]$ General survey on device behavior fingerprinting, its application scenarios, processing techniques and public datasets.\\
    \hline
    \end{tabular}
    \caption{Comparison of survey works considering device behavior fingerprinting.}
    \label{tab:survey_comp}
\end{table*}

\section{\addtxt{Motivation and Contributions}}
\label{sec:contrib}

Device behavior fingerprinting is an encouraging research field that has inspired the publication of several survey articles for the last years. In terms of device identification, in 2016, Xu et al. \cite{xu2015device} reviewed unique device fingerprinting in wireless networks. Moreover, Baldini and Steri \cite{Baldini2017Survey} published in 2017 a review on mobile phone identification based on its hardware components. Regarding the usage of device fingerprint for cybersecurity purposes, the surveys related to this study are mainly focused on Intrusion Detection Systems (IDS). In 2018, Elrawy et al. \cite{elrawy2018intrusion} published a study focused on IDS and IoT-based smart environments. Similarly, Khraisat et al. \cite{khraisat2019survey}, in 2019, published another review on general IDS-related solutions and public datasets, mostly containing network data. In \cite{mishra2017intrusion}, Mishra et al. published a survey, in 2017, where IDS analysis is addressed with a focus on cloud environments. This work explicitly considers system behavior analysis, one of the main sources to ensure a cloud system. Finally, in 2018, Liu et al. \cite{Liu2018HIDS} analyzed existing solutions and datasets covering attack detection based on system calls, with a special focus on embedded devices. 

Despite the contributions of the previous works, as illustrated in \tablename~\ref{tab:survey_comp}, none of them addresses device identification and misbehavior detection in the same study. Besides, no previous survey contemplates device behavior fingerprinting for component malfunctioning detection. \addtxtRev{In addition, there is no recent work reviewing from a broad and exhaustive perspective datasets designed both for device identification and for intrusion or malfunction detection.} \addtxt{Moreover, other surveys in domains such as digital forensics \cite{hou2019survey}, threat hunting, and threat intelligence \cite{tounsi2018survey}, relying on device identification or attack and fault detection as a basis, also considered behavior fingerprinting as an issue or challenge to cover}\addtxtRev{, motivating the importance of this work}\addtxt{. In this context}, the literature has some research questions that need to be solved. As the main relevant, we highlight:

\begin{itemize}
    \item \textit{Q1. Which scenarios, device types, and sources are present in behavior-based solutions?} Depending on the application scenario --device identification or malfunction detection-- and the problem to be solved, the devices and behavioral sources vary. However, in the literature, there is no solution detailing these elements and how they are combined.
    
    \item \textit{Q2. What and how behavior processing and evaluation tasks are used in each scenario?} Device behavior can be processed and evaluated following diverse approaches. However, the literature has not  studied these approaches from a broad perspective to have a complete view in the area.

    \item \textit{Q3. What characteristics do the most recent and representative solutions of each application scenario have?} It is required to analyze how device types and behavioral sources are utilized to solve the problems motivated by each application scenario. Furthermore, it is also needed to detect the limitations of solutions related to both scenarios.
    
    \item \textit{Q4. Which behavior datasets are available and which are their characteristics?} There is no study detailing the public datasets aligned with device behavior from a broad perspective, analyzing their characteristics, and defining in which application scenarios they can be utilized.
    
    \item  \textit{Q5. How have application scenarios evolved for the last years?} To establish the guidelines for future research, it is critical to describe how device behavior analysis is evolving in the last years and which are the current trends and open challenges of the area.

\end{itemize}

\begin{figure*}[htpb]
    \centering
    \includegraphics[width=\textwidth]{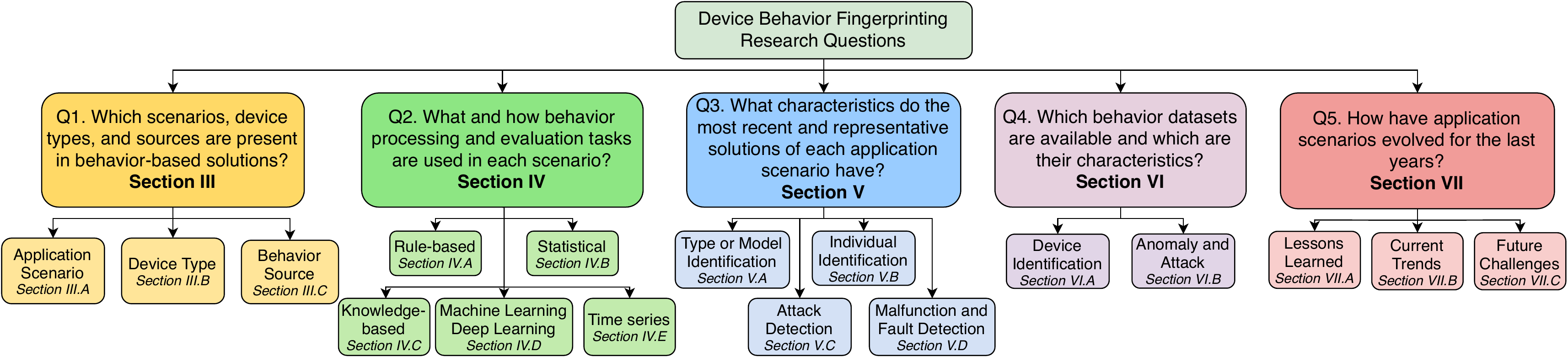}
    \caption{\addtxt{Discussed questions per article section.}}
    \label{fig:organization_scheme}
\end{figure*}

\addtxt{These research questions are closely related to each other and draw a complete picture of the existing challenges in device behavior analysis for identification and attack and malfunctioning detection. \textit{Q1} and \textit{Q2} deal with devices, data sources, and techniques used for device fingerprinting. \textit{Q3} and \textit{Q4} refer to current publications and datasets of device behavior --the key aspects of this survey and core sections of the document. While \textit{Q5} focuses on the consequences of the research done so far and its future.} \figurename~\ref{fig:organization_scheme} shows where and how the previous questions are addressed in the article at hand\addtxt{, acting as table of contents}.

To answer the previous questions and provide readers with an up-to-date vision of device behavior fingerprinting, the main contributions of this manuscript are: 

\begin{itemize}
    \item An analysis of the behavior data sources and device types utilized in the literature, paying attention to the application scenarios in which each source is contemplated (\addtxtRev{answering} \textit{Q1} \addtxtRev{in Section III}).
    \item A description and comparison of the main techniques and algorithms utilized to model and evaluate device behavior based on the morphology of the available data (\addtxtRev{answering} \textit{Q2} \addtxtRev{in Section IV}).
    \item A comprehensive review and comparison of the characteristics, advantages, and limitations of the most relevant proposals that consider device behavior to 1) identify device models or types, 2) identify individual devices, 3) detect cyberattacks, and 4) detect device/system functioning faults (\addtxtRev{answering} \textit{Q3} \addtxtRev{in Section V}).
    \item A description of the principal public datasets containing device activity and behavior. This description is divided into datasets designed for device identification and for attack or behavior anomaly detection (\addtxtRev{answering} \textit{Q4} \addtxtRev{in Section VI}).
    \item A set of lessons learned, current trends, and future challenges drawn from the device behavior works and datasets reviewed (\addtxtRev{answering} \textit{Q5} \addtxtRev{in Section VII}).
\end{itemize}

The remainder of this article is organized as follows. 
Section \ref{sec:sources} gives an analysis of device types, application scenarios, and behavior sources. Section \ref{sec:processing} reviews the main approaches and algorithms utilized to process behavioral data. Section \ref{sec:solutions} describes and compares the main solutions found in the state-of-the-art. Section \ref{sec:datasets} examines the main public datasets containing device activities. Section \ref{sec:lessons} draws a set of lessons learned, current trends, and future challenges in the research area. Finally, Section \ref{sec:conclusions} provides an insight into the conclusions extracted from the present work.

\section{Behavior Characterization Analysis}
\label{sec:sources}


With the goal of answering \textit{Q1 (Which scenarios, device types, and sources are present in behavior-based solutions?)}, this section studies the most used and promising scenarios where device behavior has been considered: device identification and misbehavior detection. After that, and aligned with these scenarios, it analyzes the main device types from which behavioral data is obtained, and the most common behavior dimensions and characteristics considered by device fingerprint solutions existing in the literature.

\subsection{Application Scenario}

According to the heterogeneous capabilities of device behavior fingerprinting, the literature has applied it in a wide variety of scenarios with different objectives. After reviewing the state-of-the-art, we highlight the following two categories as the most used and well-known: \textit{Device identification} and \textit{Misbehavior detection}.

\subsubsection{\addtxt{Device identification}}
\addtxt{It} uses the behavior of devices to identify them and their characteristics. This task can be performed from the following two perspectives.

    \textbf{Device type or model identification}. Device type identification \cite{Marchal2019Audi,Miettinen2017IoTSentinel} aims to recognize the device category such as general computer, IoT sensor, or embedded device, among others. In contrast, device model identification \cite{Ortiz2019DeviceMien,oser2018identifying} aims to differentiate between devices of the same type but different hardware and software configurations. 
    
    \textbf{Individual device identification} \cite{Jafari_Fingerprinting_Deep_Learning_2018,dong2019cpg} distinguishes between devices with identical hardware and software capabilities. This approach requires the lower level data, usually related to hardware variations during fabrication. Although device activity can also be employed to model user behavior and perform user's identification and authentication \cite{majma2017systematic,jorquera2018improving,snchez2020authcode}, user inputs and activity monitoring fall out of the scope of this study, which is focused only on device behavior analysis, without human interaction.
    

\subsubsection{\addtxt{Misbehavior detection}} \addtxt{It} seeks to identify anomalous situations based on changes in normal device behaviors. The anomalous situations are very varied; therefore, the solutions trying to recognize these situations are also heterogeneous. The next two main families of behavior anomaly detection solutions can be found in the literature.

    \textbf{Attack detection} \cite{Haefner2019CompleIoT,ali2020towards,li2019mad,Attia2015AD} intends to detect anomalies, created by cyber threats, according to the previously known normal device behavior. These solutions are commonly deployed as an IDS based on device behavior, being either Network-based (NIDS) or Host-based (HIDS). The cyberattacks detected using behavior are very diverse and depend on the monitored dimensions. These can range from impersonation and spoofing to malware execution.
    
    \textbf{Malfunction and fault detection} \cite{manco2017fault,WANG201889,samir2020detecting} tries to identify devices that are not functioning correctly because some component or service is failing. The malfunctioning could be caused by faults such as damaged hardware, a service or hardware overload, or network issues. Solutions addressing this approach assume that the fault will somehow affect the general device behavior. 

\subsection{Device Type}

Device activities, properties, and interactions can be monitored in an exhaustive range of heterogeneous devices and systems. Then, behavioral patterns can be built with diverse goals by almost any device. However, the data collection process is different depending on factors such as device hardware and software. At this point, it is important to describe the principal device and system categories used in the previous application scenarios.

    \textbf{Personal computers.} This category includes computers commonly found in homes and workplaces \cite{robinson1999personal}. We can differentiate two main kinds of personal computers, desktop devices and laptops, differentiated by power supply.
    
    \textbf{Mobile devices.} Smartphones and tablets are grouped in this category. Mobile devices are mainly constrained by battery.
 
    \textbf{Embedded systems.} These low-cost systems are designed and built to perform very specific tasks and their functionality is usually limited \change{due to}{by} processing and energy constraints \cite{jabeen2016survey}.
    
    \textbf{Industrial Control Systems (ICS)}. This family groups devices and systems that supervise and control critical services of industrial processes \cite{holm2015survey}, involving sensors and actuators. ICSs are usually deployed as supervisory control and data acquisition (SCADA) systems \cite{gomez2002survey}. 
    
    \textbf{IoT devices}. Any system with processing power and connected to the Internet can be considered as an IoT device. Typically, the IoT device concept is associated with embedded systems with connectivity capabilities such as sensors and smart-home objects\rmvtxt{, among others [38]}. \addtxt{However, it covers a wider variety of devices \cite{jabeen2016survey}, including drones, or wearable devices, among others.}
    
    \textbf{Cloud systems.} They provide the following three principal service models, in which resources can be accessed remotely and through network \cite{rittinghouse2016cloud}: Infrastructure as a Service (IaaS), Platform as a Service (PaaS), and Software as a Service (SaaS). In the last years, Cloud paradigm has evolved towards Fog \cite{osanaiye2017cloud} and Edge Computing \cite{shi2016edge}, where cloud systems are deployed closer to end-user devices, reducing latency and speeding up computations.
    
    \textbf{SDN/NFV systems.} SDN and NFV are concepts that usually appear together, although they can also be utilized separately \cite{bonfim2019integrated}. The Software Defined Networking (SDN) paradigm \cite{shin2012software} is a network architecture where network control is decoupled from the data plane, having a centralized controller managing the traffic flows and enabling network programmability and abstraction. Network Function Virtualization (NFV) paradigm \cite{mijumbi2015network} is a network architecture where network devices are vitalized using software implementations.
    
    \textbf{Containers and microservices.} Containers are software packages that include an application code and all its dependencies, allowing a lightweight deployment. Microservices \cite{pahl2016microservices} are applications with a single fixed function, commonly deployed as containers. Several microservices can be combined to build more complex applications distributedly.
    
    \textbf{Clusters.} A cluster is a set of computers, typically Linux devices \cite{vrenios2002linux}, connected closely to combine their resources and work as a single system. Then, the cluster behavior will be defined by the behavior of its components.


\subsection{Behavior Source}

Once the most representative application scenarios and devices have been explained, it is necessary to describe the behavior sources found in the literature, their pros and cons, and the solutions using each source. This description has been structured by following the next two main categories considered in the literature: \textit{externally-collected behavior sources} and \textit{in-device behavior sources}. Finally, the key aspects of the behavior data considered by each solution are compared.

\subsubsection{Externally-collected behavior sources} 

In this category, an external device is used to monitor the device behavior. Concretely, network communications and emitted electromagnetic signals are the main externally-collected sources used to model devices behavior. In the case of network-based data, data is usually collected by a proxy or a gateway, while electromagnetic signal-based data is collected by a sensor through an antenna.

\textbf{Network communications.} From the network communications perspective, a diverse set of behavioral features can be extracted by monitoring network packets. They depend on the granularity of the traffic inspection and the TCP/IP layers gathered. The main advantage of this dimension is its universality, as almost any device has network interfaces, and the possibility of monitoring many devices from a single gateway. As drawbacks, this dimension can suffer impersonation attacks and encryption makes data analysis more difficult. In this context, some solutions only focus on the amount of data sent/received and the IPs to which the device is connected \cite{Ferrando2017Network,Schmidt2018ADArima}. Other solutions also perform packet header and flow statistics analysis \cite{Miettinen2017IoTSentinel,OConnor2019HomeSnitch}. And finally, other solutions also include data related to transport or application layer protocols or payload data \cite{Thangavelu2018DEFT,Bezawada2018IoTSense}. \addtxt{Generally, payload data is protected using encryption methods, so the majority of solutions utilize header and flow-based data. However, some works focus on encrypted communication analysis for fingerprinting \cite{msadek2019iot,dai2019ssl}}. From the application usage point of view, this category is utilized for device model identification \cite{Sivanathan1,OConnor2019HomeSnitch}, device type identification \cite{selis2018classification,Marchal2019Audi}, \addtxt{\cite{hafeez2020iot}}, attack detection \cite{Blaise2020BotFP,Nguyen2019DioT,Haefner2019CompleIoT} and fault detection \cite{Spanos_IoT_anomaly_identification_2019}.

\textbf{Clock Skew}. Based on crystal oscillator imperfections that occurred during the manufacturing process, internal clock counters of different devices have slight variations. In this sense, it is possible to utilize this characteristic to differentiate devices based on their hardware behavior. The main advantage of this source is that it can be collected from outside the device. As drawback, clock skew distribution concentrates around 0, so this source cannot be applied as a unique source in large device deployments \cite{polvcak2014reliability}. Clock skew can be calculated by observing how internal device timestamps vary in time, mainly using TCP and ICMP timestamps \cite{kohno2005remote} and Wi-Fi beacon timestamps \cite{jana2009skew,lanze2012skew}, so it can be seen as a special category of network-based data. From the application perspective, clock skew has been utilized for individual device identification \cite{jana2009skew,lanze2012skew,sharma2012skew,polcak2015clock}.

\textbf{Electromagnetic signals.} This category relies on the behavior of electromagnetic signals emitted by each device. Its main advantage is the difficulty of tampering it, as it depends on emitted signal properties. In terms of disadvantages, we highlight that the data gathering process must be physically close to the monitored device, since electromagnetic signals lose intensity as the distance to the transmitter increases. Radio signals are used in the literature to distinguish \addtxt{drone models \cite{ezuma2019detection,al2019rf,allahham2020deep,basak2020drone} and to identify} physical devices \cite{Jafari_Fingerprinting_Deep_Learning_2018,Riyaz2018Radio}. However, although radio signals have been utilized to detect anomalies in the radio spectrum \cite{Rajendran2019Unsupervised}, no solution specifically focused on device behavior anomaly detection using radio signals has been found. Following a similar approach, other solutions utilize the electromagnetic signals radiated from the device components to identify physical devices \cite{Cheng2019DemiCPU}.

\tablename~\ref{tab:signal_features} compares the main characteristics of externally-collected data. As observed, features related to network communications are used both for device identification and misbehavior detection, as this source is very heterogeneous. In contrast, clock skew and electromagnetic-based features are only applied in device identification, as they are lower-level sources related to device component characteristics. 

\begin{table}[ht!]
\scriptsize
\centering
\begin{tabular}{ >{\Centering}m{1.5cm}   >{\Centering}m{1.7cm} >{\Centering}m{2.3cm} *{3}{c}}
 \hline
 \multirowcell{2}{\textbf{Feature}}&\multirowcell{1}{\textbf{\makecell[c]{Behavior\\Source}}} & \multirowcell{2}{\textbf{Device Type}} &   \multicolumn{3}{c}{\textbf{Application Scenario}} \\
 \cline{4-6}
 & & & \makecell{\textbf{DI}} & \makecell{\textbf{MD}} \\
 \hline
 Packet headers statistics & Network Communications  & Computers, IoT devices, ICS & \makecell[c]{\cite{selis2018classification} \cite{Miettinen2017IoTSentinel}\\ \cite{Ortiz2019DeviceMien} \cite{oser2018identifying} \\ \cite{formby2016s} \cite{msadek2019iot}}  & \makecell[c]{\cite{Sivanathan2} \cite{Nguyen2019DioT} \\ \cite{Blaise2020BotFP} \cite{Amouri2018NIDS} \\ \cite{Ferrando2017Network} \cite{lima2019smart} \\} \\ 
 \hline
 Network flows statistics & Network Communications & Computers, IoT devices, ICS & \makecell[c]{ \cite{radhakrishnan2014gtid} \cite{Marchal2019Audi} \\ \cite{Sivanathan1} \cite{Shahid2018IoT} \\ \cite{kotak2020iot}} & \makecell[c]{\cite{Haefner2019CompleIoT} \cite{hamza2019detecting}\\ \cite{ali2020towards} \cite{Hamad_IoT_Identification_Network_2019} \\\cite{meidan2017detection} \cite{Spanos_IoT_anomaly_identification_2019} \\ \cite{carvalho2018ecosystem} \cite{afek2019nfvbased} \\ \cite{radford2018sequence} \cite{yin2018enhancing} \\ \cite{Marir2018Distributed} \addtxt{\cite{Trimananda2020Pingpong}}} \\ 
 \hline
 Packet payload data and statistics & Network Communications & Computers, IoT devices, SDN, ICS & 
 \makecell[c]{\cite{Bezawada2018IoTSense} \cite{OConnor2019HomeSnitch} \\\cite{Thangavelu2018DEFT} \addtxt{\cite{perdisci2020iotfinder}}}& \makecell[c]{\cite{choi2018detecting} \cite{yu2019radar} \\ \cite{HASAN2019100059} \cite{dai2019ssl} \\ \cite{pacheco2018anomaly}} \\ 
 \hline
 Clock drift in time & Clock Skew & Computers, mobile and IoT devices, ICSs  & \makecell[c]{\cite{jana2009skew}\cite{lanze2012skew}\\\cite{sharma2012skew}\cite{polcak2015clock}} & \notick \\ 
 \hline
 Raw IQ samples & Electromagnetic signals & Computers, mobile and IoT devices, ICSs & \makecell[c]{\addtxt{\cite{ezuma2019detection}\cite{al2019rf}}\\\cite{Jafari_Fingerprinting_Deep_Learning_2018} \cite{Riyaz2018Radio}\\\addtxt{\cite{allahham2020deep} \cite{basak2020drone}}} & \notick \\
 \hline
 Signal frequency & Electromagnetic signals & Computers, mobile and IoT devices, ICSs & \cite{Cheng2019DemiCPU} & \notick \\
 \hline
 \end{tabular}
\caption{Externally-collected behavior characteristics. (DI: Device Identification. MD: Misbehavior Detection.)}
\label{tab:signal_features}
\end{table}

\subsubsection{In-device behavior sources} 

In this category, behavioral data monitoring is performed on the target devices. Thus, lower-level data related to the device internal functioning can be collected. This approach has the advantage of not requiring a connection to an external monitoring device. In contrast, as a drawback, if the device suffers an anomaly, such as an attack, the monitoring solution may suffer it as well.

\textbf{Hardware Events}. Hardware Performance Counters (HPC) are special registers built into modern microprocessors that store hardware-related event counters. The main advantage of this category is the precision achieved to model the device operation from a low-level perspective. In contrast, the quantity and morphology of the HPCs depend on the device CPU model, which makes it difficult to build general solutions. In the literature, some solutions \cite{wang2015confirm,Ott2019CA_HPC,Golomb2018CIoTACI} utilize HPCs to model software behavior and detect abnormal operations. In addition, \cite{Ott2019CA_HPC} also utilizes HPCs to identify and authenticate different devices.

\textbf{System processors and oscillators.} Some devices have hardware components that include a crystal oscillator. As in clock skew, the manufacturing imperfections of these components can be utilized to differentiate physical devices by comparing their counters drift in time. The main advantage of this source is its low-level, which enables to differentiate devices with the same software and hardware. However, the device should include hardware using oscillators, something unusual in resource-constrained devices. Moreover, manufacturing errors are usually small \cite{polvcak2014reliability}. In the literature, two components used for this purpose are the Real Time Clock (RTC) and the Digital Signal Processor (DSP) \cite{salo2007multi}. In addition, the time it takes \rmvtxt{the CPU }to execute a particular code or function can also be used to model system behavior. In this case, this data has been used to identify device models and the devices themselves \cite{sanchez2018clock}.

\textbf{Resource Usage}. In this category, different device components usage and status are monitored. Commonly, the monitored components are CPU, memory, disk, and network. Various parameters can be extracted from each component, such as usage percentage or input/output statistics. In terms of advantages, this source is quite general and can be monitored in many devices and systems. As drawback, continuous resource usage monitoring consumes many resources. In the literature, this data is utilized to identify devices \cite{dong2019cpg} and detect behavior anomalies caused by cyberattacks \cite{barbhuiya2018rads} or system malfunctioning \cite{GULENKO2016AD_NFV,WANG201889,Schmidt2018ADArima}.

\textbf{Software and Processes.} The software deployed in a device or system also has its particular behavior. Then, the conjunction of the isolated software behaviors can be utilized to model a global device behavior fingerprint. As advantage, software monitoring can accurately model normal device behavior. However, this source is affected by system updates and legitimate software modifications. Software can be modeled in several ways:
\begin{itemize}
    \item \textbf{System calls and logs}. They serve to monitor the interactions between the programs running on a device and its operating system. These interactions encompass process, file, and communication management operations. From the application usage point of view, system call sequences and logs have been used to characterize device behavior and detect anomalies \cite{Attia2015AD,creech2013semantic,deshpande2018hids,Mishra2020VMGuard,liu2020statistical,Nedelkoski2019ADLSTM,Kubacki2019anomalies}.

    \item \textbf{Process properties.} Device software behavior can be modeled by monitoring each process properties, such as name, status, or threads. This category also includes the resources utilized to execute a particular program or code. In the literature, this category is commonly monitored together with resource usage or system calls to detect anomalous behaviors \cite{haider2017generating}.

    \item \textbf{Software signatures}. Software snapshots (signatures) are generated for the different device executable and their configuration files using hashing algorithms. Then, the snapshots are used to detect software modifications that cause behavior anomalies \cite{samir2020detecting,He2020BoSMoS}.
\end{itemize}

\textbf{Device Sensors and Actuators.} The data collected in this dimension is very diverse and depends on the device and scenario typology. The main advantage of this source is that it can also detect environment failures or attacks. As drawback, environment knowledge is required to analyze and understand the data from this dimension, as each device may have different sensors and actuators. From the application usage point of view, sensor and actuator measurements are utilized to detect anomalies \cite{li2019mad,manco2017fault,pacheco2018anomaly,zhanwei2019abnormal,Neha2020SCADA} and model device types \cite{Sivanathan1}, while sensor hardware information is used to physically identify the devices \cite{ahmed2017hardware}.

\begin{figure}[ht!]
    \centering
    \includegraphics[width=\columnwidth]{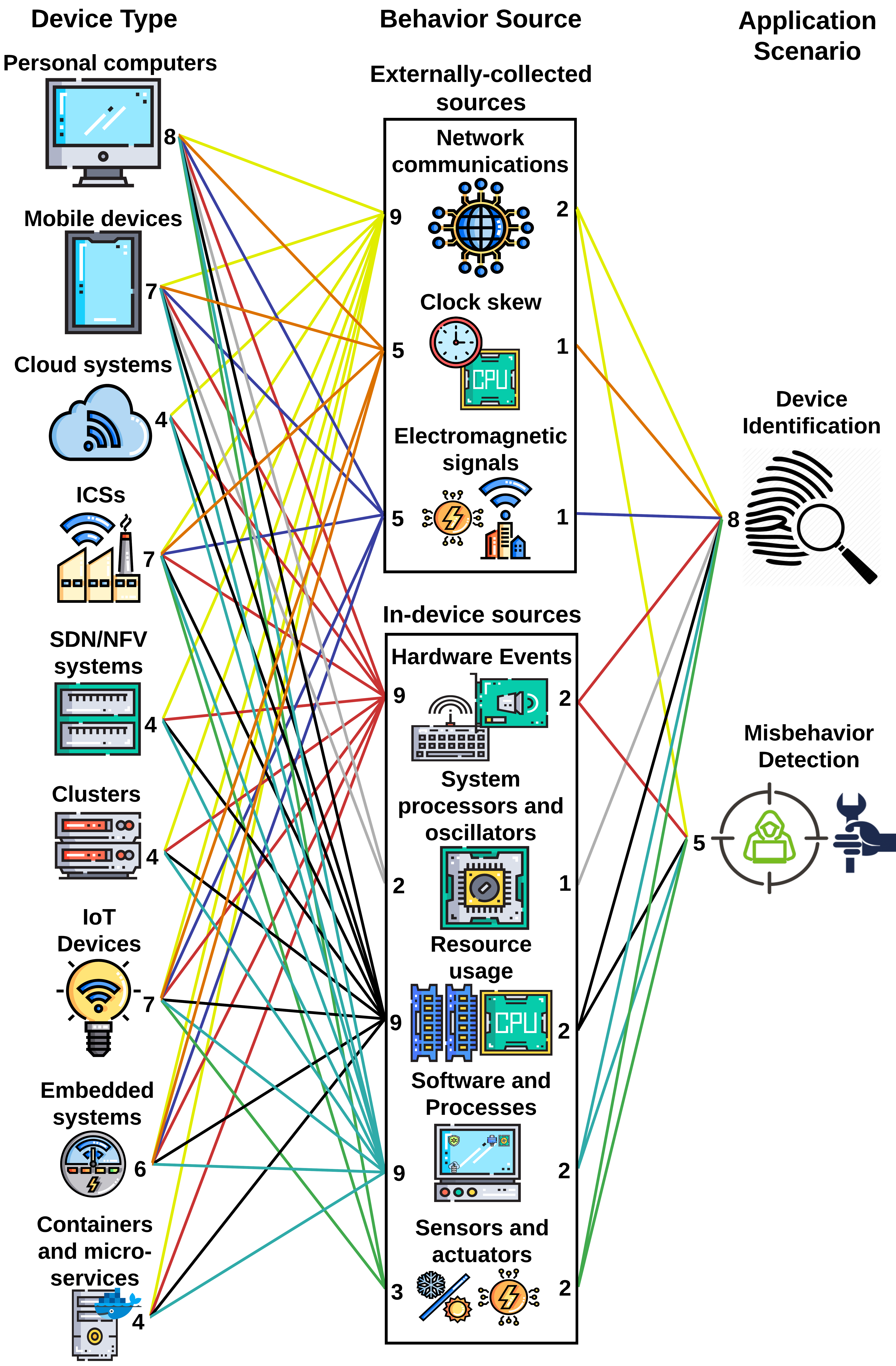}
    \caption{Behavior sources available in each device type and application scenarios. (The numbers shown for each item indicate its total number of connections.)}
    \label{fig:sources_devices}
\end{figure}

To conclude, on the one hand, \tablename~\ref{tab:in_device_features} compares the main characteristics of data directly collected from the modeled device. It can be appreciated how HPCs, CPU percentage, system calls, software signatures, and sensor values are used both for device identification and misbehavior detection. Besides, low-level information related to the system processors and sensor hardware is only employed for device identification. Finally, features related to resource usage and process properties are only employed in misbehavior detection. On the other hand, \figurename~\ref{fig:sources_devices} shows the behavior sources considered by each device type, and in which application scenario these sources are utilized. The numbers indicate the total number of connections each element has. It can be appreciated that the most extended sources, based on their generality, are network communications, hardware events, resource usage, and software and processes.

\begin{table}[ht!]
\scriptsize
\centering
\begin{tabular}{ >{\Centering}m{1.6cm}   >{\Centering}m{1.9cm} >{\Centering}m{2cm} *{3}{c}}
 \hline
 \multirowcell{2}{\textbf{Feature}}&\multirowcell{1}{\textbf{\makecell[c]{Behavior\\Source}}} & \multirowcell{2}{\textbf{Device Type}} &   \multicolumn{3}{c}{\textbf{Application Scenario}} \\
 \cline{4-6}
 & & & \makecell{\textbf{DI}} & \makecell{\textbf{MD}} \\
  \hline
  HPC & Hardware Events & Embedded systems, IoT devices & \hspace{0.2cm}\cite{Ott2019CA_HPC}\hspace{0.2cm}  & \makecell[c]{\cite{wang2015confirm} \cite{Ott2019CA_HPC} \\ \cite{Golomb2018CIoTACI}} \\
 \hline
  RTC drift & System processors and oscillators & Computers & \cite{salo2007multi} & \notick \\
  \hline
  DSP performance & System processors and oscillators & Computers & \cite{salo2007multi} & \notick \\
   \hline
 Code execution time & System processors and oscillators & Computers & \cite{sanchez2018clock} & \notick \\
  \hline
 CPU usage percentage & Resource Usage & Computers, embedded devices, microservices, cloud, NFV, and cluster systems & \cite{dong2019cpg} & \makecell[c]{ \cite{barbhuiya2018rads} \cite{WANG201889} \\ \cite{Agrawal2017AD_CC} \cite{ravichandiran2018anomaly} \\ \cite{GULENKO2016AD_NFV} \cite{Schmidt2018ADArima} \\ \cite{gulenko2018detecting} \cite{sorkunlu2017tracking} \\ \cite{samir2020detecting} \cite{du2018anomaly}}\\
 \hline
 CPU activity & Resource Usage & Microservices, NFV, cloud, and cluster systems & \notick & \makecell[c]{\cite{WANG201889} \cite{Agrawal2017AD_CC}\\ \cite{GULENKO2016AD_NFV} \cite{sorkunlu2017tracking} \\ \cite{shone2013misbehaviour}}\\
 \hline
 System storage usage & Resource Usage & Microservices, NFV, cloud, and cluster systems & \notick & \makecell[c]{\cite{WANG201889} \cite{Agrawal2017AD_CC} \\ \cite{GULENKO2016AD_NFV} \cite{gulenko2018detecting}}\\
 \hline
 System memory usage & Resource Usage & Microservices, NFV, cloud, and cluster systems & \notick & \makecell[c]{\cite{WANG201889} \cite{Agrawal2017AD_CC}\\ \cite{GULENKO2016AD_NFV} \cite{Schmidt2018ADArima} \\ \cite{gulenko2018detecting} \cite{sorkunlu2017tracking} \\ \cite{samir2020detecting} \cite{du2018anomaly} \\ \cite{shone2013misbehaviour}}\\
 \hline
 I/O throughput per network interface & Resource Usage & Microservices, NFV, cloud, and cluster systems & \notick & \makecell[c]{\cite{barbhuiya2018rads} \cite{WANG201889} \\\cite{GULENKO2016AD_NFV} \cite{Schmidt2018ADArima} \\ \cite{gulenko2018detecting} \cite{sorkunlu2017tracking} \\ \cite{du2018anomaly}} \\
 \hline
 System calls and logs & Software and Processes & Computers, resource-constrained devices, cloud and NFV systems & \makecell[c]{\cite{Attia2015AD}} & \makecell[c]{\cite{Attia2015AD}\cite{creech2013semantic}\\\cite{deshpande2018hids} \cite{Kubacki2019anomalies}\\\cite{Mishra2020VMGuard}\cite{Nedelkoski2019ADLSTM}\\\cite{liu2020statistical}}\\
 \hline
 Process properties & Software and Processes & Computers & \notick & \makecell[c]{\cite{haider2017generating} \cite{hoang2009program}\\\cite{shone2013misbehaviour}}\\
 \hline
  Software signatures & Software and Processes & IoT devices & \cite{He2020BoSMoS} & \cite{samir2020detecting} \cite{He2020BoSMoS} \\
 \hline
 Sensor measurements values & Device Sensors and Actuators & ICS\addtxt{, IoT devices} & \cite{Sivanathan1} & \makecell[c]{\cite{li2019mad}\cite{manco2017fault}\addtxt{\cite{pacheco2018anomaly}}\\\cite{zhanwei2019abnormal}\cite{Neha2020SCADA}}\\
 \hline
 Sensor hardware properties & Device Sensors and Actuators & ICS & \cite{ahmed2017hardware} & \notick\\
 \hline
\end{tabular}
\caption{In-device behavior characteristics. (DI: Device Identification. MD: Misbehavior Detection.)}
\label{tab:in_device_features}
\end{table}

\section{Behavior Processing and Evaluation Techniques}
\label{sec:processing}

Once reviewed the behavioral data monitored per type of device and application scenario, the data needs to be processed to create a fingerprint. This section deals with \textit{Q2 (What and how behavior processing and evaluation tasks are used in each scenario?)} by detailing the algorithms and techniques commonly used in the literature to create and evaluate fingerprinting profiles, highlighting their main advantages and drawbacks. The existing techniques are categorized in the following five groups: \textit{rule-based, statistical, knowledge-based, Machine Learning and Deep Learning, and time series approaches}. The previous categories are not mutually exclusive and a particular solution can belong to several categories. Furthermore, the behavior processing can be centralized, in the own device or a server, or distributed using technologies such as blockchain \cite{zheng2018blockchain}, distributed \cite{verbraeken2019survey} or federated learning \cite{yang2019federated}, among others.

\subsection{Rule-based}
This is the most straightforward approach to create behavioral profiles. It is useful for devices with a well-known behavior and a reduced set of actions. In this approach, a set of rules defines how the system should behave, that is, its behavioral fingerprint. Rules can be defined statically, based on pre-defined actions, or dynamically, based on the historical actions performed by the device. Any deviation from these rules is considered a fault or anomaly. The main advantages of this approach are its speed and simplicity. As drawbacks, it requires previous knowledge about the device behavior, and it is not suitable for changing and complex scenarios. Rule-based evaluation is utilized for device type or model definition and anomaly detection.

For device behavior evaluation, a recent approach is the usage of Manufacturer Usage Descriptions (MUDs) standard \cite{RFC8520} files, which define the normal device functioning and are commonly issued by vendors. This method is mainly utilized for IoT behavior fingerprint generation and evaluation \cite{afek2019nfvbased,hamza2019detecting}. Another rule-based approach is to explicitly define the software that the device can execute \cite{He2020BoSMoS} or thresholds for resource usage \cite{Alcarria2017rule}.

\subsection{Statistical}

In this approach, relatively basic statistical data processing techniques are utilized to extract inferences (properties) from data samples. This approach is usually considered in data pre-processing and anomaly detection. The main advantage of this approach is its simplicity and that these algorithms do not require large datasets. However, it does not handle well multi-dimensional data, and consistent evaluation decisions require previous knowledge in the area.

For pre-processing, it is common to infer features using statistical functions such as average, standard deviation, quartiles, maximum, or minimum, among others. Regarding evaluation, in some solutions \cite{barbhuiya2018rads}, the interquartile range (IQR) is used as a statistical measure representing the presence of outliers and anomalies based on data variability (dispersion). In the same line, Euclidean Distance is used by some approaches \cite{Spanos_IoT_anomaly_identification_2019, pacheco2018anomaly, Ferrando2017Network} to determine anomaly values based on the distance between two data measurements. Finally, some works \cite{jana2009skew,manco2017fault} utilize \textit{Expectation Maximization} algorithm for clustering and parameter estimation based on statistically-inferred latent variables.

\subsection{Knowledge-based}

This approach aims to represent knowledge extracted from received data and build a reasoning system capable of inferring new knowledge. Commonly, the knowledge is built based on a set of ontologies, and the decision-making process is based on if-then derivation rules. The main advantages of this approach are the explainability of the inferred solutions and that it can solve problems involving incomplete data. As drawbacks, this approach takes longer time, and it has reduced scalability, as the system could become too complex if large amounts of data are utilized.

Knowledge-based approaches are utilized mainly for behavioral anomaly detection, being the main ones look-ahead algorithms and finite state machines. \textit{Look-ahead algorithms} are commonly combined or used to make decisions in more complicated approaches, such as state machines. Furthermore, these algorithms are also directly used to detect anomalies \cite{Attia2015AD}. \textit{Finite state machines}, such as \textit{Markov Models} \cite{gagniuc2017markov} and \textit{n-gram models}  \cite{wressnegger2013close}, describe the sequential logic followed by a certain entity and predict its future status based on the previous ones. In the literature, they are widely applied for behavior anomaly detection \cite{Attia2015AD,samir2020detecting,hamza2019detecting,Golomb2018CIoTACI}.

\subsection{Machine Learning and Deep Learning}

In recent years, and based on the increase of processing power and available data, Machine Learning (ML) \cite{alpaydin2020introduction} and Deep Learning (DL) \cite{lecun2015deep} algorithms have gained enormous relevance in almost every industrial or research area\addtxt{, becoming the dominating trend for data processing and evaluation}. The main advantages of ML/DL based approaches are their capacity to detect complex data patterns, handle multi-dimensional and multi-variate data, and adapt themselves to dynamic and heterogeneous scenarios using massive data. As disadvantages, the model decisions are usually hardly explainable, based on the black-box nature of the generated models. Besides, these algorithms, especially in DL, require large amounts of data to be trained, and the algorithm training can take much time and resources. Also, most algorithms require parameter tuning, which implies repeating the training process several times. Since ML and DL techniques are very diverse, they have been widely used for device behavior fingerprint generation and evaluation, both for device identification \cite{Sivanathan1,OConnor2019HomeSnitch,selis2018classification,Bezawada2018IoTSense,kotak2020iot,radhakrishnan2014gtid,Jafari_Fingerprinting_Deep_Learning_2018,Riyaz2018Radio,Cheng2019DemiCPU} and misbehavior detection \cite{Blaise2020BotFP,Nguyen2019DioT,Sivanathan2,Amouri2018NIDS,Hamad_IoT_Identification_Network_2019,meidan2017detection,HASAN2019100059,Nedelkoski2019ADLSTM}.

According to the morphology of the data they receive and the type of predictions they make, ML/DL algorithms applied in behavior analysis are distinguished into two main categories: Supervised Learning and Unsupervised Learning.

The goal of \textit{Supervised learning} is to infer a model capable of predicting the output of data vectors based on training labeled data \cite{alpaydin2020introduction}. Supervised algorithms are mainly divided into classification and regression techniques.

\begin{itemize}
    \item \textit{Classification} algorithms try, based on the training data, to predict the class to which unseen data vectors belong. Additionally, anomaly detection can be performed using classification algorithms by labeling the data as normal/anomaly. Common ML classification algorithms are \textit{Decision Tree (DT)} \cite{safavian1991survey}, \textit{Random Forest (RF)} \cite{liaw2002classification}, \textit{Logistic Regression (LR)} \cite{kleinbaum2002logistic}, \textit{Naive Bayes (NB)} \cite{friedman1997bayesian} or \textit{Support Vector Machine (SVM)} \cite{steinwart2008support}. These algorithms are widely utilized for behavior evaluation in device identification \cite{selis2018classification,Marchal2019Audi,Miettinen2017IoTSentinel,Sivanathan1,Bezawada2018IoTSense,Shahid2018IoT,OConnor2019HomeSnitch,oser2018identifying,formby2016s,Thangavelu2018DEFT,Cheng2019DemiCPU} and behavioral anomaly recognition \cite{deshpande2018hids,du2018anomaly,Hamad_IoT_Identification_Network_2019,meidan2017detection,Amouri2018NIDS,HASAN2019100059,carvalho2018ecosystem,lima2019smart,Marir2018Distributed}.
    
    \item Regarding \textit{Regression} algorithms, the output is a continuous number and not a class\rmvtxt{, like in classification techniques}. Usual ML regression algorithms are \textit{Linear and Polynomial Regression} \cite{weisberg2005applied}, which are applied in behavior analysis to evaluate device behavior and its fluctuation \cite{Amouri2018NIDS}.
\end{itemize}

In \textit{Unsupervised learning} \cite{alpaydin2020introduction}, data vectors are not labeled, so feature vectors only contain input data. This kind of algorithm is used to extract patterns by modeling probability densities on the given data. The three main applications of Unsupervised learning are dimensionality reduction, clustering, and anomaly detection.

\begin{itemize}
    \item \textit{Dimensionality Reduction} algorithms aim to reduce the number of variables or features under consideration by obtaining a set of principal variables from the input data. In behavior-based solutions, \textit{Principal Component Analysis (PCA)} \cite{wold1987principal} and \textit{t-Distributed Stochastic Neighbor Embedding (t-SNE)} \cite{maaten2008visualizing} are utilized to speed up computations and derive new features \cite{Sivanathan2,Shahid2018IoT,hamza2019detecting}. Moreover, dimensionality reduction is combined with statistical algorithms for anomaly evaluation \cite{WANG201889,Spanos_IoT_anomaly_identification_2019,Agrawal2017AD_CC,sorkunlu2017tracking}.
    
    \item \textit{Clustering} algorithms have the objective of grouping the input vectors into a different set of objects based on their similarities. In device behavior fingerprinting, \textit{k-means} \cite{krishna1999genetic} and \textit{Density-based spatial clustering of applications with noise (DBSCAN)} \cite{tran2013revised} are usually applied to infer device classes or types \cite{Marchal2019Audi}, \addtxt{\cite{hafeez2020iot}}, \cite{Thangavelu2018DEFT,gulenko2018detecting,shone2013misbehaviour}.
    
    \item \textit{Anomaly Detection} algorithms seek to identify rare items, events, or observations based on a set of unlabeled data points and the assumption that most of the training data is normal. From this approach, \textit{One-Class SVM (OC-SVM)} \cite{li2003improving} and \textit{Isolation Forest (IF)} \cite{liu2008isolation} are widely used in the literature \cite{Haefner2019CompleIoT,ali2020towards,Perales2019AD_ICS}.
    
\end{itemize}

From a DL perspective, \textit{Artificial Neural Networks (ANN)} \cite{lecun2015deep} are frequently used in the above approaches. However, a type of architecture cannot be related to a specific use due to neural networks flexibility, as layers, neurons, and their connections can be organized in many ways depending on the problem to be solved. The main types of networks applied in behavior processing are: \textit{Multi-Layer Perceptrons (MLP)}, utilized for device identification \cite{oser2018identifying,kotak2020iot} and anomaly type classification \cite{carvalho2018ecosystem}; \textit{Autoencoders}, applied for behavior anomaly detection \cite{Ortiz2019DeviceMien} and dimensionality reduction purposes; \textit{Recurrent Neural Networks (RNN)}, such as \textit{Long Short-Term Memory networks (LSTM)} and \textit{Gated Recurrent Unit networks (GRU)}, applied from a time series perspective for device identification \cite{Ortiz2019DeviceMien,Jafari_Fingerprinting_Deep_Learning_2018} and behavior anomaly recognition \cite{Nguyen2019DioT,radford2018sequence,Nedelkoski2019ADLSTM,Neha2020SCADA,li2019mad}; and \textit{Convolutional Neural Networks (CNN)}, utilized for physical device identification based on signal processing from a time series approach \cite{Jafari_Fingerprinting_Deep_Learning_2018,Riyaz2018Radio}.

The previous network topologies can be combined to perform more complex tasks. For example, some solutions \cite{Ortiz2019DeviceMien} utilize LSTM layers to build an autoencoder, while other approaches \cite{yin2018enhancing} combine different neural networks to build \textit{Generative Adversarial Networks (GAN)} \cite{metz2016unrolled}.

\begin{table*}[bp]
\scriptsize
\centering
\begin{tabular}{  >{\Centering}m{1.8cm} >{\Centering}m{1.2cm}   >{\Centering}m{1.2cm} >{\Centering}m{1.5cm} >{\Centering}m{1.5cm} >{\Centering}m{1.5cm} >{\Centering}m{1.5cm} >{\Centering}m{1.5cm} >{\Centering}m{1.5cm} >{\Centering}m{1.5cm}}
 \hline
 \textbf{Approach} & \textbf{Simplicity} & \textbf{Expert knowledge required} & \textbf{Fast computation / Low resource} & \textbf{Large datasets required} & \textbf{Large training time} & \textbf{Multi-dimensional data} & \textbf{Decision explainability}& \textbf{Adaptability} & \textbf{Complex feature correlations}\\
 \hline
 Rule-based & \yestick & \yestick & \yestick & \notick & \notick & \notick & \yestick & Dynamic approaches & \notick \\
 \hline
 Statistical & \yestick & \yestick & \yestick & \notick & \notick & \notick & \notick & \notick & \notick \\
 \hline
 Knowledge-based & Partial & \notick & \notick & \notick & \notick & \notick & \yestick & \notick & Partial \\
 \hline
 \makecell[c]{ML/DL-based} & \notick & \notick & \notick & \makecell[c]{Mainly DL} & \makecell[c]{Mainly DL} & \yestick & Partial & \yestick & \yestick \\
 \hline
 Time series & \notick & \notick & \notick & \yestick &\yestick & ML/DL-based & \notick & ML/DL-based & ML/DL-based \\
 \hline
\end{tabular}
\caption{Behavioral processing approaches comparison.}
\label{tab:algorithm_comparison}
\end{table*}

\subsection{Time Series}


Time series analysis utilizes data measurements as a sequence of values where each measurement is related to the previous and the next ones. It includes a wide variety of algorithms and models, including the ones based on ML/DL or statistical algorithms. This approach is utilized both for device identification and anomaly detection, directly in the model generation or as data pre-processing. The main advantages of this approach are its improved performance over single-value processing approaches. However, it requires a large amount of data to detect the temporal patterns, and the processing is time-consuming.

Time series analysis methods are divided into two different types, \textit{frequency-based} methods, which analyze data as a signal with a certain frequency, and \textit{time-based} methods, which analyze data evolution with respect to time. 

In terms of frequency-based methods, \textit{Fourier Transform (FT)} \cite{bracewell1986fourier}, and derived functions, are applied as pre-processing to obtain the frequencies that form the value signal \cite{Ott2019CA_HPC,Marchal2019Audi}. From time-based methods, \textit{AutoRegressive Moving Average (ARMA)} and derived algorithms are used in behavior prediction applications \cite{Schmidt2018ADArima,Ferrando2017Network}. In addition, \textit{Dynamic Time Warping} algorithm is also utilized in device behavior evaluation \cite{dong2019cpg}, directly comparing the values of two time series.

Besides, as stated before, Deep Learning has been applied in behavioral data evaluation from a time series perspective utilizing RNNs \cite{Ortiz2019DeviceMien,Nguyen2019DioT,radford2018sequence,Nedelkoski2019ADLSTM,Neha2020SCADA,li2019mad} and CNNs \cite{Jafari_Fingerprinting_Deep_Learning_2018,Riyaz2018Radio}.

\begin{figure}[!ht]
    \centering
    \includegraphics[trim=0cm 0cm 0cm 0cm, clip=true,width=\columnwidth]{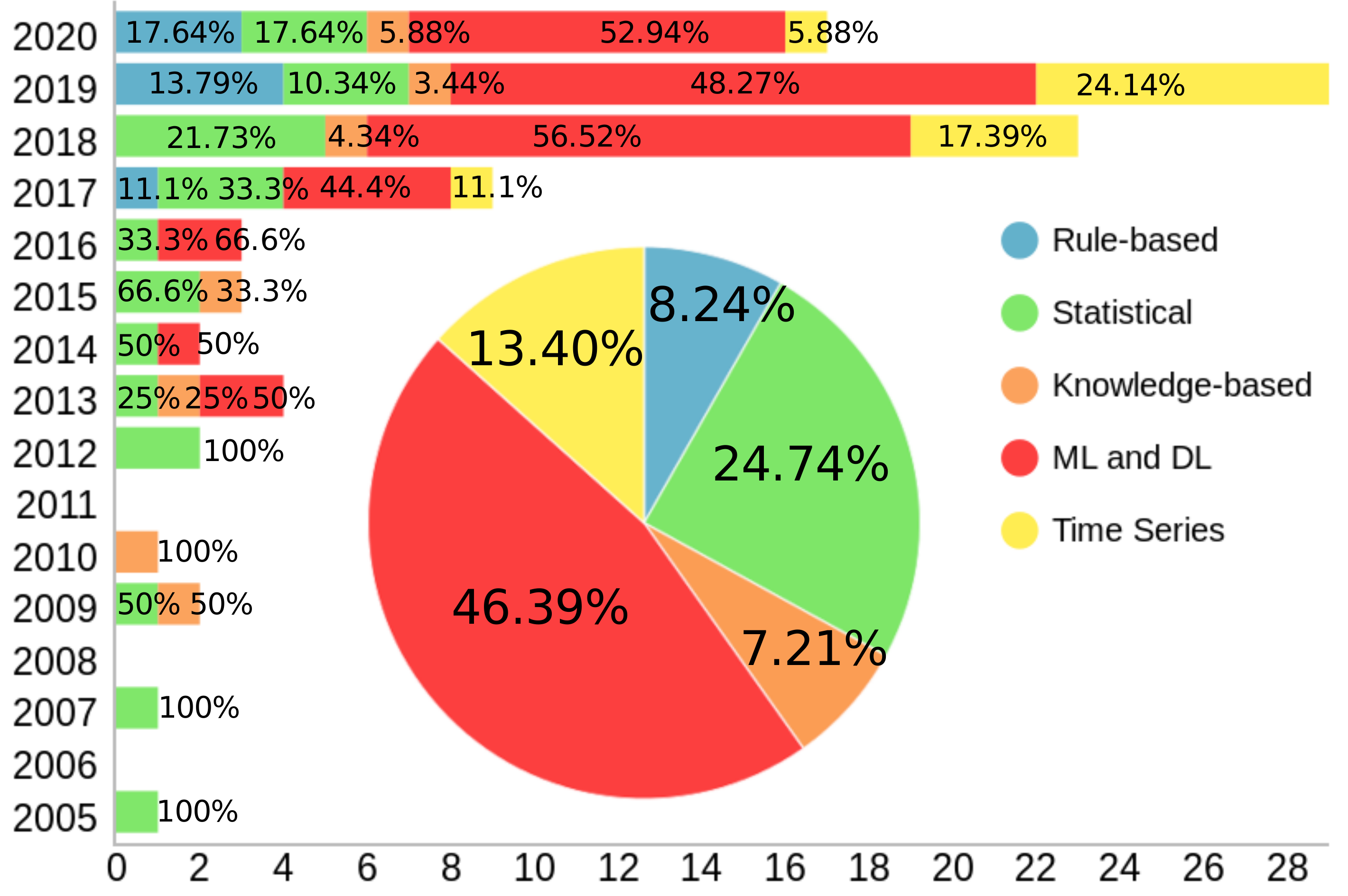}
    \caption{Yearly and global distribution of processing techniques used by device behavior fingerprinting solutions.}
    \label{fig:technique_year}
\end{figure}

\tablename~\ref{tab:algorithm_comparison} compares the main properties of the five behavior processing approaches identified in the literature analysis. As general conclusion, when the behavior of the device is composed of a limited and known number of actions and there is not a large number of dimensions in the data, the appropriate approaches would be those based on rules and statistical algorithms, given their reduced complexity and resource consumption. However, when the data features maintain complex relationships between them, the most suitable solutions are those based on knowledge and ML/DL approaches. Finally, when there is a relationship between the different measurements based on their order, a time series approach may provide improved results. Depending on the amount of data, the available resources, and the complexity of the feature correlations, some particular algorithms are better than others. For example, a simple IoT device, like a bulb, with a limited and known set of actions, can be modeled with a rule-based approach, leveraging its limited resources. In contrast, a cloud service that executes different tasks would be hard to model using rules, instead, an ML/DL-based approach exploiting the correlations in the sources available would be more successful. Overall, \figurename~\ref{fig:technique_year} shows the global and per year distribution of works using each technique, note that some works may utilize techniques belonging to more than one category. \addtxt{ML and DL rise as the leading group of processing techniques applied to device behavior fingerprinting, as it is already the main trend in the area and is still gaining even more prominence.}

Additionally, to properly evaluate and compare the solutions performance, it is critical to define relevant metrics. Then, independently of the evaluation approach followed, there is a set of common metrics utilized in the majority of behavior-based solutions. \tablename~\ref{tab:metrics} shows these common metrics. In the case of classification approaches, these metrics are based on the values present on a confusion matrix, while in the case of regression approaches, the metrics are based on prediction errors \cite{Sivanathan3,ravichandiran2018anomaly}. Moreover, some solutions also consider factors such as detection time or resource usage.

\begin{table}[ht!]
\scriptsize
\centering
\begin{tabular}{  >{\Centering}m{1.5cm} >{\Centering}m{3.6cm}   >{\Centering}m{2.9cm}}
 \hline
 \textbf{Metric name} & \textbf{Description} & \textbf{Equation} \\
 \hline
 Accuracy &  Total number of correct predictions over the total made & \(\displaystyle \frac{TP+TN}{TP+FP+TN+FN} \)  \\
 \hline
 Precision & Ratio of actual positives over all the elements predicted as positives & \(\displaystyle \frac{TP}{TP+FP} \)  \\
 \hline
  Recall, Sensitivity or True Positive Rate (TPR) & Proportion of actual positives correctly identified & \(\displaystyle \frac{TP}{TP+FN} \)  \\
 \hline
  Specificity or True Negative Rate (TNR) & Proportion of actual negatives correctly identified & \(\displaystyle \frac{TN}{FP+TN} \)  \\
 \hline
   False Positive Rate (FPR) or False Acceptance Rate (FAR) & Proportion of the elements wrongly determined as positive among the actual negatives & \(\displaystyle \frac{FP}{FP+TN} \)  \\
 \hline
   False Negative Rate (FNR) or False Rejection Rate (FRR) & Proportion of the elements wrongly determined as negatives among the actual positives & \(\displaystyle \frac{FN}{TP+FN} \)  \\
 \hline
   F1-Score & It is the harmonic mean of precision and recall. Also known as F-Score or F-measure & \(\displaystyle \frac{2 \times precision \times recall}{precision + recall} \)  \\
 \hline
   Equal Error Rate (EER)  & Threshold that equals the FAR and FRR. & \(\displaystyle FAR=FRR \)  \\
 \hline
    Area Under Curve (AUC) & Area covered by the plot of TPR and FPR (ROC Curve) at different threshold values between 0 and 1 & \(\displaystyle \int ROC \)  \\
 \hline
    Mean Squared Error (MSE) & Average of the squares of the prediction errors. It is utilized in regression & \(\displaystyle \frac{1}{n}\sum_{i=1}^{n}(y_{i} - x_{i})^{2} \)  \\
\hline
     Root Mean Squared Error (RMSE) & Root of the average of the squares of the prediction errors. It is utilized in regression & \(\displaystyle \sqrt{(\frac{\sum_{i=1}^{n}(y_{i} - x_{i})^{2}}{n})} \)  \\
\hline
     Mean Absolute Error (MAE) & Absolute average of the prediction errors. It is utilized in regression & \(\displaystyle \frac{1}{n}\sum_{i=1}^{n}|y_{i} - x_{i}| \)  \\
\hline
      Root Relative Squared Error (RRSE) & Error relative to a simple predictor that always returns the average of the actual values & \(\displaystyle \sqrt{(\frac{\sum_{i=1}^{n}(y_{i} - x_{i})^{2}}{\sum_{i=1}^{n}(x_{i} - X)^{2}})}, \newline X= \frac{1}{n} \sum_{i=1}^{n} x_{i} \)  \\
\hline
      Detection or modeling time & Period elapsed between an attack or anomaly starts and the monitoring system detects it, or the time elapsed to model the device behavior accurately \cite{Miettinen2017IoTSentinel,Nguyen2019DioT} & \(\displaystyle - \)  \\
\hline
      Processing overhead or resource consumption & Resource usage of behavior monitoring and processing, which is particularly relevant in resource-constrained devices \cite{wang2015confirm,Golomb2018CIoTACI,GULENKO2016AD_NFV} & \(\displaystyle - \)  \\
 \hline
\end{tabular}
\begin{tablenotes}
    \scriptsize
    \item TP: True Positive, TN: True Negative, FP: False Positive, FN: False Negative.
\end{tablenotes}
\caption{Common evaluation metrics considered by device behavior fingerprinting solutions.}
\label{tab:metrics}
\end{table}

\section{Behavior-based Solutions and Applications}
\label{sec:solutions}

\addtxt{After analyzing the processing and evaluation techniques used in device fingerprinting (\textit{Q2}), and the scenarios, devices, and data sources (previously, with \textit{Q1}), we have the background needed to review and understand device behavior-based solutions. In this sense, this} section performs an in-depth review of the most relevant works of the literature that deal with behavioral fingerprinting to answer \textit{Q3 (What characteristics do the most recent and representative solutions of each application scenario have?)}. The analysis of each solution considers the application scenario, device type, behavior source, data monitored, processing and evaluation algorithms, and results criteria. \addtxt{We give particular importance to IoT devices because of their role in current real-world deployments. Still, it is important to note that other devices could be fingerprinted considering the same data sources.} Below, the approach followed by each solution is detailed and grouped by application scenario and behavior source.

\subsection{Device Type or Model Identification}

In this application scenario, we review solutions whose objective is to identify device models or types. Devices belonging to the same model or type are treated as equals by the literature. \change{and their main characteristics are compared in TABLE VI}{The main characteristics, algorithms and performance of each solution are compared in \tablename~\ref{tab:model_identification_table}}.

\subsubsection{\addtxt{Network-based identification}}

\addtxt{Many} works in the area of device type or model behavior fingerprinting address the identification problem from a network analysis perspective\addtxt{, deriving statistical features for ML/DL technique application}. Furthermore, they are mainly focused on IoT and ICS devices differentiation, as this section shows. In this context, the authors of~\cite{formby2016s}, proposed two \rmvtxt{different }fingerprinting methods for ICS device \change{s}{models}. The first was based on the response time between a TCP acknowledgment and the application layer response, once the data had been processed. The second method used \rmvtxt{the} physical operation times\rmvtxt{to develop a unique signature for each device model} by measuring the time elapsed to apply some actions in an actuator. \rmvtxt{The fingerprint classification accuracy reached by both methods was 99\% (using an ANN) and 92\% (using NB), respectively. }In \cite{Miettinen2017IoTSentinel}, Miettinen et al. proposed IoT Sentinel, an IoT device type identification approach based on device setup network communications. The main goal of this work was to \rmvtxt{identify device types connected to the network in order to }recognize potentially vulnerable \change{ones}{device types} and enhance their security based on rules. Packet headers were analyzed to derive \rmvtxt{23 }features \addtxt{resilient to traffic encryption.}\rmvtxt{, which were used as input for Random Forest classification. The average device identification accuracy was 81.5\% in 27 tested devices, being over 95\% in 17 devices. The device classification process took 157.7 ms on average.}

Bezawada et al. \cite{Bezawada2018IoTSense} also presented a network-based methodology to perform \rmvtxt{device }behavioral fingerprinting and device type identification inspired \rmvtxt{by previous works }in SIP-based \rmvtxt{device }fingerprinting \cite{francois2009automated,francois2010machine}. A behavior model data was divided into static, based on the \change{set of}{header} protocols used by the IoT device, and dynamic, based on \rmvtxt{the packet }flow sequences \addtxt{and packet payloads}.\rmvtxt{Different features were derived from the previous data, some of them relative to headers (network/transport/application protocols, IP options), and others focused on the packet payload (entropy, TCP payload length, and TCP window size). Gradient boosting was applied to a dataset generated by the authors to classify different device models.}\rmvtxt{The authors reported 86-99\% identification rate (TPR) and 99\% accuracy.} By following the same direction, Shahid et al. \cite{Shahid2018IoT} identified different IoT device types using bidirectional flow characteristics. Four different device types were utilized: sensor, camera, bulb, and plug. \rmvtxt{t-SNE was used for dimensionality reduction, and typical ML and DL classification algorithms were utilized for evaluation. This solution achieved 99.9\% accuracy using Random Forest.}

Also dealing with device type or model identification, the authors of~\cite{selis2018classification}, utilized ping operations to generate a fingerprint of different IoT devices to distinguish real embedded machines from virtual and emulated embedded systems. Several devices were grouped in each category to make them diverse enough \change{for the classifiers to recognize}{to model} previously unseen devices. For each ping, \rmvtxt{timing information was collected, such as ping response time and system timestamp. Then, 14}\addtxt{time-based }statistical features \rmvtxt{(e.g., average, max, min, variance, mode, or median) }were calculated using ping requests separated \rmvtxt{in time }by 0.2 seconds.\rmvtxt{Finally, a classification approach was performed using typical ML algorithms. Random Forest, combined with Extremely Randomized Trees for feature selection, achieved a detection rate of 99.5\% using only 25 pings (5 seconds) and 99.9\% using 200 (40 s).} Oser et al. \cite{oser2018identifying} utilized TCP timestamps to measure the clock skew of different IoT device models and identify them. \rmvtxt{In IoT devices, the authors commented that crystal imperfections produce a drift of about 8.64 seconds per day, which could be enough to identify different physical crystals. }562 devices of 51 different models were utilized for classification-based testing. Using only clock skew\change{as the unique feature for device classification, the algorithm}{, the system} could not identify most of the devices. Then, the authors decided to utilize 12 additional features derived from the timestamps gathered to calculate the clock skew.\rmvtxt{Using these derived features, Random Forest achieved 97.03\% precision, 94.64\% recall, and 99.76\% accuracy when classifying device models.} Thangavelu et al. \cite{Thangavelu2018DEFT} proposed DEFT, a distributed device fingerprint and identification system. \rmvtxt{The system was designed for SDN applications, but it could be used in other environments. }In this approach, \addtxt{SDN} network gateways performed device monitoring and classification locally, while a centralized control entity generated and distributed the classifiers. \rmvtxt{The gateways and the controller were synchronized to identify new device types and share data. }\change{Network }{Statistical} features were extracted based on packet headers \change{, DNS queries, and HTTP URIs. Selected features were related to statistical information about common IoT}{and} application layer protocols\rmvtxt{(DNS, HTTP/S, SSDP, QUIC, MQTT, STUN, NTP, and BOOTP)}, and grouped in 15-minutes sessions. To identify new device types, clustering algorithms (k-means) were applied. \rmvtxt{In the classification experiments of known devices, Random Forest achieved 98\% accuracy. When recognizing unknown devices, accuracy was 97\%, having +97\% F1-Score in 14 of the 16 tested devices.} \addtxt{Similarly, Perdisci et al. \cite{perdisci2020iotfinder} analyzed DNS application protocol to derive IoT model fingerprints following a document retrieval-based approach.}

Another relevant work in the scenario of IoT device model identification was proposed by Marchal et al. \cite{Marchal2019Audi}. The authors presented AuDI (Autonomous IoT Device-Type Identification), a system designed to identify IoT device type by passively analyzing its periodic network communications\change{. The system did not use pre-defined labels, instead, network data was grouped}{, grouping them} using clustering algorithms. To recognize periodic flows, Discrete Fourier Transform (DFT) was applied to candidate periods, transforming time domain to frequency domain. Then, 33 different features\rmvtxt{, grouped in 4 different categories (periodic flows, period accuracy, period duration, and period stability),} were calculated for each period. \rmvtxt{Finally, k-NN was used to classify the cluster-labeled data, achieving $>$90\% F1-Score for 21 of the 23 labels, and 98.2\% overall accuracy. }\rmvtxt{The authors claimed that the collected dataset will be published in the future. }Similarly, Arunan Sivanathan et al. \cite{Sivanathan1,Sivanathan3} worked on device type classification. In this case, \change{the network data was captured using \textit{tcpdump}, and packets and flows were used to extract behavior features. These}{packet and flow-based statistical} features were utilized to perform device classification\rmvtxt{, achieving an average accuracy of 99.88\% and RRSE of 5.06\%,} and behavioral monitoring tasks\rmvtxt{, achieving 97.5\%, 97.3\%, 97.4\% average weighted precision, recall, and F1-Score, respectively}. \addtxt{Using the same dataset, Msadek et al. \cite{msadek2019iot} focused on encrypted traffic analysis to identify IoT device models. In this work, the authors derived statistical features from headers using a sliding window.} 

In the same direction, OConnor et al. \cite{OConnor2019HomeSnitch} proposed HomeSnitch, a framework designed to classify home IoT devices communication by semantic behavior (e.g., firmware update/check, audio/video recording, data uploading). \rmvtxt{This framework tried to enhance network security by recognizing known behaviors and alerting about unknown ones. }To build application-level models from packet headers, HomeSnitch used \textit{adudump} \cite{adudump} traffic analysis tool. After that, 13 different features were extracted to describe application data exchanges. The authors used YourThings dataset \cite{yourthings} for solution testing\rmvtxt{, with Random Forest giving the best results: 99.69\% accuracy, 93.93\% F1-Score, 96.82\% TPR, and 11.96\% UBMR (Unknown Behavior Miss Rate)}. \rmvtxt{To force network access control based on the classification, the system was built upon the SDN paradigm.} \addtxt{Similarly, Trimananda et al. \cite{Trimananda2020Pingpong} proposed Ping-Pong, a tool designed to extract packet-level signatures for events (e.g., light bulb turning ON/OFF) based on device model. This work covered traffic encryption and unknown proprietary protocols by applying a clustering-based approach over statistical packet analysis. Furthermore, Hafeez et al. \cite{hafeez2020iot} proposed IoT-KEEPER, a system for both identify device types and detect malicious activities using an unsupervised approach based on fuzzy C-means clustering and interpolation.} 

\change{In the last group of works dealing with device type or model identification, we can find the work of Ortiz et al.}{Applying more sophisticated DL-based solutions, Ortiz et al. \cite{Ortiz2019DeviceMien} presented DeviceMien, a probabilistic framework for device identification which considered stacked LSTM-autoencoders to automatically learn features and classes from raw TCP packets}. Then, the system modeled, using DBSCAN\rmvtxt{optimized through Bayesian Modeling}, each device as a distribution of the generated classes. For testing, the authors used two different datasets, one public, \cite{Sivanathan1}, and another private. \rmvtxt{Previously seen devices classification reached over 99\% accuracy for devices when using at least 50 samples. The system could also distinguish between IoT and Non-IoT devices by examining the average number of flow classes observed over a set of samples. Besides, the correct class of unseen devices was inferred with over 82\% average F1-Score and 70\% accuracy by using a combination of OC-SVMs. }\rmvtxt{Finally, }Kotak and Elovici \cite{kotak2020iot}\rmvtxt{also performed IoT device type identification based on network traffic. However}, as a novel approach, \rmvtxt{the authors }performed a pre-processing step that converted the TCP network traffic (pcap format) to grayscale images. Then, an MLP was utilized to classify different device flows based on the device type. The dataset utilized was from \cite{Sivanathan1}\rmvtxt{, and this solution achieved over 99\% accuracy when identifying 10 different types of network flows (9 classes for IoT devices and 1 class for non-IoT traffic)}.

\addtxt{Another research line covering device identification is based on the analysis of deployment scenarios such as Smart Homes or agriculture networks \cite{guth2018detailed,boyes2018industrial}. Kumar et al. \cite{kumar2019all} analyzed home networks in order to perform device type identification and security analysis. In total, 83M devices deployed in 16M households were collected, analyzing their distribution and known vulnerability issues. In a device subset, a 96\% accuracy was achieved using expert rules and an ensemble of RF classifiers trained using data from different application layer protocols. Another smart home device analysis was performed by Huang et al. \cite{huang2020iot}. However, device categories were only manually standardized in this study, mentioning device type identification and anomaly detection as future work paths.}

\addtxt{Digital forensics has also leveraged device identification when it helps in forensic investigations, as the increasing number of devices generates new challenges and motivates to work on more advanced identification methods \cite{macdermott2018iot,yaqoob2019internet}. One example of these scenarios is Amazon Alexa ecosystem forensics \cite{chung2017digital,li2019iot}, where the behavior-based identification of the devices present in the scenario is a highly valuable asset. Moreover, the digital forensics field is also leveraging new technologies such as blockchain when dealing with large scenarios such as IoT environments \cite{ryu2019blockchain}.}

\subsubsection{\addtxt{Radio-based identification}}

\addtxt{Drone model identification is the main research area where radio behavior fingerprinting is employed for type or model identification. Although this problem has been traditionally addressed based on physical characteristics, such as images \cite{ruiz2018idrone}, RADAR and LIDAR \cite{coluccia2020detection}, or sound \cite{bernardini2017drone}(out of the scope of this study), there is an emerging research line based on radio analysis and fingerprinting. A relevant work presented by Ezuma et al. \cite{ezuma2019detection} analyzed controller signals to classify unmanned aerial vehicles (UAV). In the same line, Al-Sa'd et al. \cite{al2019rf} used DNNs to classify drone models based on their radio communications. Using the same dataset, Allahham et al. \cite{allahham2020deep} improved the previous results using a 1D CNN. Similarly, Basak et al. \cite{basak2020drone} also applied CNNs for drone identification but using their own dataset, which will be published in the near future.} 

\tablename~\ref{tab:model_identification_table} compares the solutions focused on device type and model identification. From the previous solution analysis, we can observe that the device type and model identification application scenario has been \addtxt{mainly} covered from a \change{communication network}{network communication} perspective. Moreover, it is noticed that most of the solutions in this area are focused on IoT, as the heterogeneous nature of IoT devices motivates the usefulness of solutions capable of distinguishing devices according to their type and model. Many solutions achieve classification results over 99\% in accuracy and F1-Score metrics, which indicates that this area is relatively covered by approaches with good performance. \addtxt{Besides, drone identification is the main application of radio-based fingerprinting for model identification. Here, further research is still required to achieve similar performance to network-based identification.} 

\begin{table*}[htpb]
    \centering
    \scriptsize
    \begin{tabular}{ >{\Centering}m{0.6cm} >{\Centering}m{0.6cm}  >{\Centering}m{1.2cm} >{\Centering}m{1.4cm} >{\Centering}m{1.9cm} >{\Centering}m{1cm} >{\Centering}m{1.9cm} >{\Centering}m{1cm} >{\Centering}m{1.5cm} >{\RaggedRight\arraybackslash}m{4.2cm} } 
    \hline
    \textbf{Work} & \textbf{Year} & \textbf{Device Type} & \textbf{Approach} & \textbf{Algorithms} & \textbf{Behavior Source} & \textbf{Features} & \textbf{Dataset} & \textbf{Classes} & \makecell[c]{\textbf{Results}} \\
    \hline
    \hline
    \cite{formby2016s} & 2016 & ICS & Classification & ANN, NB & Network & Response delay times & Private & Device Model & 99\% and 92\% accuracy for response and operation time recognition, respectively.\\
    \hline
    \cite{Miettinen2017IoTSentinel} & 2017 & IoT Devices & Classification & RF & Network & Packet header-based & \cite{Miettinen2017IoTSentinel} & Device Type & 81.5\% average accuracy on 27 devices, over 95\% for 17 of them.\\
    \hline
    \cite{Bezawada2018IoTSense} & 2018 & IoT Devices & Classification & Gradient boosting, k-NN, DT & Network & Header and payload statistics & Private & Device Type & 99\% average accuracy and 86-99\% TPR \\
    \hline
    \cite{Shahid2018IoT} & 2018 & IoT Devices & Classification & t-SNE, RF & Network & Flow statistics & Private & Device Type & 99.9\% accuracy differentiating sensor, camera, bulb, and plug devices. \\
    \hline
    \cite{selis2018classification} & 2018 & IoT Devices & Classification & RF & Network & Ping timestamps & Private & \addtxt{Real / Virtual} Device & Detection rate of 99.5\% using 25 pings and 99.9\% using 200 pings. \\
    \hline
    \cite{oser2018identifying} & 2018 & IoT Devices & Classification & RF, SVM, MLP & Network & Clock skew and timestamp features & Private & Device Model & 97.03\% precision, 94.64\% recall and 99.76\% accuracy identifying 51 models. \\
    \hline
    \cite{Thangavelu2018DEFT} & 2018 & IoT Devices & Classification & k-means, RF & Network & IoT protocol flows statistics & Private & Device Type & 97\% accuracy, +97\% F1-Score (14/16 classes)\\
    \hline
    \cite{Marchal2019Audi} & 2019 & IoT Devices & Classification & Clustering + k-NN & Network & Flow periods (DFT) & To be published & Device Type & F1-Score above 90\% for 21/23 labels and 98.2\% overall accuracy.\\
    \hline
    \cite{Sivanathan1} & 2019 & IoT Devices & Classification & RF & Network & Flow and packet statistics & \cite{Sivanathan1} & Device Model & 99.88\% accuracy 5.06\% RRSE. \\
    \hline
    \cite{msadek2019iot} & 2019 & IoT Devices & Classification & AdaBoost & Network & Encrypted flow statistics & \cite{Sivanathan1} & Device Model & 95.5\% accuracy and F1-Score.\\
    \hline
    \cite{OConnor2019HomeSnitch} & 2019 & IoT Devices & Classification & RF, k-NN, Gradient Boosting & Network & Data exchange statistics & \cite{yourthings} & Device Behavior Type & 99.69\% accuracy, 93.93\% F1-Score and 96.82\% TPR\addtxt{, and 11.96\% UBMR}.\\
    \hline
    \addtxt{\cite{kumar2019all}} & \addtxt{2019} & \addtxt{IoT Devices} & \addtxt{Classification} & \addtxt{Rules + RF} & \addtxt{Network} & \addtxt{Header and app-layer statistics} & \addtxt{Private} & \addtxt{Device Type} & \addtxt{96\% accuracy on a manually labeled subset.} \\
    \hline
    \cite{Ortiz2019DeviceMien} & 2019 & IoT Devices & Classification & LSTM-autenc., DBSCAN, OC-SVM & Network & Derived using LSTM-autoencoders & Private / \cite{Sivanathan1} & Device Model & Seen devices: 99\% accuracy. Unseen devices: 82\% F1-Score and 70\% accuracy.\\
    \hline
    \addtxt{\cite{Trimananda2020Pingpong}}& \addtxt{2020} & \addtxt{IoT Devices} & \addtxt{Classification} & \addtxt{DBSCAN, State machine} & \addtxt{Network} & \addtxt{Packet sequence statistics} & \addtxt{Private} & \addtxt{Device Activities} & \addtxt{97.05-97.48\% avg detection and 0.18-0.32\% avg FPR in actions of 19 devices.}\\
    \hline
    \cite{kotak2020iot} & 2020 & IoT Devices & Classification & DNN & Network &  Images generated from raw data & \cite{Sivanathan1} & Device Type & 99\% accuracy identifying 10 network flow types (9 IoT and 1 non-IoT).\\
    \hline
    \addtxt{\cite{hafeez2020iot}} & \addtxt{2020} & \addtxt{IoT Devices} & \addtxt{Classification} & \addtxt{C-means and interpolation} & \addtxt{Network} & \addtxt{Flow and header statistics} & \addtxt{\cite{hafeez2020iot}} & \addtxt{Device Type / Anomalies} & \addtxt{99\% accuracy for device type identification and 98\% TPR, 4\% FPR and 98\% F1-Score for attack detection.}\\
    \hline
    \addtxt{\cite{perdisci2020iotfinder}}&\addtxt{2020}&\addtxt{IoT Devices} & \addtxt{Classification} & \addtxt{Statistical based on term frequency} & \addtxt{Network} & \addtxt{DNS analysis} & \addtxt{To be published} & \addtxt{Device model} & \addtxt{$\approx$95\% avg AUC and 0.01\% max FPR on 53 models.} \\
    \hline
    \hline
    \addtxt{\cite{ezuma2019detection}} & \addtxt{2019} & \addtxt{IoT Devices} & \addtxt{Classification} & \addtxt{k-NN} & \addtxt{Radio signals} & \addtxt{IQ samples} & \addtxt{Private} & \addtxt{Drone model} & \addtxt{98.13\% accuracy identifying 15 UAV controllers.} \\
    \hline
    \addtxt{\cite{al2019rf}} & \addtxt{2019} & \addtxt{IoT Devices} & \addtxt{Classification} & \addtxt{DNN} & \addtxt{Radio signals} & \addtxt{IQ samples} & \addtxt{\cite{allahham2019dronerf}} & \addtxt{Drone model}& \addtxt{99.7\% accuracy using 2 classes, 84.5\% using 10 classes.} \\
    \hline
    \addtxt{\cite{allahham2020deep}} & \addtxt{2020} & \addtxt{IoT Devices} & \addtxt{Classification} & \addtxt{1D CNN} & \addtxt{Radio signals} & \addtxt{IQ samples} & \addtxt{\cite{allahham2019dronerf}} & \addtxt{Drone model}& \addtxt{94.6\% accuracy for 10 drone classes.} \\
    \hline
    \addtxt{\cite{basak2020drone}} & \addtxt{2020} & \addtxt{IoT Devices} & \addtxt{Classification} & \addtxt{CNN} & \addtxt{Radio signals} & \addtxt{IQ samples} & \addtxt{Private} & \addtxt{Drone model}& \addtxt{$\approx$99\% accuracy for 10 drones and controllers when SNR is 0 dB.} \\
    \hline
    \hline
    \end{tabular}
    \caption{Device type or model identification solutions based on device behavior fingerprinting.}
    \label{tab:model_identification_table}
\end{table*}

\subsection{Individual Device Identification}

\change{It}{This section} analyzes behavior-based solutions focused on identifying the device itself. It means that they differentiate devices with the same hardware/software. At this point, it is important to note that these approaches will also be able to distinguish different device types and models (the previous category), and this fact is also considered and evaluated in some of them. In these solutions, features usually have a lower level, related to hardware components, trying to differentiate fabrication variations on the device components. \tablename~\ref{tab:identical_device_table} compares the \addtxt{main characteristics, algorithms applied and performance of} solutions detailed in this subsection.

\subsubsection{\addtxt{Processor-based identification}}

In this category, Salo's \cite{salo2007multi} proposed a fingerprinting software method capable of differentiating identical personal computers using quartz crystals characteristics. Concretely, the author utilized the CPU Time-Stamp Counter (TSC), the Real-Time Clock (RTC), and the Sound Card Digital Signal Processor (DSP). The solution aimed to verify how accurate the RTC\rmvtxt{(\textit{/dev/rtc})} and DSP\rmvtxt{(\textit{/dev/dsp})} were in terms of CPU cycles \change{. To measure this accuracy, the solution measured}{by measuring} the one-second ticks of the RTC and the time needed by the DSP to process one second of audio. \rmvtxt{A test was launched for one hour to store repeated measurements of thirty-eight computers from the same lab with the same CPU and software. }\change{The}{Then,} statistical analysis \change{results showed that RTC measurements were able to differentiate the 98.5\% of pair machines, and DSP measurements were able to distinguish the 93.3\%.}{was applied to distinguish computer pairs between them.} Also exploiting processor differences, but based on execution time, Sanchez-Rola et al. \cite{sanchez2018clock} proposed CryptoFP, a novel approach to identify machines with the same software and hardware through the generation of a fingerprint using the time taken to execute a specific function. \change{This fingerprint was generated locally without network traffic or external time stamps. CryptoFP was composed of two phases, the generation of a fingerprint from the timing of the code execution and the determination of whether two fingerprints belong to the same machine through fingerprint comparison. The fingerprinting process was based on the generation of a matrix \textit{n}x\textit{m}, where \textit{n} is the number of function calls to measure and \textit{m} is the number of times this process is repeated.}{This fingerprint was generated by executing the same function many times, repeating different parameters to model its time variability.} In the fingerprint comparison, the tool \change{considered}{compared} the most frequent (mode) time values for each call parameter over all iterations. The authors conducted several experiments to test long-term fingerprint stability, and CPU workload and temperature impact in the fingerprint generation. \rmvtxt{The solution was able to differentiate two sets of identical machines, one with 89 computers and the other with 176, with 100\% uniqueness in host execution. Besides, the solution achieved +80\% uniqueness in web browser execution.} \addtxt{For future work, t}he authors considered \rmvtxt{as }solution \rmvtxt{drawback that the proposal can generate }scalability \rmvtxt{problems }as fingerprints are compared one by one. \addtxt{Finally, Lorenz et al. \cite{lorenz2020fingerprinting} considered embedded circuits of IoT sensors for unique fingerprinting. To perform the fingerprinting, predefined voltage sequences were supplied to the sensor, monitoring how its output varies. Fingerprints were evaluated directly comparing output sequences and using RMSE as error measure. Results in individual identification varied according to sensor model, meaning that some models have more fabrication variability that others.}

\subsubsection{\addtxt{Clock-based identification}}

Based on clock skew capabilities, Jana and Kasera \cite{jana2009skew} worked on uniquely differentiate wireless access points (AP) based on the clock skew calculated from their beacon frame timestamps. \rmvtxt{The objective was to detect unauthorized wireless devices based on their beacon frame timestamps. }This work utilized the uw/sigcomm2004 dataset \cite{uw-sigcomm2004-20061017}. \rmvtxt{From a set (50-100) of timestamp differences between APs and a receiver, Linear Programming Method (LPM) and Least Square Fitting (LSF) were utilized to generate a clock skew value. }The results, using Expectation Maximization statistical algorithm to compare AP frames, indicated that clock skew seems to be an efficient and robust fingerprinting method capable of detecting different WLAN APs. Similar results to the previous ones were presented by Sharma et al. in \cite{sharma2012skew}. In this case, the authors utilized TCP and ICMP timestamp headers to calculate the clock skew between two devices\change{. The authors utilized the work of Kohno et al. [59] as basis for clock skew calculation, validating it}{, validating the work of Kohno et al. \cite{kohno2005remote}}. They tested their approach with 210 different devices, some of them identical, finding that \rmvtxt{at least 70 packets were needed to generate a consistent skew measurement. }they were able to distinguish both different and identical devices. Besides, they also tested clock skew stability based on the measurement methodology and on several environmental factors, such as temperature or operating system. \rmvtxt{Finally, this work checked how clock skew does not variate significantly when the device operating system varies or with NTP updates (only for TCP timestamps).}Based on these results, the authors concluded that this approach is suitable for moderate size networks.

Focused on wireless unique device identification, Lanze et al. \cite{lanze2012skew} considered clock skew stability and uniqueness. To measure the clock skew, the authors took \rmvtxt{two kinds of timestamps, }the timestamps from a wireless AP (sender) sent in wireless beacons and the timestamps from the measuring wireless client (receiver). To carry out their experiments, they gathered clock skews using five different laptops \rmvtxt{with different Wi-Fi chipsets }from 388 different APs. \rmvtxt{In addition, they ran the experiment in different areas for increasing the number of samples. }Through their experiments, they concluded that all clock skews were in a rather short range \change{between -30 ppm and +30 ppm}{($\pm$30 ppm)} due to restrictions of the suppliers' quality specifications. \rmvtxt{It comes out that clock skew alone cannot serve as a unique fingerprint for wireless access points. }Therefore, although the clock skew restricts the set of possible devices, it cannot serve as a unique fingerprint for a wireless access point and has to be enriched with other features to achieve uniqueness. In the same line, Radhakrishnan et al. \cite{radhakrishnan2014gtid} published GTID, a system for individual wireless device and device type fingerprinting based on clock skew. This approach utilized clock skew and communication patterns to generate device signatures \change{based on}{from} a \addtxt{DL-based} time series approach. The system was tested using a previous dataset of the team \cite{gatech-fingerprinting-20140609,uluagac2013passive}, collected from 37 different devices, including some repeated models. \rmvtxt{To evaluate the signatures, ANNs were utilized achieving from 99 to 95\% average accuracy and 74\% average recall on device ID classification, and 86\% average accuracy and 68\% average recall on device type classification.} Similarly to \cite{lanze2012skew} and \cite{radhakrishnan2014gtid}, Pol{\v{c}}{\'a}k et al. \cite{polvcak2014reliability,polcak2015clock} also discussed clock skew performance when uniquely identifying different devices. Here, the authors concluded that clock skew is not completely stable. Besides, based on the clock skew distribution of the evaluated devices, the authors claimed that clock \rmvtxt{manufacturers pretend to achieve 0 ppm clock skew, so }skews are distributed close to 0 ppm. These factors prevent a quick fingerprint technique to be capable of uniquely differentiate devices in large scenarios. Finally, the authors also discussed and demonstrated the possibility of masquerading or falsifying the clock skew. The authors concluded that this technique might be suitable for small networks or in combination with additional data.

\subsubsection{\addtxt{Resource usage-based identification}}

\change{A solution exploiting resource usage was proposed in }{Resource usage was exploited for individual identification in }\cite{dong2019cpg}. In this work, the authors developed a fingerprinting method based on the CPU usage graph when the device is executing a fixed task. For this purpose, a benchmark program that included several read/write operations and calculations was developed. \rmvtxt{To fingerprint each device, 128 CPU usage measurements, one every 0.2 seconds (25.6 seconds in total), were utilized to generate a usage graph. For testing, ten identical PCs were used.}In the evaluation process, the graph was compared to the previous ones of the same device using the Dynamic Time Warping algorithm. The percentage of stable fingerprints was calculated using the Shannon entropy and stability measurement, achieving a 93.43\% of unique fingerprints.

\subsubsection{\addtxt{Electromagnetic signal-based identification}}

Other works solved the identical device identification problem using electromagnetic signals as data source. Using radio signals, Jafari et al. \cite{Jafari_Fingerprinting_Deep_Learning_2018} used \change{MLPs, CNNs, and LSTMs}{DL techniques} to identify wireless devices and distinguish among identical wireless devices from the same manufacturer. The authors used ZigBee devices from which a historical radio frequency trace dataset was obtained. In total, six identical devices were employed in the tests\change{. Accuracy results were: 96.3\% for MLP, 94.7\% for CNN, and 75\% for LSTM. Finally, the authors concluded}{, concluding} that it was possible to identify devices based on their radio frequency traces, even if they were from the same model. A similar approach was addressed in \cite{Riyaz2018Radio}, where Riyaz et al. utilized raw radio samples to build a unique device signature using Software Defined Radio (SDR) transmissions. This solution \change{achieved 98\% accuracy when identifying 5 identical devices using a CNN classifier}{was tested on 5 identical devices}. \rmvtxt{Other algorithms, such as SVM and Logistic Regression were also tested, achieving worse results. }In addition, the authors analyzed how detection accuracy is impacted by measuring distance, concluding that classification performance starts to degrade at 34 feet. Finally, Cheng et al. proposed in \cite{Cheng2019DemiCPU} a method capable of identifying identical laptops and smartphone devices (also different models) based on the electromagnetic signals radiated from the CPU. \rmvtxt{Using Extra-Trees classifier (a variant of Random Forest), the authors achieved 99.1\% average precision and recall for all devices tested (70), and more than 98.6\% precision and recall for 30 identical devices, using one round fingerprint. With multi-round, results were enhanced to 99.9\% in the previous metrics. The authors mentioned }As a drawback\change{that}{,} this solution requires the use of an external sensor to measure the CPU radiated signals within a 16 mm range.

\tablename~\ref{tab:identical_device_table} compares the solutions focused on individual device identification. As a general view of individual device identification solutions, it can be appreciated that solutions are focused on general computers and wireless devices. This ensures solution universality, but opens the door to future perspectives focused on more specific device types such as IoT or ICS. It is also noticed the lower-level nature of the behavior sources utilized, which in this case are mainly based on clock and processor properties, and electromagnetic signals. Many solutions achieved high individual identification performance. However, many of these approaches noticed scalability issues in large device deployments, as fabrication variations are limited within determined quality standards.

\begin{table*}[htpb]
    \centering
    \scriptsize
    \begin{tabular}{ >{\Centering}m{0.6cm} >{\Centering}m{0.6cm}  >{\Centering}m{1.4cm} >{\Centering}m{1.4cm} >{\Centering}m{1.6cm} >{\Centering}m{1.6cm} >{\Centering}m{1.6cm} >{\Centering}m{0.8cm} >{\Centering}m{1.7cm} >{\RaggedRight\arraybackslash}m{4cm} } 
    \hline
    \textbf{Work} & \textbf{Year} & \textbf{Device Type} & \textbf{Approach} & \textbf{Algorithms} & \textbf{Behavior Source} & \textbf{Features} & \textbf{Dataset} & \textbf{Classes} & \makecell[c]{\textbf{Results}} \\
    \hline
    \hline
    \cite{salo2007multi} & 2007 & General computers & Classification & Statistical & System processors and oscillators & RTC and DSP drift compared to the TSC & Private & Different physical devices & 98.5\% and 93.3\% of differentiation \addtxt{by RTC and DSP in 38 PCs}, respectively. \\
    \hline
    \cite{sanchez2018clock} & 2018 & General computers & Classification & Statistical (Mode) & System processors & Matrix of code execution times & Private & Different physical devices & 100\% host-based and +80\% web-based device identification in two sets of 89 and 176 PCs. \\
    \hline
    \addtxt{\cite{lorenz2020fingerprinting}} & \addtxt{2020} & \addtxt{IoT Devices} & \addtxt{Classification} & \addtxt{Statistical} & \addtxt{System circuits} & \addtxt{Outputs based on voltage} & \addtxt{Private} & \addtxt{IoT sensors} & \addtxt{0\% to 94.3\% FPR in individual sensor and 0\% FPR in model identification.}\\
    \hline
    \hline
    \cite{jana2009skew} & 2009 & Wireless access points & Classification & Expectation Maximization & Clock skew & Wi-Fi beacons timestamps & \cite{uw-sigcomm2004-20061017} & Known APs & Clock skew is a robust method and can detect different WLAN APs. \\
    \hline
    \cite{sharma2012skew} & 2012 & General computers & Classification & Statistical & Clock skew & TCP and ICMP timestamp & Private & Different physical devices & \addtxt{Both }identical and different devices correctly identified. \\ 
    \hline
    \cite{lanze2012skew} & 2012 & Wireless devices & Data analysis & Statistical & Clock skew & Wi-Fi beacons timestamps & Private & Different physical devices & Clock skew is not enough to uniquely identify a large set of devices. \\
    \hline
    \cite{radhakrishnan2014gtid} & 2014 & Wireless devices & Classification & ANN & Clock skew + Network & Communi-cation skew and patterns & \cite{gatech-fingerprinting-20140609} & Individual devices and device type & From 99 to 95\% accuracy and 74\% recall on ID, and 86\% accuracy and 68\% recall on type classification. \\
    \hline
    \cite{polcak2015clock} & 2015 & General computers & Data analysis & Statistical & Clock skew & TCP timestamps & Private & Different physical devices & Clock skew is only suitable for small networks or combined with other data.\\
    \hline
    \hline
    \cite{dong2019cpg} & 2019 & General computers & Classification & Dynamic Time Warping & Resource usage & CPU usage- based graph & Private & Physical devices & 93.43\% of uniqueness in the generated fingerprints of 10 identical devices. \\
    \hline
    \hline
    \cite{Jafari_Fingerprinting_Deep_Learning_2018} & 2018 & Wireless devices & Classification & MLP, CNN, LSTM & Electromagnetic signals &  Radio frequency IQ samples & Private & Different physical devices & 96.3\% accuracy for MLP, 94.7\% for CNN and 75\% for LSTM when identifying 6 identical ZigBee devices.\\
    \hline
    \cite{Riyaz2018Radio} & 2018 & Wireless devices & Classification & CNN & Electromagnetic signals & Raw frequency IQ samples & Private & Different physical devices & 98\% accuracy is achieved when identifying 5 identical devices. \\
    \hline
    \cite{Cheng2019DemiCPU} & 2019 & Laptops and Smartphones & Classification & Extra-Trees & Electromagnetic signals & CPU radiated magnetic signals & Private & Different physical devices & 99.1\% average precision and recall for all devices (70), and $>$98.6\% precision and recall for 30 identical devices. \\
    \hline
    \hline
    \end{tabular}
    \caption{Individual device identification solutions based on device behavior fingerprinting (works are grouped by behavior source, using double horizontal lines to separate them, and sorted by year).}
    \label{tab:identical_device_table}
\end{table*}

\subsection{Attack Detection}

The third main scenario where behavior fingerprinting is highly relevant is attack detection. Abnormal situations can have a wide range of forms, such as network attacks, malware, malicious firmware modifications, or unauthorized user interactions. Detection can be performed either modeling normal device behavior and detecting deviations, from an anomaly detection standpoint, or collecting normal and abnormal labeled data and performing classification tasks. \tablename~\ref{tab:attack_anomaly_table} compares the \addtxt{main characteristics, algorithms applied and performance of} solutions detailed in this subsection.

\subsubsection{\addtxt{Network-based attack detection}}

The most exploited source in terms of behavior-based attack detection is network monitoring. Many solutions, mainly focused on IoT \cite{meidan2017detection,Ferrando2017Network,Amouri2018NIDS,yu2019radar,Haefner2019CompleIoT,hamza2019detecting,Nguyen2019DioT,Hamad_IoT_Identification_Network_2019,HASAN2019100059,fernandez2019intelligent,Sivanathan2,ali2020towards,Blaise2020BotFP} but also on SDN/NFV \cite{carvalho2018ecosystem,afek2019nfvbased} and general computers \cite{yin2018enhancing,Marir2018Distributed,radford2018sequence,lima2019smart}, have utilized this source for attack detection.

\addtxt{One of the leading research lines focuses on detecting attacks that deploy unauthorized devices in the environment. }In \cite{meidan2017detection}, the authors worked on unauthorized IoT device detection using white lists and classification ML algorithms. TCP/IP flows were used to \change{identify}{extract features capable of characterizing} nine different types of devices (17 distinct IoT devices were used). \rmvtxt{In total, 274 features extracted from application, transport, and network layers served to classify the device type using Random Forest.}\rmvtxt{This classification was performed over 20 consecutive network sessions, and then the majority rule was applied over the classification results to decide the device type. IoT device types not white-listed were correctly detected as unknown in 96\% of cases (on average), and white-listed device types were correctly classified as their actual type in 99\% of cases.} This work also discussed the \addtxt{system} resilience to \change{adversarial }{cyber}attacks\rmvtxt{, concluding that the system would be resilient to malware infections}. \addtxt{Similarly, }in \cite{Hamad_IoT_Identification_Network_2019}, the authors used \change{network-flow}{packet headers and payload} data to extract flow-based features capable of creating device type fingerprints. Then, unknown or suspicious devices with abnormal behavior could be identified, and their communication \rmvtxt{was }restricted for further monitoring. \rmvtxt{To create a fingerprint for each device, the authors extracted unique behavioral and flow-based features from the header and payload of network packets. }The dataset used for testing came from IoT Sentinel \cite{Miettinen2017IoTSentinel}. \rmvtxt{Several ML classification algorithms were tested to distinguish device types, achieving an average accuracy of 90.3\% with Random Forest. }In the same line, Ferrando and Stacey \cite{Ferrando2017Network} built a behavior profile of IoT devices based on entropy and dispersion of metrics related to IP directions, ports, bytes received/sent, and latency. Anomalies were detected based on the distance between the average values and the \change{current ones}{ones being evaluated}. \rmvtxt{The authors proposed different evaluation approaches such as Euclidean Distance and AutoRegressive Integrated Moving Average (ARIMA), but no performance metrics were given using actual data.}

\addtxt{In contrast, the majority of works in this area cover the detection of direct cyberattacks, both common ones such as flooding or port scans, and more sophisticated ones like DDoS, botnets or ransomware.} Amouri et al. \cite{Amouri2018NIDS} proposed an IDS based on IoT device network behavior. This system had a distributed architecture composed of traffic sniffers in the local network and a central super node. Device behavior was built on packet counters determined by MAC and network layer data. The proposed architecture \rmvtxt{had two levels, a first one where traffic sniffers }applied DT algorithm to classify network instances, and \addtxt{then} \rmvtxt{a second one where a super node applied }Linear Regression to generate time-based device profiles relying on the measure of behavior fluctuation. \rmvtxt{After 3000s (3 reports to the super node), the system achieved 100\% detection (TPR).}

\addtxt{Also from an ML-based perspective, }Sivanathan et al. \cite{Sivanathan2} \rmvtxt{also} addressed behavioral changes and attack monitoring \change{. This proposal relied}{based} on flow and packet network analysis \change{to perform traffic modeling based on}{and} clustering. \change{The authors applied PCA for dimensionality reduction and k-means for clustering. This solution achieved 84.3\%, 89.4\%, 91.3\%, and 86.2\% of average detection rate for ARP Spoofing, Ping of Death, TCP SYN flooding, and Fraggle, respectively. For reflection attacks, the rate of detection for Smurf, SNMP, SSDP, and TCP SYN reflection attacks was 99.1\%, 58.8\%, 88.5\%, and 92.0\%, respectively.}{The authors tested both direct network attacks (ARP Spoofing, Ping of Death, TCP SYN flooding, and Fraggle) and reflection attacks (Smurf, SNMP, SSDP, and TCP SYN reflection).} A similar approach was followed in \cite{HASAN2019100059}, where the authors performed attack and anomaly classification \change{using the}{using MQTT protocol traces gathered from} DS2OS dataset \cite{pahl2018all}.\rmvtxt{. This dataset is based on a virtual IoT environment where devices communicate with each other using the MQTT protocol. Different ML and DL classification algorithms were tested, being Random Forest the algorithm that best performed when classifying normal traces and different attacks with a 99\% F1-Score.} In the same line, Lima et al. \cite{lima2019smart} presented an approach for detecting DoS/DDoS attacks using ML techniques. The authors built a customized attack dataset based on several public datasets (CIC-DoS, CIC-IDS2017, and CIC-IDS2018 \cite{CICIDS}) to benchmark normal traffic and different DoS/DDoS classification. \rmvtxt{Random Forest achieved an online detection rate (DR) of attacks above 96\%, with high F1-Score (99.5\%) and low false alarm rate (0.2\%) using a sampling rate (SR) of 20\% of network traffic.} The solution presented in \cite{radford2018sequence} also considered \rmvtxt{unsupervised anomaly detection based on }network traffic data extracted from the CIC-IDS 2017 \cite{CICIDS} dataset\addtxt{, but in this case for an unsupervised anomaly detection approach.} \change{This work was focused on cybersecurity and attack detection. }{Here, }traffic sequences were modeled in sliding windows that were fed to \change{RNNs. Concretely, the proposed architecture had an embedding layer that projects the sequence values into a dense 50-dimensional vector. This layer was followed by two LSTM layers with dropout, and finally, a dense layer.}{an LSTM network.}\rmvtxt{This approach achieved over 71\% AUC in all the evaluated attacks and 87\% AUC average. The authors draw that the results obtained may not be encouraging enough, and further research is needed in this area.} \addtxt{Similarly,} traffic-based anomaly detection is covered by a wide variety of other works using anomaly detection approaches \cite{carvalho2018ecosystem,yu2019radar}.

A different view was provided by Yin et al. \cite{yin2018enhancing}, who applied DL for botnet behavior modeling and detection. This solution was based on a GAN that generates simulated data, augmenting the model trained with the original data. The authors utilized network flows \change{as device behavior source, deriving statistical features such as flow duration, packet length, or total bits transmitted. The authors utilized}{derived from} ISCX botnet dataset \cite{beigi2014towards} as benchmark\rmvtxt{, achieving 74.04\% precision, 71.17\% accuracy, 70.59\% F1-Score, and 15.59\% TPR}. Also focused on botnet attacks, Blaise et al. \cite{Blaise2020BotFP} presented a bot detection technique based on \rmvtxt{the }host behavior. This solution was divided into three steps: characterizing the host behavior based on network signatures (aggregated attribute frequency distribution), inferring benign host behavior using clustering algorithms (DBSCAN), and classifying new hosts based on previously labeled instances\rmvtxt{(assigning the closest cluster center to new instances)}. \rmvtxt{Concretely, nine features regarding IPs, ports, and headers were extracted from network flows using TCP, UDP, and ICMP packets. }To validate the approach, the authors used the CTU-13 dataset \cite{GARCIA2014CTU13}\rmvtxt{, where a 100\% TPR and 0.9\% FPR were obtained when detecting botnet activities}. On similar research paths, Fern\'andez et al. \cite{fernandez2019intelligent} analyzed ransomware detection based on behavior analysis in Medical Cyber-Physical Systems\rmvtxt{(MCPS)}. This work analyzed network flows \change{in 10-second windows and extracted}{extracting} different statistical features. Then, anomaly detection and classification ML models \change{are}{were} combined to evaluate the live generated vectors. \rmvtxt{OC-SVM was utilized for anomaly detection and NB for classification. Then, based on the model outputs, different rules were defined. The system achieved 95.96\% F1-Score, 92.32\% precision, 99.97\% recall, and 4.6\% FPR for anomaly detection, and +99\% accuracy for ransomware classification.}

\change{Hamza et al. [10] proposed an approach based on Manufacturer Usage Descriptions (MUD) to enhance IoT security. In this case, }{In another line, some authors have proposed the usage of Manufacturer Usage Descriptions (MUDs) to enhance IoT security. In Hamza et al. \cite{hamza2019detecting},} \change{the solution was based on an SDN architecture.}{flow counters were used to generate feature vectors, applying PCA and k-means for dimensionality reduction and clustering, respectively.} \rmvtxt{The authors generated and published a dataset with benign network activity and traces of several network attacks such as ARP spoofing, TCP SYN, and UDP flooding or reflection attack. Flow counters were used to generate feature vectors, applying PCA and k-means for dimensionality reduction and clustering, respectively. }\change{Then, an anomaly detection approach based on boundary detection and Markov Chains achieved 94.9\% accuracy, 89.7\% TPR, and 5.1\% FPR, improving previous methods such as Snort IDS [160]}{Then, an approach based on boundary detection and Markov Chains was applied for MUD monitoring and anomaly detection, testing it on several network attacks such as ARP spoofing, TCP SYN, and UDP flooding or reflection attack.} Another approach using MUD to improve IoT security was proposed by Afek et al. in \cite{afek2019nfvbased}. From an NFV perspective, this \rmvtxt{work implemented a framework to be deployed on a service provider level. This }proposal presented a hybrid approach where MUD compliance checking is a service implemented as a virtual network function (VNF), and traffic monitoring is implemented on the network gateway to ensure P2P communications. For devices with no MUD, the authors used the algorithm proposed in \cite{Hamza2018MUD} for \change{generating MUD files from network flows.}{MUD generation.}

\change{Similarly,}{Additionally, other works also apply trust-based approaches to their solutions, increasing the granularity of the evaluation.} Haefner and Ray presented ComplexIoT in \cite{Haefner2019CompleIoT}, a behavioral framework designed to evaluate each traffic flow in an IoT device and calculate a trust score for it. The authors collected traffic of 25 devices approximately (general computers, smartphones, IoT devices).\rmvtxt{Flows were aggregated in 30 minutes (15 seconds without a packet ends the flow).} Based on the Flow Trust Score of each connection, calculated using IF, different policies and rules are applied to mitigate possible attacks. This solution is deployed on an enforcement architecture as an SDN environment based on OpenFlow. 

From a distributed perspective, the authors of \cite{Nguyen2019DioT} used federated learning to build DÏoT, an autonomous self-learning distributed system for detecting compromised IoT devices. The system created communication profiles for each device based on network packets and flows\rmvtxt{, being able to identify different device types}. Then, an anomaly detection-based approach was applied to detect changes in the device behavior \change{, detecting}{caused by} network attacks \addtxt{(Mirai botnet)}. The architecture was deployed using a network gateway (router) as the Anomaly Detection component. Besides, an IoT Security Service was in charge of maintaining a repository of GRU models.\rmvtxt{A number of 30 IoT devices and an installation of the Mirai botnet were employed to test the platform, obtaining a 95.6\% attack detection rate and fast ($\approx$257 ms) compromised device detection. The authors claimed that the models and datasets will be made public in the future} Another \addtxt{DL-based} distributed solution was proposed in \cite{ali2020towards}, in which Ali et al. submitted an IoT device behavior capturing system powered by blockchain and designed to enable trust-level confidence to outside networks. The authors deployed a Trusted Execution Environment (TEE) \cite{Sabt2015TEE} to provide a secure execution environment for sensitive code and blockchain data. The data came from the N-BaIoT dataset \cite{meidan2018BaIoT} and contained network features related to benign and botnet attack flows. \rmvtxt{ANNs were utilized for behavior modeling and to continuously detect abnormal behavior. This method was compared to other ML anomaly detection algorithms, such as IF and OC-SVM. For testing, a Mirai botnet-based DDoS attack was applied, achieving 99.2\% TPR and 175 $\pm$ 230ms detection time.} \change{In }{Also from a distributed perspective, in }\cite{Marir2018Distributed}, the authors proposed a behavior anomaly detection system based on network traffic. \change{The system gave a distributed vision of large networks, so}{Here,} the data was stored using a Hadoop Distributed File System (HDFS), and the processing was based on distributedly training \addtxt{a Deep Belief Network (DBN) and a stacked layer SVM} using Apache Spark. \rmvtxt{The traffic flows were processed using a Deep Belief Network (DBN) for dimensionality reduction technique, and then a stacked layer SVM was used as classifier.}The system was tested using different datasets, \change{including}{(}KDD99 \cite{KDD99}, NSL-KDD \cite{NSL-KDD}, UNSW NB-15 \cite{UNSW2017Dataset}, CIC-IDS 2017 \cite{CICIDS}\addtxt{)}. \rmvtxt{The achieved F1-Score ranged from 93\% to 97\%.}

\subsubsection{\addtxt{Sensor-based attack detection}}

Regarding sensor measurements to detect attacks, the main solutions based on this approach are applied to IoT and ICS environments \cite{pacheco2018anomaly,li2019mad,zhanwei2019abnormal,Neha2020SCADA}. \addtxt{Pacheco and Hariri \cite{pacheco2018anomaly} focused on IoT sensor behavior analysis to detect common attacks such as DoS, Flooding or Impersonation. This approach recognized previously known and unknown attacks by calculating Euclidean distance from normal sensor measurements.} The authors of \cite{li2019mad} performed anomaly detection in cyber-physical systems (CPS), using GANs and time series data. From this perspective, the authors built an unsupervised GAN framework based on LSTM networks \addtxt{, which was tested using SWaT dataset \cite{Mathur2016SWAT}, WADI dataset \cite{goh2016dataset}, and KDD99 dataset \cite{KDD99}}. \rmvtxt{This approach achieved 99.99\% precision, 99.98\% recall, and 77\% F1-Score using SWaT dataset [166], 46.98\% precision, 99.99\% recall and 37\% F1-Score using the WADI dataset [167], and 94.92\% precision, 96.33\% recall, and 94\% F1-Score using the KDD99 dataset [163]. Note that the previous results were obtained in different executions choosing the given metric to optimize.}

\change{The previous results were improved by}{Similarly,} Neha et al. \cite{Neha2020SCADA}\change{, where}{ proposed} a behavioral-based IDS for ICSs, in this case for SCADA systems\rmvtxt{, was proposed}. This approach applied RNNs to detect cyber-physical attacks. The model received sensor measurements gathered from the SWaT dataset \cite{Mathur2016SWAT}\rmvtxt{, achieving 98.05\% accuracy and 97\% TPR when classifying normal and injected data}. Zhanwei and Zenghui \cite{zhanwei2019abnormal} \addtxt{also }proposed an anomaly detection system for ICSs\addtxt{, but }based on the behavior of the data sequences from the industrial control Modbus/TCP network traffic. The authors tested their system both in a simulated water tank scenario and in a real chemical mixing infrastructure\change{. This approach utilized}{, utilizing} sensor measurements to generate a behavior model and predict future behavior. \rmvtxt{Results showed 5.5-6.4\% FPR and 11-17\% FNR when detecting different tampering and MitM attacks through a linear model.}

\subsubsection{\addtxt{System calls, logs and software signature-based attack detection}}

\change{Some o}{O}ther solutions rely on system calls, execution logs, and software signatures to model device activity and detect attack situations \cite{creech2013semantic,Attia2015AD, deshpande2018hids,liu2020statistical,He2020BoSMoS}. These solutions cover a wide range of device types, including resource-constrained devices, general computers, and cloud systems. 

\addtxt{Based on system call collection and processing, }Gideon Creech \cite{creech2013semantic} developed an IDS based on system call patterns. The authors utilized a semantic approach over the system call traces to understand running programs and detect anomalies \addtxt{utilizing an Extreme Learning Machine (ELM)}. A Linux system was monitored under different types of vulnerability exploitation attacks, and the dataset was made publicly available as ADFA-LD \cite{creech2013semantic}\rmvtxt{. Several tests were carried out utilizing an Extreme Learning Machine (ELM) and the semantic features extracted from the system calls, achieving 100\% TPR and 0.6\% FPR}. Also covering cloud intrusion detection using system calls, in \change{[97]}{\cite{Mishra2020VMGuard}}, the authors developed a HIDS for \rmvtxt{IaaS }cloud \change{solutions}{environments} that utilized system calls \rmvtxt{and Hidden Markov Models (HMM) }to build a normal behavior profile \addtxt{based on Term Frequency-Inverse Document Frequency (TF-IDF). Then, ML-based classifiers were employed to recognize the attacks}. \rmvtxt{Then, the HMM was used as classifier achieving 97\% accuracy, 100\% detection rate, and 5.66\% FPR. }\addtxt{Following similar paths, Liu et al. \cite{liu2020statistical} developed a general IDS based on system call TF-IDF statistical patterns derived from n-gram models.} In \cite{deshpande2018hids}, Deshpande et al. \addtxt{also} faced \rmvtxt{with }cloud computing intrusion detection based on system calls \change{. The authors gathered system calls using \textit{audit} framework and aggregated them in time windows to calculate call frequency vectors.}{using ML classifiers and call frequency vectors.} \rmvtxt{Then, a k-NN classifier was used to decide if a vector was abnormal. The solution achieved 90\% accuracy and 96\% TPR. However, TNR was only 42.5\%.}

From a different perspective, Attia et al. proposed in \cite{Attia2015AD} an adaptive host-based anomaly detection framework for resource-constrained devices. The designed use case targeted the detection of malicious updates on Android applications. \rmvtxt{The framework collected the system calls of the monitored applications by using \textit{Strace}. }\change{Then, it}{It} generated a normal behavioral model for each monitored application\rmvtxt{. This normal profile was defined using short sequences of system calls} using n-gram language models. \rmvtxt{Then, look-ahead, n-gram tree, and varied-length n-grams algorithms were tested for anomaly detection.}\rmvtxt{Performance varies depending on the algorithm, look-ahead achieved the best detection rate of $\approx$70\% and zero FPR, while n-gram tree achieved the best results in CPU and RAM consumption. The resource consumption of this solution is 20-50\% CPU and $<$8\% RAM.}Additionally, \rmvtxt{for IoT security improvement, }He et al. \cite{He2020BoSMoS} proposed BoSMoS, a distributed software status monitoring system \addtxt{for Industrial IoT (IIoT)} enabled by blockchain. \rmvtxt{The system was designed for Industrial IoT (IIoT) and aimed to detect malicious behaviors based on software modifications. }To accomplish its goal, the system \change{generated}{stored} a snapshot of the device software \addtxt{in the blockchain} and \addtxt{then} monitored its system file calls.\rmvtxt{Blockchain was used as a trusted decentralized database to store trusted software snapshots. Then, in each IIoT device a monitoring module was deployed to generate system software snapshots based on the executable and users' profiles.This module also monitored file calls. Hence, when target software was accessed, the module checked its authenticity instantly.} \change{The system performance was measured based on the delay to detect modified files. It}{This solution} was executed in 300s intervals, so modified software did not run for more than these 300s. Finally, the authors also tested solution scalability, performance, and security.

\subsubsection{\addtxt{Hardware event-based attack detection}}

Apart from the behavioral data considered by the previous solutions, other works such as \cite{wang2015confirm,Golomb2018CIoTACI,Ott2019CA_HPC} used Hardware Performance Counters (HPC) to model system behavior. These solutions focused on resource-constrained devices such as embedded systems and IoT devices. In \cite{wang2015confirm}, the authors presented ConFirm, a technique to identify device behavior and detect malicious modifications in the firmware of embedded systems\rmvtxt{. This technique is based on the monitoring of the number of low-level hardware events that occur during firmware execution} using HPCs. \rmvtxt{To avoid the disablement of the system, it was installed as a legacy bootloader extension. }Deviations, based on execution paths, were calculated to evaluate the system performance. The proposal was tested on ARM and PowerPC embedded processors, verifying that the solution was able to detect all the tested modifications with low resource overhead. In \cite{Golomb2018CIoTACI}, Golomb et al. proposed CIoTA\rmvtxt{ (Collaborative IoT Anomaly Detection)}, a lightweight framework using blockchain to perform distributed and collaborative anomaly detection in resource-constrained devices. In this solution, an Extended Markov Model (EMM) captured an application control-flow asynchronously using HPCs. Attack informing blocks were submitted to the blockchain (validated by neighbor devices) to ensure that an attacker cannot exploit a large number of devices within a short period of time. The system was tested in an IoT platform composed of 48 Raspberry Pi simulating smart cameras and lights. An exploit was executed to simulate a bot behavior in some devices. \rmvtxt{Results showed that using 20 models, consensus can easily detect the attack, achieving a zero false positive ratio. }The authors also mentioned some countermeasures, such as alerts, service restart, or poweroff. 

Ott and Mahapatra \cite{Ott2019CA_HPC} utilized HPCs and their occurrence frequency to enable continuous authentication of embedded software. For this purpose, the HPCs streams were processed using Short-Time Fourier Transforms (STFT) to extract frequency information. The authors discussed the usage of classifiers; however, they considered these models too heavy for embedded systems and chose to build their own authentication algorithm \change{. This algorithm started with a 512 data point window, then a Hanning window function was used, and its output was given to the STFT algorithm. Then, a threshold was defined to transform the frequencies to 256 bits. Finally, a cyclic redundancy check (CRC-8) function reduced the output to 8 bits. These 8 bits were used to build the system authentication state machine, which was responsible for performing the authentication process, achieving}{ based on cyclic redundancy check (CRC) function and state machines.} \rmvtxt{This authentication approach achieved 97\% TPR and 1.5\% FPR in a Linux system.}

\subsubsection{\addtxt{Resource usage-based attack detection}}

An alternative approach to detect anomalies caused by attacks consists in resource usage monitoring \cite{shone2013misbehaviour,barbhuiya2018rads,ravichandiran2018anomaly}, \addtxt{applied} mainly in cloud and container systems. Shone et al. proposed in \cite{shone2013misbehaviour} a misbehavior monitoring solution for \addtxt{DoS detection in }cluster-based systems. This solution utilized resource usage metrics together with process and file modification monitoring to model the system behavior. Anomaly detection was addressed based on thresholds, clustering, and statistical similarity calculation. \rmvtxt{In a simulated environment, the authors achieved 0.11\% FPR and 0\% FNR detecting DoS attacks, consuming 0.5\% RAM and 14\% of CPU.} Similarly, Barbhuiya et al. proposed in \cite{barbhuiya2018rads} a DDoS and cryptomining attack detection framework for cloud data centers. The solution, called RADS (Real-time Anomaly Detection System), monitored CPU and network utilization \change{over time to detect resource usage anomalies. The anomaly detection process was done using the CPU usage percentage and network usage as a time series.}{as a time series for anomaly detection.} Then, different window-based approaches were applied to perform attack identification \change{. Raw data was collected with a 5 seconds frequency and grouped in one-minute windows, calculating the measurements average and standard deviation. Attack detection was performed using Spike detection analysis based on the IQR.}{based on IQR Spike detection analysis.} \addtxt{For testing, }a real-world \rmvtxt{testing }dataset was gathered from \addtxt{Bitbrains data center} \cite{Shen2015Dataset}.\rmvtxt{, containing measurements from Bitbrains data center. Evaluation results achieved 90-95\% F1-Score and 0-3\% FPR when detecting DDoS and cryptomining attacks.}

\change{On the other hand,}{Additionally,} \addtxt{some works have also covered attack countermeasure actions. In this line, } the authors of~\cite{ravichandiran2018anomaly} presented an anomaly detection mechanism based on resource behavior designed to identify when a cloud system should be auto-scaled. \rmvtxt{The system design considered CPU, network, and disk usage. However, in the testing deployment, only CPU resources were used.} To detect anomalies, an AutoRegressive (AR) model was trained \addtxt{using CPU usage statistics}, and the prediction error \rmvtxt{(MSE, RMSE, MAE)}on the test dataset was used as anomaly measurement. The system was only tested using two DoS and stress \rmvtxt{example }attacks, detecting both of them. \rmvtxt{No additional experiments are carried out to evaluate system performance.}

The main characteristics of the attack detection solutions are summarized in \tablename~\ref{tab:attack_anomaly_table}. Based on the attack detection solution analysis, we can claim that attack detection is the most varied behavior application scenario. Although network is the most used source, others such as system calls or resource usage also have notable relevance. The same heterogeneous distribution can be observed regarding processing and evaluation approaches, having a balance between classification and anomaly detection. The concrete sources and techniques applied are related to the type of attacks addressed. Thus, although many solutions achieved successful results, the rapid evolution of attack techniques leads to the need for new future solutions in this area. 

\begin{table*}[htpb]
    \centering
    \scriptsize
    \begin{tabular}{ >{\Centering}m{0.6cm} >{\Centering}m{0.6cm}  >{\Centering}m{1.2cm} >{\Centering}m{1.4cm} >{\Centering}m{1.7cm} >{\Centering}m{1.2cm} >{\Centering}m{1.6cm} >{\Centering}m{1.3cm} >{\Centering}m{1.7cm} >{\RaggedRight\arraybackslash}m{4.0cm} } 
    \hline
    \textbf{Work} &\textbf{Year} & \textbf{Device Type} & \textbf{Approach} & \textbf{Algorithms} & \textbf{Behavior Source} & \textbf{Features} & \textbf{Dataset} & \textbf{Attack Type} & \makecell[c]{\textbf{Results}} \\
    \hline
    \hline
    \cite{meidan2017detection} & 2017 & IoT Devices & Classification & RF & Network & Flow-based statistics & Private & Untargeted / targeted attacks & \change{99-96\% accuracy}{99\% accuracy in white-listed devices and 96\% in not white-listed.}\\
    \hline
    \cite{Ferrando2017Network} & 2017 & IoT Devices & Classification & ARIMA, Euclidean distance & Network & Header statistics & Private & Unusual changes and attacks & Anomalies visualized based on behavioral distance\addtxt{, no performance metrics were given}. \\
    \hline
    \cite{Amouri2018NIDS} & 2018 & IoT Devices & Classification & DT, Linear Regression & Network & Mac and network layer counters & Private & Traffic anomalies & 100\% detection (TPR) after 3000s (3 reports).\\
    \hline
    \cite{yin2018enhancing} & 2018 & General computers & Classification & GAN & Network & Traffic flow statistics & \cite{beigi2014towards} & Botnet behavior & 74.04\% precision, 71.17\% accuracy, 70.59\% F1-Score, 15.59\% TPR for botnet activity detection. \\
    \hline
    \cite{Marir2018Distributed} & 2018 & General computers & Classification &  (Spark) DBN and SVM & Network & Traffic flow statistics & \cite{KDD99}, \cite{NSL-KDD}, \cite{UNSW2017Dataset}, \cite{CICIDS} & Network attacks & 93-97\% F1-Score in the tested datasets. \\
    \hline
    \cite{carvalho2018ecosystem} & 2018 & SDN & Anomaly Detection & SVM, kNN, MLP & Network & Traffic statistics & Private & DDoS, port-scan and flash crowd & Attacks were detected and mitigated\\
    \hline
    \cite{radford2018sequence} & 2018 & General networks & Anomaly Detection & LSTM & Network & Traffic flows & \cite{CICIDS} & Common network attacks & 87\% AUC average, over 71\% AUC in all attacks. \\
    \hline
    \cite{yu2019radar} & 2019 & IoT Devices & Anomaly Detection & RPNI + RANSAC & Network & Application-layer series & Private & IoT anomalies & The attacks are discovered with high accuracy. \\
    \hline
    \cite{Haefner2019CompleIoT} & 2019 & PCs, IoT Devices & Anomaly Detection & IF & Network & Flow statistics & Private & DDoS and botnets & Different device confidence based on behavior Flow Trust Score.\\
    \hline
    \cite{hamza2019detecting} & 2019 & IoT Devices & Anomaly Detection & PCA, k-means, Markov Chains & Network & Flow counters &  \cite{hamza2019detecting} & Network attacks & 94.9\% accuracy, 89.7\% TPR, and 5.1\% FPR. \\
    \hline
    \cite{afek2019nfvbased} & 2019 & NFV & Anomaly Detection & While-listing (MUD) & Network & Traffic flows & Private & Unauthorized connections & Unknown connections forbidden\\
    \hline
    \cite{Hamad_IoT_Identification_Network_2019} & 2019 & IoT Devices & Classification & RF & Network & Flow-based statistics & \cite{Miettinen2017IoTSentinel} & Attack prevention & 90.3\% accuracy using RF, outperforming other ML algorithms. \\
    \hline
    \cite{HASAN2019100059} & 2019 & IoT Devices & Classification & SVM, RF, ANN, LR & Network & MQTT-traces features & \cite{pahl2018all} & DoS, control, Scan & 99\% F1-Score classifying normal and attack traces.\\
    \hline
    \cite{lima2019smart} & 2019 & General computers & Classification & RF & Network & TCP/IP header statistics & \cite{lima2019smart} & DoS/DDoS & 96.5\% attack detection rate, 99.5\% F1-Score, 0.2\% FAR\\
    \hline
    \cite{fernandez2019intelligent} & 2019 & CPSs & Anomaly Detection / Classification & OC-SVM / NB & Network & Flow statistics & Private & Ransomware attacks & 95.9\% F1-Score, 4.6\% FPR in anomaly detection, and +99\% classification accuracy.\\ 
    \hline
    \cite{Nguyen2019DioT} & 2019 & IoT Devices & Anomaly Detection & (Fed. Learn.) GRU & Network & Header statistics & To be published & IoT attacks & 95.6\% attack detection rate and fast ($\approx$257 ms) attack detection. \\
    \hline
    \cite{Sivanathan2} & 2020 & IoT Devices & Anomaly Detection & PCA, k-means & Network & Header statistics & \cite{Sivanathan1,hamza2019detecting} & Network attacks & 91.3\%-84.3\% average detection rate for direct network attacks, and 99.1\%-58.8\% for reflection attacks.\\
    \hline
    \cite{ali2020towards} & 2020 & IoT Devices & Anomaly Detection  & (Blockchain) Neural Network & Network & Flow statistics & \cite{meidan2018BaIoT} & DDoS attacks & 99.2\% TPR and 175 $\pm$ 230ms to attack detection. \\
    \hline
    \cite{Blaise2020BotFP} & 2020 & IoT Devices & Classification & DBSCAN & Network & TCP, UDP, ICMP headers & \cite{GARCIA2014CTU13} & Botnet detection (and attacks) & 100\% TPR, 0.9\% FPR\\
    \hline
    \hline
    \addtxt{\cite{pacheco2018anomaly}} & \addtxt{2018} & \addtxt{IoT Devices} & \addtxt{Classification} & \addtxt{Euclidean Distance} & \addtxt{Sensors} & \addtxt{Sensor measurements} & \addtxt{Private} & \addtxt{Common network attacks} & \addtxt{98\% accuracy for known attacks and up to 97.4\% for unknown attacks.}\\
    \hline
    \cite{li2019mad} & 2019 & ICSs & Anomaly Detection & LSTM-based GAN & Sensors & Measurement value sequences & \cite{Mathur2016SWAT}, \cite{goh2016dataset}, \cite{KDD99} & Cyber-physical attacks & 99.99\%-46.98\% precision, 99.98\%-96.33\% recall and 94\%-37\% F1-Score, depending on the dataset.\\
    \hline
    \cite{zhanwei2019abnormal} & 2019 & ICSs & Anomaly Detection & Linear model & Sensors & Sensor measurements & Private & Tampering and MitM& 5.5-6.4\% FPR and 11-17\% FNR \\
    \hline
    \cite{Neha2020SCADA} & 2020 & ICSs & Classification & RNN & Sensors & Sensor value sequences & \cite{Mathur2016SWAT} & Cyber-physical attacks & 98.05\% accuracy and 97\% TPR when classifying normal and injected data.\\
    \hline
    \hline
    \cite{creech2013semantic} & 2013 & General computers & Anomaly Detection & ELM & System calls & Semantic features & \cite{creech2013semantic} & Vulnerability exploitation & 100\% TPR and 0.6\% FPR.\\
    \hline
    \cite{Attia2015AD} & 2015 & Mobile devices & Anomaly Detection & Look-ahead, N-gram tree & System calls & \change{Strace tool}{n-gram sequences} & Private & Malicious app updates & $\approx$70\% detection rate and 0\% FPR. \addtxt{20-50\% CPU and $<$8\% RAM.} \\
    \hline
    \cite{deshpande2018hids} & 2018 & Cloud systems & Classification & k-NN & System calls & System call traces (audit) & Private & Anomalous call sequences & 90\% accuracy, 96\% TPR, 42.5\% TNR\\
    \hline
    \change{[97]}{\cite{Mishra2020VMGuard}} & \change{2013}{2020} & Cloud systems & Classification & \change{HMM}{RF} & System calls & \change{System calls identifiers}{Frequency statistics} & \change{Private}{\cite{warrender1999detecting}} & Anomalous system calls & \change{97\% accuracy, 100\% detection rate, and 5.66\% FPR}{100-94\% TPF, 6.2-0\% FPR and 6-0\% FNR.} \\
    \hline
    \addtxt{\cite{liu2020statistical}} & \addtxt{2020} & \addtxt{General computers} & \addtxt{Classification} & \addtxt{IF, LOF, OC-SVM, k-NN} & \addtxt{System calls} & \addtxt{n-gram sequence stats.} & \addtxt{\cite{creech2013semantic}, \cite{Creech2014Dahc}, \cite{haider2017generating}} & \addtxt{Anomalous system calls} & \addtxt{73.7\% overall best AUC. $<$ 110s for evaluation.} \\
    \hline
    \cite{He2020BoSMoS} & 2020 & IoT Devices & \rmvtxt{Distributed }Anomaly Detection & Hash equality checking \rmvtxt{(Blockchain)} & Software signatures& \change{Executable and configurations}{File} snapshots & Private \makecell[c]{(Simulated)} & Software modification & Executable modification detection within 300 seconds\rmvtxt{ in the performed tests}. \\
    \hline
    \hline
    \cite{wang2015confirm} & 2015 & Embedded systems & Anomaly Detection & Execution path deviation & Hardware Events & HPCs & Private & Firmware modifications & The system is plactical with low overhead\\
    \hline
    \cite{Golomb2018CIoTACI} & 2018 & IoT Devices  & Anomaly Detection & (Blockchain) EMM & Hardware Events  &  HPCs app control-flow  & Private & Adversarial attacks & Exploit execution easily identified, enhancing network overall security. \\
    \hline
    \cite{Ott2019CA_HPC} & 2019 & Embedded systems & Continuous Authenticat.\rmvtxt{ion} & Own (Window + Fourier + CRC) & Hardware Events & HPCs & Private & Abnormal software & 97\% TPR, 1.5\% FPR in the authentication of embedded software. \\
    \hline
    \hline
    \cite{shone2013misbehaviour} & 2013 & Cluster systems & Anomaly Detection & Threshold+ k-means + statistical & Resource usage & Hardware, process and file info & Private & DoS attacks & 0.11\% FPR and 0\% FNR detecting DoS attacks, consuming only 0.5\% RAM and 14\% of CPU. \\
    \hline
    \cite{barbhuiya2018rads} & 2018 & Cloud data centers & Anomaly Detection & IQR & Resource usage & CPU, network & \cite{Shen2015Dataset} & DDoS, Cryptomining & 90-95\% F1-Score and FPR of 0-3\% \\
    \hline
    \cite{ravichandiran2018anomaly} & 2018 & Cloud systems & Anomaly Detection & Autoregressive (AR) model & Resource usage & CPU & Private & DoS, service stress attack & Attacks are fully detected\\
    \hline
    \hline
    \end{tabular}
    \caption{Main attack detection solutions based on device behavior fingerprinting (works are grouped by behavior source, using double horizontal lines to separate them, and sorted by year).}
    \label{tab:attack_anomaly_table}
\end{table*}

\subsection{Malfunction and Fault Detection}

The last behavior application scenario identified is malfunction and fault detection. In these solutions, the purpose is to detect faulty devices or malfunctioning components based on device behavior changes. This approach has been applied to several device types, such as IoT \cite{choi2018detecting,Spanos_IoT_anomaly_identification_2019}, ICSs \cite{manco2017fault}, NFV systems \cite{Nedelkoski2019ADLSTM,GULENKO2016AD_NFV,Schmidt2018ADArima,gulenko2018detecting}, general computers \cite{Kubacki2019anomalies}, cloud systems \cite{WANG201889,Agrawal2017AD_CC}, and containers \cite{sorkunlu2017tracking,samir2020detecting,du2018anomaly}. \tablename~\ref{tab:malfunction_table} compares the solutions detailed in this subsection.

\subsubsection{\addtxt{Network-based fault detection}}

Choi et al. \cite{choi2018detecting} addressed faulty IoT device identification based on behavior fingerprinting from sensor data and its correlation. This solution was named DICE, and it was installed in the network gateway to \change{collect sensor data and extract some context from it}{extract context from application-layer communications and generate}\rmvtxt{. Using} statistical features for a vector distance-based evaluation\rmvtxt{, the system achieved an average precision of 94.9\% and 92.5\% recall, and 3 minute average time to detect faults}. In the same line, Spanos et al. \cite{Spanos_IoT_anomaly_identification_2019}\rmvtxt{, under EU H2020 Project GHOST,} proposed a security solution based on the generation of behavioral templates using the IoT device network communications. \change{After a dimensionality reduction using PCA, clustering algorithms (DBSCAN)}{PCA dimensionality reduction and DBSCAN clustering} were applied to the network data to detect abnormal devices. Based on Euclidean distance, devices located far from a cluster center generated an alert and triggered some mitigation actions. This proposal was validated under simulated physical damage and mechanical exhaustion anomalies. \rmvtxt{Besides, }

\subsubsection{\addtxt{Sensor-based fault detection}}

\addtxt{Sensor data has also been applied in the literature for fault detection. In this line, }Manco et al. \cite{manco2017fault} explored ICS fault detection based on sensor stream data analysis. The system performed window-based processing to obtain statistical features, and then clustering to build classes from unlabeled data. Finally, outlier detection was performed to distinguish failures using Expectation Maximization algorithm. This approach was tested in train door failure detection\rmvtxt{, achieving 89.5\% AUC}.

\subsubsection{\addtxt{System log-based fault detection}}

From \change{a}{the} system log perspective, in \cite{Nedelkoski2019ADLSTM}, the authors applied a multimodal LSTM network approach to perform anomaly detection in NFV microservices based on distributed execution traces. \rmvtxt{They obtained over 90\% accuracy using real-word cloud traces.}Kubacki et al. \cite{Kubacki2019anomalies} explored abnormal behavior detection based on system logs related to performance metrics\rmvtxt{such as system interrupts rate per second, data transfer rate, CPU queue length, and memory usage}. The authors performed a pulse-oriented time series analysis to characterize periodical behaviors and detect anomalies\rmvtxt{. The evaluation was performed} using a self-developed algorithm called PANAL\rmvtxt{, which is based on statistical analysis}. The correlation between metrics was \rmvtxt{also }evaluated on real logs, finding a high correlation during \rmvtxt{certain }anomalous situations such as truncated cyberattacks or data backups. \rmvtxt{As this was a data analysis work, the authors did not provide metrics regarding system performance when detecting anomalous behaviors.}

\subsubsection{\addtxt{Resource usage-based fault detection}}

\change{When it comes to malfunction and fault detection,}{In the malfunction and fault detection scenario,} the most common data source is resource usage, especially for fault finding in cloud and container systems. In this context, Gulenko et al. \cite{GULENKO2016AD_NFV} proposed an anomaly detection architecture for large-scale NFV systems. In this proposal, different resource usage metrics were collected from each host\rmvtxt{, including CPU and RAM usage, disk I/O operations, and network I/O activity. To keep a low resource consumption, the solution collected between 130 and 180 metrics easily accessed on a typical Linux machine, parsing the \textit{/proc} file system} in short time intervals\rmvtxt{(300 ms)}. To process the data, the architecture used techniques based on online unsupervised clustering and classification algorithms\rmvtxt{capable of handling continuous data streams}. \rmvtxt{Multiple analysis steps were chained together and executed on different hosts to achieve scalability.} The authors claimed that the preliminary evaluation \rmvtxt{of the collected data }showed a high degree of reliable recognition of pre-defined failure scenarios\rmvtxt{, exceeding 95\%}. In addition, Sorkunlu \change{, Chandola, and Patra}{et al.} \cite{sorkunlu2017tracking} published a method to track the behavior of a cluster system based on its resource usage. \rmvtxt{The used resource usage metrics were CPU, disk I/O, HPCs, network I/O, and virtual memory.}Data was \change{grouped into}{organized into} three-dimensional tensors (compute nodes, usage metrics, and time). To measure behavior changes, data was grouped in ten-minute time windows and dimensionality reduction algorithms were applied. Finally, the reconstruction error was measured. \rmvtxt{The experiments used the \textit{TACC\_Stats} monitor, giving 86 different resource metrics, and all anomalies were correctly detected.}In \cite{Schmidt2018ADArima}, by the same team as \cite{GULENKO2016AD_NFV}, the authors proposed an unsupervised detection approach using the Online ARIMA \rmvtxt{[173]} forecasting algorithm \addtxt{\cite{OnlineArima}}. This model was based on predicting the next expected values and comparing them with the actual ones. \rmvtxt{The used data included CPU percentage, disk-io time and load, memory usage and percentage, network-io bytes, packets, errors, and dropped packets. Concretely, each metric was collected in a 500ms loop. }The authors introduced controlled anomalies, such as disk pollution, or HDD, CPU, and memory stress and leak\change{. Results showed up to a 100\% accuracy in the anomaly detection}{, being able to recognize all of them}. This team also addressed black-box service modeling \cite{gulenko2018detecting} based on clustering to detect functioning anomalies like in the previous work. The used clustering algorithm was BIRCH \cite{zhang1996birch}. \rmvtxt{In this work, almost all anomalies were detected perfectly, except for the memory leak and CPU stress anomalies, which achieved 83\% detection rate.}

Following a similar approach, Wang et al. \cite{WANG201889} proposed a self-adaptive monitoring architecture for online anomaly detection in cloud computing. The system gathered performance metrics from different sources such as CPU, Network, Memory, and Disk. \rmvtxt{Then, PCA was applied over these metrics, followed by a sliding window to cache monitoring data. The evaluated faults were CPU hog, network congestion, memory leak, and disk interference. }To calculate anomalies, the PCA-based eigenvector of the \rmvtxt{evaluated} metrics was compared to the standard eigenvector. The adaptability could be achieved by adjusting \change{the}{a} sliding window based on the estimated anomaly degree. A similar line to this work was covered by Agrawal et al. \cite{Agrawal2017AD_CC}, where similar features were collected and PCA was used as dimensionality reduction algorithm. \rmvtxt{Here, the authors achieved 88.54\% accuracy and 86\% F1-Score. }Besides, Du et al. \cite{du2018anomaly} proposed a framework to monitor and classify anomalous behaviors in microservices and containers. \change{The framework had a monitoring component that gathers data about CPU, memory, and network resources and groups the measurements in 30 second windows. Then, d}{D}ifferent anomalies, such as high CPU consumption or memory leak, were injected, and the generated data was labeled\change{. In the experiments, k-NN achieved the best results with an F1-Score between 97\% and 93\%}{ for using ML classifiers}. Finally, Samir and Pahl \cite{samir2020detecting} utilized hierarchical hidden Markov models (HHMM) to detect anomalies in container clusters.\rmvtxt{Anomalies were detected based on CPU and memory utilization. To test the system, anomalies based on resource exhaustion and workload contention were injected.} HHMM model was compared with Dynamic Bayesian Network and Hierarchical Temporal Memory \addtxt{to detect resource exhaustion and workload contention}, achieving the best results in three different generated datasets\rmvtxt{: 95\% F1-Score and 19\% FAR for Dataset A, 95\% F1-Score and 27\% FAR for Dataset B, and 90\% F1-Score and 31\% FAR for Dataset C}. 

\begin{figure*}[bp]
    \centering
    \begin{subfigure}{0.50\textwidth}
    \begin{center}
    \includegraphics[trim=0.1cm 0.1cm 0.1cm 0.1cm, clip=true,width=0.72\columnwidth]{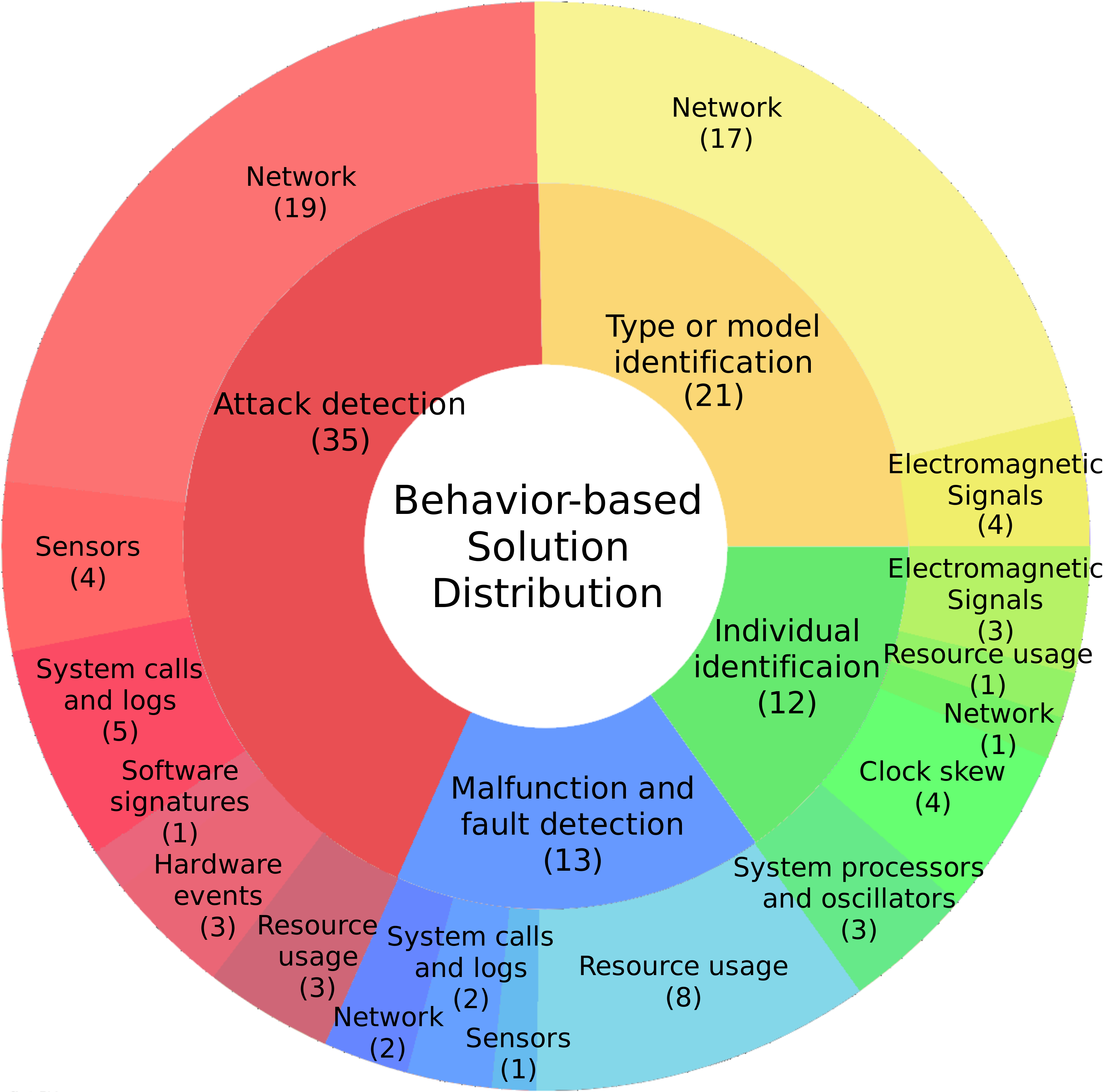}
    \end{center}
    \caption{Scenario and source distribution scheme. (Internal ring: Application Scenario. External ring: Behavior Source.)}
    \end{subfigure}
    \begin{subfigure}{0.49\textwidth}
    \begin{center}
    \includegraphics[trim=0.7cm 1cm 0.3cm 0.8cm, clip=true,width=\columnwidth]{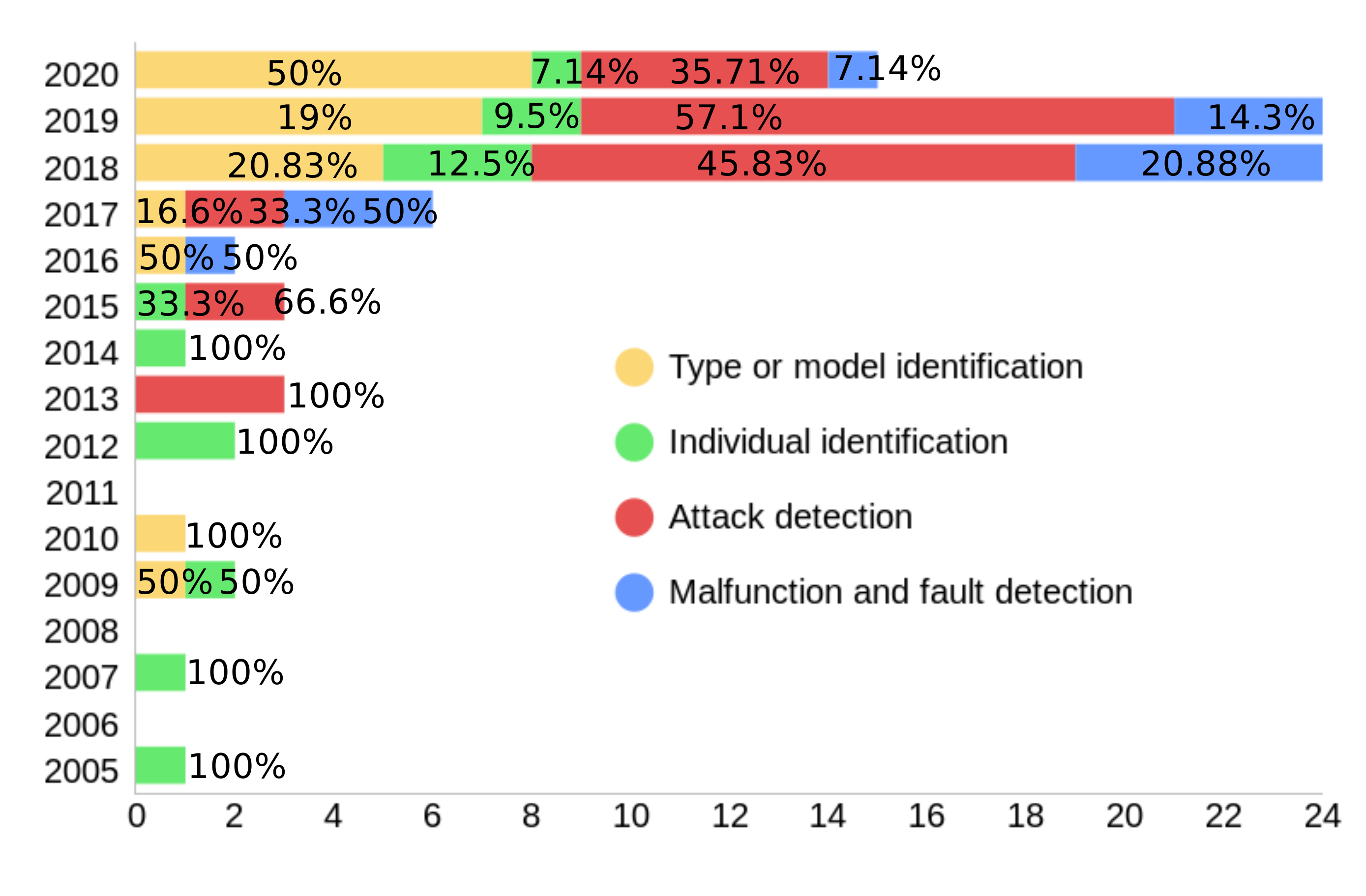}
    \end{center}
    \caption{Yearly solution distribution.}
    \end{subfigure}
    \caption{Distribution graphs of device behavior fingerprinting solutions.}
    \label{fig:solution_distribution}
\end{figure*}

\tablename~\ref{tab:malfunction_table} compares the main characteristics and results of the solutions focused on fault and malfunction detection. From the description of the previous solutions, we can observe that resource usage and system logs are the most used behavior source for fault detection, especially in NFV, cloud, containers, and microservice systems. In contrast, IoT devices and ICSs faults have been solved based on a network and sensor-based perspective. Moreover, most of the solutions are focused on anomaly detection-based evaluation, instead of using labeled data. Finally, \figurename~\ref{fig:solution_distribution} shows the distribution of the analyzed solutions regarding their application scenario and behavior source, and their publication year.

\begin{table*}[ht!]
    \centering
    \scriptsize
    \begin{tabular}{ >{\Centering}m{0.6cm} >{\Centering}m{0.6cm}  >{\Centering}m{1.4cm} >{\Centering}m{1.4cm} >{\Centering}m{1.7cm} >{\Centering}m{1.3cm} >{\Centering}m{1.6cm} >{\Centering}m{0.8cm} >{\Centering}m{1.5cm} >{\RaggedRight\arraybackslash}m{4.4cm} } 
    \hline
    \textbf{Work} & \textbf{Year} & \textbf{Device Type} & \textbf{Approach} & \textbf{Algorithms} & \textbf{Behavior Source} & \textbf{Features} & \textbf{Dataset} & \textbf{Anomaly} & \makecell[c]{\textbf{Results}} \\
    \hline
    \hline
    \cite{choi2018detecting} & 2018 & IoT Devices & Anomaly Detection & Vector distance & Network & Sensor values statistics & \cite{casas_datasets,activity_datasets}& Faulty IoT sensors & 94.9\% and 92.5\% average precision and recall, respectively. 3 mins for detection.  \\
    \hline
    \cite{Spanos_IoT_anomaly_identification_2019} & 2019 & IoT Devices & Classification & PCA, DBSCAN, Euclidean distance & Network & Statistical features & Private & Physical and mechanical errors & Successful threat detection regarding physical damage and mechanical exhaustion.\\
    \hline
    \hline
    \cite{manco2017fault} & 2017 & ICSs & Anomaly Detection & Expectation Maximization & Sensors & Sensor values statistics & Private & System Faults & 89.5\% AUC detection train door failures.\\
    \hline
    \hline
    \cite{Nedelkoski2019ADLSTM} & 2019 & NFV systems & Anomaly Detection & LSTM & System logs & \rmvtxt{Microservice }Execution traces statistics & Private & \change{Service }{Microservice} anomalies & $>$90\% accuracy using real-word cloud traces.\\
    \hline
    \cite{Kubacki2019anomalies} & 2019 & General computers & Statistical Analysis & PANAL (\rmvtxt{pulse-oriented }time series) & System logs & Performance metrics & Private & Anomalous \rmvtxt{system }behavior & \rmvtxt{Exploratory }Study on metric correlations regarding performance, event, and process logs.\\
    \hline
    \hline
    \cite{GULENKO2016AD_NFV} & 2016 & NFV systems & Anomaly Detection & Clustering and Classification \rmvtxt{(Not Specified)} & Resource usage & CPU, memory, disk, network \rmvtxt{(Linux /proc)} & Private & NFV anomalies & 95\% recognition of pre-defined anomalous scenarios. \\
    \hline
    \cite{sorkunlu2017tracking} & 2017 & Cluster systems & Anomaly Detection & PCA & Resource usage & CPU, memory, disk, network & Private & Cluster anomalies & Anomalies correctly detected.\\
    \hline
    \cite{Agrawal2017AD_CC} & 2017 & Cloud systems & Anomaly Detection & Robust PCA & Resource usage & CPU, memory, disk  & Private & Cloud faults & 88.54\% accuracy and 86\% F1-Score\\
    \hline
    \cite{Schmidt2018ADArima} & 2018 & NFV systems & Anomaly Detection & Online ARIMA & Resource usage & CPU, memory, disk, network  & Private & NFV Resource anomalies & 100\% accuracy detecting controlled HDD, CPU and memory anomalies. \\
    \hline
    \cite{WANG201889} & 2018 & Cloud systems & Anomaly Detection & PCA, eigenvector & Resource usage & CPU, memory, disk, network & Private & Cloud faults & The system detects injected test faults. \\
    \hline
    \cite{gulenko2018detecting} & 2018 & NFV systems & Anomaly Detection & BIRCH & Resource usage & CPU, memory, disk, network  & Private & System anomalies & \rmvtxt{Almost }All anomalies \rmvtxt{perfectly }detected, except 83\% detection for memory leak and CPU stress. \\
    \hline
    \cite{du2018anomaly} & 2018 & Microservices Containers& Classification & SVM, RF, k-NN, NB & Resource usage & CPU, memory, network & Private & Container anomalies & 97-93\% F1-Score using k-NN as classifier. \\
    \hline
    \cite{samir2020detecting} & 2020 & Container clusters & Anomaly Detection & HHMM & Resource usage & CPU, memory & Private & Resource exhaustion & 95-90\% F1-Score and 19-31\% FAR. \\
    \hline
    \hline
    \end{tabular}
    \caption{Main malfunction and fault detection solutions that use device behavior fingerprinting (works are grouped by behavior source, using double horizontal lines to separate them, and sorted by year).}
    \label{tab:malfunction_table}
\end{table*}

\section{Public Datasets}
\label{sec:datasets}

To address \textit{Q4 (Which behavior datasets are available and which are their characteristics?)}, this section reviews the main public datasets containing device behavior activities and characteristics found in the literature. \addtxt{Specifically, it analyzes datasets contemplating the scenarios, devices and sources discussed in \textit{Q1}, allowing validating most of the techniques presented in \textit{Q2}, and studying the scope of the solutions analyzed in \textit{Q3}.} Each dataset is described by taking into account the devices and sources monitored, and data morphology. Below, the analysis is organized according to the two main application scenarios stated in Section \ref{sec:sources}, which are Device identification and Misbehavior detection --attack and anomaly detection.

\subsection{Device Identification Datasets}

Several datasets published in recent years and collecting device behavior are conceived to perform device model, type, or individual identification. In 2006, Maya Rodrig et al. published the uw/sigcomm2004 dataset \cite{uw-sigcomm2004-20061017}. The main purpose of this dataset is to analyze how Wi-Fi networks work and how they can be improved. This dataset contains 70 GB of both wired and wireless traces. The wireless traces were collected for five days using three computers in monitor mode near access points. Selcuk Uluagac published in \cite{gatech-fingerprinting-20140609} the dataset associated with his research work on network-based individual device identification \cite{radhakrishnan2014gtid,uluagac2013passive}. This dataset contains the inter-arrival time of network traffic packets collected from 30 wireless devices. 1.5 GB of data was collected both actively, directly communicating with the devices, and passively, sniffing the communications. This dataset can be used to generate network-based fingerprints and derive parameters such as approximated clock skew.

With a similar goal, but focused on IoT, Miettinen et al. published the IoT Sentinel dataset~\cite{Miettinen2017IoTSentinel}. This dataset contains the traffic generated during the setup of 31 IoT devices of 27 different types (4 types have 2 devices). To avoid anomalies and have data variety, the device setup process was collected at least 20 times for each device, generating a total of 64 MB of data. Another dataset dealing with IoT devices is the Yourthings dataset \cite{yourthings}, which contains raw network traffic from 45 different smart-home IoT devices. The data was collected for 10 days in March and April of 2018. Each day data contains from 10 to 13 GB. \rmvtxt{In [136], the authors utilized the collected network flows to characterize each device model and evaluate its security properties. }Following the same \change{environment}{approach}, in \cite{Sivanathan1}, Sivanathan et al. published a dataset collected for IoT device classification under IoT Traffic Traces name. The data was collected in 2016 for 20 days from 28 different IoT devices, including cameras, lights, plugs, sensors, appliances, and health-monitors. In addition, this dataset also includes captures from non-IoT devices such as laptops and smartphones. In total, $\approx$9.5 GB of raw pcap files are available. As additional content, post-processing tools to obtain IP, NTP, and DNS flows are also enclosed. \addtxt{Regarding radio frequency, Allahham et al. \cite{allahham2019dronerf} published DroneRF in 2019, a dataset containing 3.8 GB of radio data collected from 3 different drones during functioning. This dataset has been designed for drone detection, identification and tracking.} More recently, Hagelskjær et al. published in 2020 a dataset designed for IoT device identification based on radio spectrum monitoring \cite{hagelskjaer2020dataset}. The dataset contains +50 GB of 863-870 MHz band raw spectrum measurements with a sampling frequency of 10 MSPS collected in November 2018. The published dataset contains both raw spectrum captures and pre-processed features extracted with PCA. \rmvtxt{The raw data from different device locations are available, such as in the same room, in different rooms, or upstairs.}

\begin{table*}[htpb]
    \centering
    \scriptsize
    \begin{tabular}{ >{\Centering}m{2.8cm}  >{\Centering}m{0.5cm} >{\Centering}m{1.8cm} >{\Centering}m{1.5cm} >{\Centering}m{2.0cm} >{\Centering}m{1cm} >{\RaggedRight\arraybackslash}m{6.5cm} }
        \hline
        \textbf{Dataset} & \textbf{Year} & \textbf{Device Type} & \textbf{Data Source} & \textbf{Data} & \textbf{Size} & \makecell[c]{\textbf{Details}} \\
        \hline
        The uw/sigcomm2004 dataset \cite{uw-sigcomm2004-20061017} & 2006 & Wireless and wired devices & Network & Raw traces & 70 GB & This dataset includes the traces collected by wireless and wired monitoring using tcpdump.\\
        \hline
        The gatech/fingerprinting dataset  \cite{gatech-fingerprinting-20140609} & 2014 & Wireless devices & Network & Inter-arrival time information & 1.5 GB & Inter-arrival time information collected from 30 wireless devices to generate unique fingerprints. \\
        \hline
        IoT Sentinel \cite{Miettinen2017IoTSentinel} & 2017 & IoT devices & Network & Raw traces and processed features & 64 MB & Network communications dataset collected during the setup process of 31 devices.\\   
        \hline
        Yourthings \cite{yourthings} & 2018 & IoT devices & Network & Raw traces and processed features & +110 GB & 10 days of network traffic collected from 45 different smart-home IoT devices. Flows utilized to evaluate security. \\
        \hline
        IoT Trace Dataset \cite{Sivanathan1} & 2018 & IoT devices & Network & Raw traces and processed features & $\approx$9.5 GB & Network flows collected during 20 days from 28 different IoT Devices. The source includes tools to derive flow statistics.\\
        \hline
        \addtxt{DroneRF \cite{allahham2019dronerf}}& \addtxt{2019} & \addtxt{IoT devices (Drones)} & \addtxt{Radio spectrum} & \addtxt{Radio segments} & \addtxt{3.8 GB} & \addtxt{227 segments collected from 3 different drones during functioning.}\\
        \hline
        Device spectrum identification \cite{hagelskjaer2020dataset} & 2020 & IoT devices & Radio Spectrum & Raw spectrum and processed features & +50 GB & 863-870 MHz radio spectrum measurements collected in diverse scenarios, like in the same room and different rooms.\\
         \hline
    \end{tabular}
    \caption{Most relevant device identification datasets that use device behavior fingerprinting.}
    \label{tab:dataset_identification}
\end{table*}

\tablename~\ref{tab:dataset_identification} summarizes the public datasets previously described, paying attention in their publication year, monitored devices, and data sources collected. Most of the datasets (5 of \change{6}{7}) contain network traces or network-based features. It could be due to the facility to monitor from outside the device behavior without modifying its software. Furthermore, this source is quite generic as almost every device has at least one network interface. Additionally, the \change{only dataset}{two datasets} not based in network communications contain\rmvtxt{s} spectrum measurements, another externally-collected source. In this context, there is a missing spot for device identification datasets containing sources such as clock skew, system logs or events, and resource usage metrics.

\subsection{Anomalous Behavior and Attack Datasets}

The second dataset category is based on public datasets containing anomalous device behavior, either based on attacks or other exceptional situations. Note that most of these datasets also contain normal or benign device behavior, which can be utilized to model normal device behavior and identify it, like in the previous subsection. Next, the main datasets found in the literature will be detailed.

The family of datasets that considers network communications to create device behavior fingerprints is extensive. One of the most representative is the CTU-13 dataset \cite{GARCIA2014CTU13}, a botnet traffic activity dataset collected in 2011. 13 different botnet samples were captured during different attack conditions such as Command and Control (C\&C) connection and the launching of diverse attacks --DDoS, or port scanning, among others. Additionally, the dataset also contains normal and background network traffic. In total, this dataset contains +140 hours of network traffic with a total size of $\approx$700 GB. \addtxt{Besides, the dataset has been updated in the last years to include IoT malware captures.} A set of relevant datasets, IDS 2017 and 2018 datasets \cite{CICIDS}, was created by the Canadian Institute of Cybersecurity (CIC). They contain raw network traces and derived features obtained during different network attacks. Concretely, the monitored attacks were FTP and SSH Brute Force, DoS, Heartbleed, Web Attacks, Infiltration, Botnet, and DDoS. In addition, these datasets also contain benign traffic. The 2017 dataset was collected from 25 users and contains 51.1 GB of data, while the 2018 dataset contains 220 GB of traffic from 500 different devices. The previous datasets were collected and processed by Lima et al. \cite{lima2019smart} to extract $\approx$40 MB of vectors with 73 features relative to IP headers of the traffic flows. Then, the dataset was published together with a research article. Also from CIC, the ISCX botnet dataset \cite{beigi2014towards} contains raw network captures of 16 different botnet malware. This dataset is generated by combining previous CIC datasets containing botnet activity. In total, the dataset contains 5.3 GB of training traces and 8.5 GB for testing. Aligned with the previous datasets, in \cite{Kangdataset}, the authors provided a novel network dataset, published in September 2019, which contains several types of attacks in an IoT environment. The dataset is composed of $\approx$ 1.5 GB of real and simulated attacks, such as port scanning, flooding, brute force, or ARP spoofing, among others. In the case of real attacks, the network packets were obtained from Mirai botnet. To identify the network behavior of the devices infected, packets were captured while simulating attacks through tools such as NMAP.

Anomalous behavior or attacks affecting IoT devices is another cutting edge field where several datasets have been created and published. In this sense, the N-BaIoT dataset \cite{meidan2018BaIoT} contains more than 7 million vectors, with 115 features each, giving around 20 GB, obtained by processing the network communications of 9 different IoT commercial devices under attack. Vectors contain 11 labels, 10 for different botnet attacks, produced by Mirai and BASHLITE, and 1 for benign traffic. Similarly, the DS2OS dataset \cite{pahl2018all} contains 61 MB of features obtained from application layer traces collected from simulated IoT devices such as light controllers, thermometers, movement sensors, washing machines, batteries, thermostats, smart doors, and smartphones. This dataset is designed for anomaly detection in IoT node communications. In the same line, the USNW IoT Benign and Attack Traces Dataset \cite{hamza2019detecting} monitored network communications of 27 devices for 30 days, being 10 of these devices victims of network attacks such as ARP spoofing, TCP/UDP flooding, and packet reflection. In total, more than 64 GB of data is available. This dataset also provides the source code to derive vectors with 238 features using packet counters and traffic flows. Another relevant dataset is the NGIDS-DS dataset \cite{haider2017generating}, which consists of 6.7 GB of labeled network and device operating system logs collected on a simulated critical infrastructure. The dataset is designed for host-based intrusion detection and contains normal and attack scenarios. The authors used the \textit{IXIA Perfect Storm} tool to generate a wide variety of network attacks. The data was obtained from a machine running \textit{Ubuntu 14.04} and different common services such as \textit{Apache}. The OS logs contain the date, process id, system call, event id, and the network data consist of raw traffic. 

A similar approach was followed to generate the UNSW-NB15 dataset \cite{UNSW2017Dataset}. This dataset contains 100 GB of raw traffic flows and derived features from several attacks launched using \textit{IXIA Perfect Storm}. This attack set includes the same type of attacks as NGIDS-DS dataset. The Aposemat IoT-23 dataset \cite{Aposemat_IoT_23}, published in January 2020 by the same team as for CTU-13 \cite{GARCIA2014CTU13}, is another labeled dataset containing 23 captures of malicious and benign IoT network traffic. Concretely, 20 captures include malware activity, while 3 include normal network activity of 3 IoT device types. The dataset includes 11.3 GB of pcap files and 8.7 GB of network log files. The authors utilized known malware, such as Mirai, Okiru, or Torii botnets, port scanning, DDoS, C\&C connections. \addtxt{In the same direction, IoT-KEEPER dataset \cite{hafeez2020iot} was published in 2020. This dataset contains 11.8GB of pcap files collected from several IoT devices affected by common attacks such as port scanning, botnet execution, DoS, or malware injection. Besides, it also contains network activity from real computers, replicating a real edge network environment. Finally, LITNET-2020 dataset \cite{damasevicius2020litnet} contains feature vectors generated during 12 attacks on general computers deployed on an academic network. In total, this dataset contains 26.9 GB of vectors with 85 processed flow features extracted using Netflow.}

Focused on application layer communications of general computers, ECML-PKDD 2007 \cite{ECML2007Dataset} and HTTP CSIC 2010 \cite{gimenez2010http} datasets are available. ECML-PKDD 2007 \cite{ECML2007Dataset} contains 80 MB of application layer requests in XML format. There are 25000 valid and 15000 attack requests, the attack requests include SQL Injection, LDAP Injection, cross-site scripting (XSS), and command execution, among others. The data includes web requests and also context information such as server operating system, services, etc.     \change{Also dealing with the communication application layer, t}{T}he HTTP CSIC 2010 dataset \cite{gimenez2010http} includes 56 MB of normal and abnormal HTTP requests. It was published by the Spanish Research National Council (CSIC) to test web application attack protection systems. The dataset is divided into 36000 normal and 25000 anomalous requests. The anomalous requests are divided into three types of attacks: static, dynamic, and unintentional illegal requests. Concretely, static attacks try to gather hidden resources, while dynamic attacks are SQL injections, XSS, etc. This dataset is usually used as benchmark for HTTP \rmvtxt{layer }anomalous behavior detection solutions.

From the system calls and execution traces perspective, it is worth commenting the ADFA Intrusion Detection Datasets for Linux \cite{creech2013semantic} and Windows \cite{Creech2014Dahc}. These datasets contain 9 MB of Linux system call identifiers and 13.6 GB of Windows XML system call traces of DLL libraries. Both datasets include normal and attack system calls. Attacks include HydraFTP, HydraSSH, Meterpreter, Webshell, and a poisoned executable. Currently, these are widely used for benchmarking solutions based on system call traces \cite{aghaei2019host,sudqi2019lightweight}. The Firefox-SD dataset \cite{murtaza2013host} is also based on system calls, but in this case made by \textit{Firefox} browser in Linux. The dataset contains +1 TB of normal activity traces, collected while executing seven browser testing frameworks, and attack-based traces, generated under attacks using known exploits such as memory consumption, integer overflow, or null pointer exploit. 

Dealing with ICSs and anomaly detection, one of the reference datasets is the Secure Water Treatment (SWaT) dataset \cite{Mathur2016SWAT}. This dataset was collected in 2016 from a real water treatment testbed managed by a SCADA system. It contains 11 days of continuous operation, 7 of them normal and 4 under attack by 36 different data injections. This dataset contains $\approx$16 GB of traffic logs and 361 MB of measurements obtained from 51 sensors and actuators\rmvtxt{ deployed in the scenario}. Additionally, SWaT dataset was updated in December 2019 with 45 GB of raw traffic and 6 MB of measurement logs, collected during 3 hours of normal traffic and 1 hour in which 6 attacks were launched. Similarly, the Water Distribution (WADI) dataset \cite{goh2016dataset} contains 575 MB of labeled sensor and actuator logs collected in the same water treatment plant. In this case, the dataset contains data from 123 sensors and actuators collected during 16 days of operation, having 14 days of normal traffic and 2 days with 15 data injection attacks launched in total. Also in the ICS field, in \cite{Perales2019AD_ICS}, Perales et al. developed a dataset called Electra, based on a railway electric traction substation. The monitored network protocols were Modbus TCP and S7Comm, common in SCADA systems. This dataset contains 1.7 GB of derived features originating from raw captures. \rmvtxt{Besides, the authors perform classification and anomaly detection (RF, SVM, DNN, OC-SVM, IF) using the published data, achieving 99-93\% F1-Score. In this same work, the authors also perform a comparison between attack datasets focused on traditional networks [155],[157],[162],[163] and in ICSs [165][166].}

Regarding resource usage monitoring, the GWA-T-12 Bitbrains dataset \cite{Shen2015Dataset} contains performance metrics collected from 1750 virtual machines located in Bitbrains data center. Resource usage metrics are collected in five-minute samples, the monitored resources are the CPU usage, memory usage, disk read/write throughput, and network received/transmitted throughput. In total, 2.7 GB of traces are available, divided into two sets of machines (1250 VMs used for fast storage and 500 with lower performance). Although BEHACOM \cite{sanchez2020behacom} dataset is focused on user activity monitoring (keyboard and mouse interactions), it also contains resource usage metrics regarding active applications, CPU, and memory. This data was collected from the computers of 12 users over 55 days. In total, this dataset contains 6.1 GB of features derived from user activity. Also dealing with resource usage monitoring but from the mobile devices prism, CIC has released two different datasets on dynamic smartphone behavior and its relationship with malware. The first one is CIC-AAGM (CIC Android Adware and General Malware) \cite{lashkari2017towards}, which contains +20 GB of traffic flows generated when installing 1900 different applications, being 250 adware, 150 malware, and 1500 benign. The second is InvesAndMal2019 \cite{taheri2019extensible} dataset, which includes device status, traffic flows, permissions, API calls, and logs generated by 426 malware and 5065 benign Android applications. In total +275 MB of logs and features are available.

\change{At this point is important to mention that o}{O}ther existing datasets are more than 20 years old, which makes them outdated with regard to current scenarios. This is the case of DARPA 1998/1999 \cite{DARPA1998,DARPA1999}, KDD99 \cite{KDD99}, and NSL-KDD \cite{NSL-KDD} datasets. The original datasets, DARPA 1998 and 1999, are composed of $\approx$ 10 GB of network traffic and system logs collected by MIT Lincoln Laboratory. The aim of these datasets was to build a generic evaluation dataset for intrusion detection. 56 different attacks were recorded, including different DoS, buffer overflow, and reconnaissance attacks, among others. The network traces were stored in tcpdump format and the system logs as BSM/NT audit data. Afterward, KDD99 dataset was derived from DARPA traffic by extracting 1.2 GB of features from the traffic flows. Besides, NSL-KDD is a refinement of KDD99 were duplicated entries are deleted and classes are more balanced, reducing the dataset to around 60 MB. These datasets have become some of the most popular datasets for intrusion detection evaluation. However, as commented before, they are outdated compared to current networks and attacks.

\begin{table*}[htpb]
    \centering
    \scriptsize
    \begin{tabular}{ >{\Centering}m{2.5cm}  >{\Centering}m{0.8cm} >{\Centering}m{1.8cm} >{\Centering}m{1.7cm} >{\Centering}m{2.3cm} >{\Centering}m{1cm} >{\RaggedRight\arraybackslash}m{6cm} }
        \hline
         \textbf{Dataset} & \textbf{Year} & \textbf{Device Type} & \textbf{Data Source} & \textbf{Data} & \textbf{Size} & \makecell[c]{\textbf{Details}} \\
        \hline
        DARPA \cite{DARPA1998,DARPA1999} & 1998-1999 & General computers & Network and system logs & Raw network packets and logs (bsm) & $\approx$ 10 GB  & Attack and normal network and system activity. One of the most used IDS datasets, but it is outdated.\\
        \hline
        KDD99 \cite{KDD99} & 1999 & General computers & Network & Connection record features & 1.2 GB & Derived features based on DARPA 1998/1999 network traffic.\\
        \hline
        UNM dataset \cite{warrender1999detecting} & 1999 & General computers & System calls & System calls and process IDs & $\approx$500 KB & System call identifiers collected during normal behavior and under some attacks. \\
        \hline
        ECML-PKDD 2007 \cite{ECML2007Dataset} & 2007 & Web systems & Network & Requests and contextual information & 80 MB & 25000 valid and 15000 attack XML web queries, including context information such as server OS. \\
        \hline
        NSL-KDD \cite{NSL-KDD} & 2009 & General computers & Network & Connection record features & 60 MB & Based on KDD99 data, but with additional processing like filtering duplicated data.\\
        \hline
        HTTP CSIC 2010 \cite{gimenez2010http} & 2010 & Web systems & Network & HTTP requests & 56 MB & 61000 normal and anomalous HTTP requests. It includes diverse attacks and also unintentional illegal requests. \\
        \hline
        CTU-13 \cite{GARCIA2014CTU13} & 2011 & General computers & Network & Raw captures and flows & $\approx$700 GB & 13 different scenarios were botnet activity is combined with normal traffic. \\
        \hline
        Firefox-SD \cite{murtaza2013host} & 2013 & Application (Firefox) & System calls & Raw system calls & +1 TB & Firefox browser system calls while normal activity and under different attacks. \\
        \hline
        ADFA-LD \cite{creech2013semantic} & 2013 & General computers & System calls & Linux system logs & 9 MB  & System calls collected on 60 different attack sets. \\
        \hline
        ADFA-WD \cite{Creech2014Dahc} & 2014 & General computers & System calls & XML Windows DLL traces & 13.6 GB & System call dataset composed by virtual kernel calls done by DLL libraries. \\
        \hline
        CIC-ISCX \cite{beigi2014towards} & 2014 & General computers & Network & Raw captures & 13.8 GB & Botnet activity dataset collected from 16 real botnet malware.\\
        \hline
        GWA-T-12 Bitbrains \cite{Shen2015Dataset} & 2015 & Distributed data centers (Cloud) & Resource usage & CPU, Memory, Disk and Network statistics & 2.7 GB & Performance metrics (CPU, memory, disk and network) collected from 1750 VMs each 5 mins.\\
        \hline
        SWaT \cite{Mathur2016SWAT} & 2016 & ICSs & Network, and sensors/actuators & Network and sensor/actuator logs & $\approx$16.3 GB & 7 days of normal activity and 4 days of data injection attacks in a real water treatment testbed.\\
        \hline
        WADI \cite{goh2016dataset} & 2016 & ICSs & Sensors / actuators & Sensor/actuator logs & 575 MB & 16 days of logs of 123 industrial sensors and actuators. 15 attacks launched over 2 days.\\
        \hline
        NGIDS-DS \cite{haider2017generating} & 2017 & Critical infrastructure & Network and system logs & Raw network packets and audit logs & 6.7 GB & Critical infrastructure attacks simulated on an Ubuntu 14.04 machine using IXIA PerfectStorm tool. \\
        \hline
        UNSW-NB15 \cite{UNSW2017Dataset} & 2017 & General computers & Network & Raw captures and processed features & 100 GB & IDS dataset, attacks generated using IXIA PerfectStorm tool. \\
        \hline
        CIC-IDS 2017\cite{CICIDS} & 2017 & General computers & Network & Raw captures and processed features & 51.1 GB & IDS dataset based on 25 users activity, it contains common network attacks.\\
        \hline
        CIC-AAGM \cite{lashkari2017towards} & 2017 & Mobile devices & Network & Raw captures and processed features & +20 GB & Flows generated by 1900 different applications (250 adware, 150 malware, 1500 benign).\\
        \hline
        DS2OS \cite{pahl2018all} & 2018 & IoT devices & Network & Application \rmvtxt{layer }traces & 61 MB & IoT smart home devices normal and abnormal activity. \\
        \hline
        N-BaIoT \cite{meidan2018BaIoT} & 2018 & IoT devices & Network & Processed features & $\approx$20 GB & Botnet (Mirai and BASHLITE) activity collected from 9 IoT devices. \\
        \hline
        CIC-IDS 2018 \cite{CICIDS} & 2018 & General computers & Network & Raw captures and processed features & 220 GB & IDS dataset collected in 500 devices which contain common network attacks.\\
        \hline
        Smart-Detection \cite{lima2019smart} & 2019 & General computers & Network & Processed features & $\approx$40 MB & DoS detection based on previous datasets (CIC-DoS, CIC-IDS 2017 and CIC-IDS 2018). \\
        \hline
        ELECTRA \cite{Perales2019AD_ICS} & 2019 & ICSs & Network & \rmvtxt{Modbus/ S7Comm }Processed features & 1.7 GB & Data collected from attacks to an electric traction system. \\
        \hline
        USNW IoT Benign and Attack Traces \cite{hamza2019detecting} & 2019 & IoT devices & Network & Raw captures and processed features & +64 GB & IoT benign and attack network traces. Attacks include ARP spoofing, TCP/UDP flooding and packet reflection. \\
        \hline
        IoT network intrusion dataset \cite{Kangdataset} & 2019 & IoT devices & Network & Raw captures & $\approx$1.5 GB & Network captures of real and simulated attacks to IoT and non-IoT devices.\\
        \hline
        InvesAndMal2019 \cite{taheri2019extensible} & 2019 & Mobile devices & System logs and Network & Processed logs and features & +275 MB & Device status, traffic flows, API calls and logs generated from +5500 apps (426 malware and 5065 benign). \\
        \hline
        BEHACOM \cite{sanchez2020behacom} & 2020 & General computers & Resource usage & CPU and memory statistics &  6.1 GB & Active application, CPU and memory statistics collected from 12 users over 55 days.\\
        \hline
        IoT-23 \cite{Aposemat_IoT_23} & 2020 & IoT devices & Network & Raw captures & 20 GB & By the same team that CTU-13. 20 attack and 3 benign traces. Attacks simulated using infected Raspberry Pis. \\
        \hline
        \addtxt{IoT-KEEPER dataset \cite{hafeez2020iot}} & \addtxt{2020} & \addtxt{IoT devices} & \addtxt{Network} & \addtxt{Raw captures} & \addtxt{11.8 GB} & \addtxt{Network captures collected from IoT devices under common attacks.}\\
        \hline
        \addtxt{LITNET-2020 \cite{damasevicius2020litnet}}& \addtxt{2020} & \addtxt{General Computers} & \addtxt{Network} & \addtxt{Processed features} & \addtxt{26.9 GB} & \addtxt{Dataset collected on an academic network under 12 different attacks.}\\
        \hline
    \end{tabular}
    \caption{Most relevant anomalous behavior and attack datasets that use device behavior fingerprinting.}
    \label{tab:dataset_anomalies}
\end{table*}

The same issue occurs with the system call dataset of the University of New Mexico (UNM) \cite{warrender1999detecting}. This dataset was collected in 1999 and contains $\approx$500 kB of system call and process identifiers. The collected system calls contain normal activity and different attacks such as buffer overflows and trojans. This dataset has been widely used as benchmark for system call anomalies-based attack detectors \cite{hoang2009program}. However, the system call arguments are not available and it is outdated regarding modern attacks.

\tablename~\ref{tab:dataset_anomalies} gives an overview of the public datasets with focus on behavior anomaly and attack detection. It can be appreciated how most of the datasets are focused on network, followed by system calls and logs. The datasets monitoring the previous sources are varied and cover several device types such as IoT, ICSs, mobile devices, or general computers. However, other sources such as resource usage or HPCs are under-exploited regarding public datasets for anomaly detection. \addtxt{Datasets monitoring IoT device communications during attack and malware execution have gained importance for the last years and nowadays are the dominant type of anomalous behavior datasets. The most common IoT malware families, such as Mirai botnet, have been largely monitored from a network-based perspective in datasets such as CTU, USNW IoT Traces or IoT-23. However, there is no IoT-based dataset containing in-device behavior sources, something highly useful for modeling how malware works and what changes occur within the device functioning itself.}

\begin{figure*}[ht!]
    \centering
    \begin{subfigure}{0.50\textwidth}
    \begin{center}
    \includegraphics[trim=0.8cm 0.5cm 0.8cm 0.8cm, clip=true,width=0.75\columnwidth]{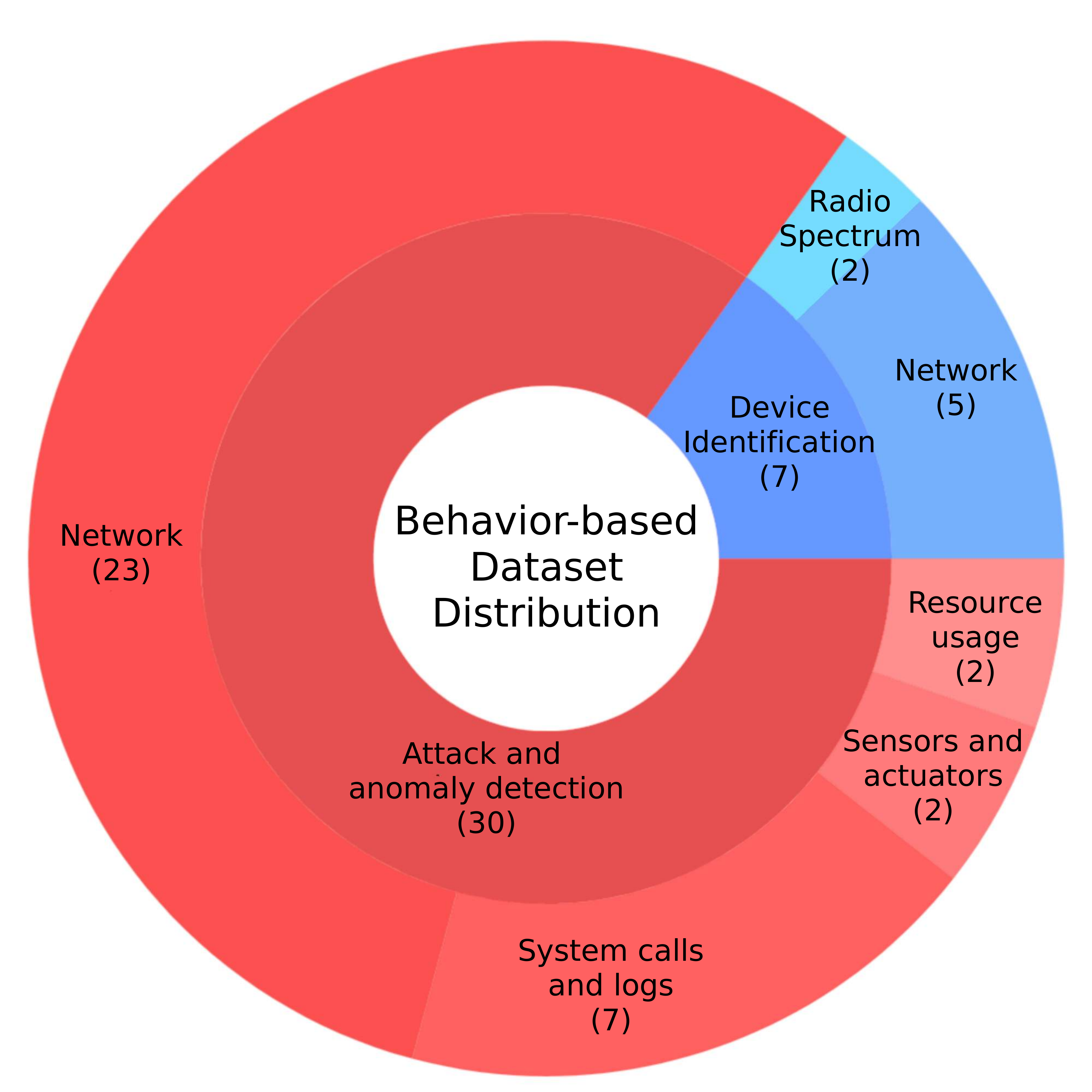}
    \end{center}
    \caption{Scenario and source dataset distribution. (Internal ring: Application scenario. External ring: Behavior Source.)}
    \end{subfigure}
    \begin{subfigure}{0.49\textwidth}
    \begin{center}
    \includegraphics[width=\columnwidth]{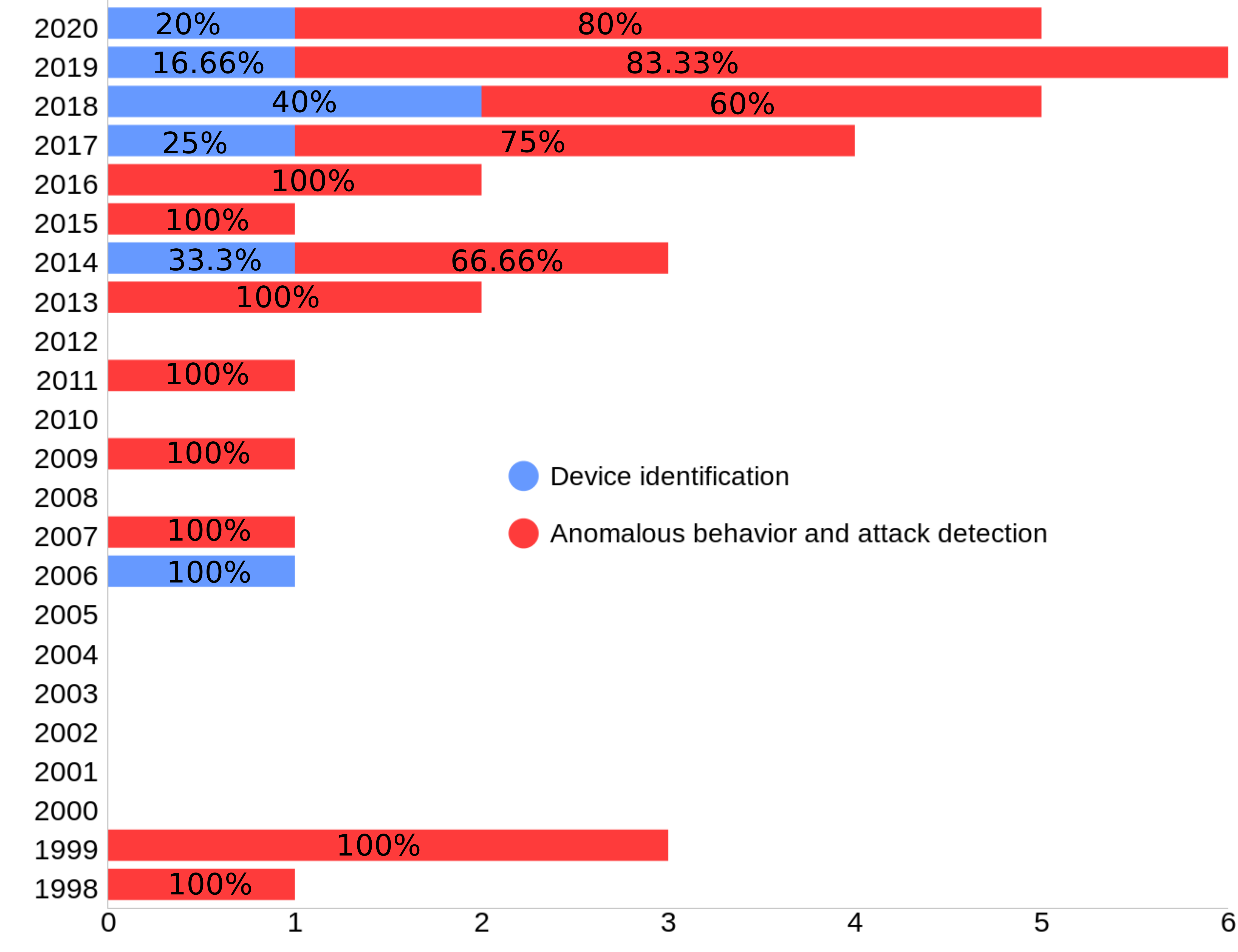}
    \end{center}
    \caption{Yearly dataset distribution.}
    \end{subfigure}
    \caption{Distribution graphs of device behavior fingerprinting dataset.}
    \label{fig:dataset_chart}
\end{figure*}

\figurename~\ref{fig:dataset_chart} shows the dataset distribution regarding main application scenarios and behavior source collected, and their publication year. Note that some datasets can contain several sources at the same time, for example, network communications and system logs. As final section thoughts, we notice that when it comes to developing a behavior evaluation solution, a key aspect is data availability, as the underlying solutions depend on it. Many works utilize self-collected private datasets to validate their approaches. However, to have a proper performance comparison, it is worth having public datasets allowing to cross-verify the proposed solutions. Furthermore, some teams do not have enough resources to collect enough data but have good processing and evaluation ideas. Therefore, having public datasets is essential to make diverse and well-performing behavior-based proposals possible.

\section{Lessons Learned, Trends and Challenges}
\label{sec:lessons}

\addtxt{Based on the different aspects of behavior fingerprinting analyzed through questions \textit{Q1-Q4},} this section \changeRev{is in charge of responding}{responds} \textit{Q5 (How have application scenarios evolved for the last years?}). To this end, it summarizes the main lessons learned, trends, \addtxtRev{and} \addtxt{open challenges} \rmvtxtRev{and conclusions }extracted from the present study of device behavior fingerprinting. \rmvtxt{In addition, it presents some research challenges in the behavioral fingerprinting field.} 

\subsection{Lessons Learned}

After reviewing and analyzing the state-of-the-art, we were able to identify the following main lessons:

\textbf{Network communications are the most exploited source.} As \figurename~\ref{fig:solution_distribution} shows, it is utilized in \change{100}{85}\% of works focused on device models or type identification, and in \change{56.25}{54.28}\% of attack detection solutions. However, this source is less exploited in individual device identification (\change{9.09}{8.33}\% of the solutions) and malfunction detection (15.38\%). This is because the data obtained from the network communication perspective is not sensitive enough as required for these scenarios, e.g., two devices of the same model deployed with the same purpose will have almost identical network communications.
    
\textbf{Clustering is widely applied for inferring classes}. As \tablename~\ref{tab:model_identification_table} shows, in device type or model identification approaches, many solutions combine unlabeled data with clustering to group data samples and derive device classes, and then apply ML/DL classification approaches. Besides, some attack and malfunction detection techniques also rely on this approach (see \tablename~\ref{tab:attack_anomaly_table} and \tablename~\ref{tab:malfunction_table}). This fact shows the viability of clustering techniques for deriving classes from unlabeled behavioral data.

\addtxtRev{\textbf{ML and DL are the favorite approaches for both classification and anomaly detection.} \tablename~\ref{tab:model_identification_table}, \ref{tab:identical_device_table}, \ref{tab:attack_anomaly_table}, and \ref{tab:malfunction_table} show that ML and DL are the main solutions applied for data processing, no matter if the objective is classification (of device types/models or attacks) or anomaly detection, either to detect attacks or faults. This fact shows the enormous flexibility and capabilities of these techniques inferring complex data patterns, outperforming traditional processing methods.}

\textbf{Individual \addtxtRev{device} identification is one of the most complex application scenarios}. Only some lower-level features, such as system clocks, code execution time, clock skew, or electromagnetic signals are sensitive enough to detect minimum physical differences that occurred during the device manufacturing processes. Thus, these are the ones required for individual identification. However, the monitoring of these sources is usually complex.
    
\textbf{There is no consensus in misbehavior detection solutions}. \changeRev{A}{As \tablename~\ref{tab:attack_anomaly_table} and \ref{tab:malfunction_table} show, a}ttack and malfunction detection is addressed from heterogeneous perspectives. The selection of data sources and processing techniques depends on the type of anomalies that will be detected. Although network is the most used source, many solutions use system calls and logs, hardware events, or resource usage. 

\textbf{Public datasets are mainly focused on network, system calls, and logs.} \figurename~\ref{fig:dataset_chart} shows that there are \change{32}{35} \changeRev{elements}{datasets} containing these sources (note that some datasets contain both sources at the same time, so they are counted both as network and calls/logs source). Moreover, \tablename~\ref{tab:dataset_identification} and \ref{tab:dataset_anomalies} show that in most cases the datasets contain raw data instead of processed information or features.

\subsection{Current Trends}

The main approaches expected in future works, based on the evolution of the proposals published in recent years,
are:

\textbf{ML and DL algorithms \addtxtRev{are gaining} prominence}. As \figurename~\ref{fig:technique_year} shows, ML and DL are the most usual techniques, with \change{45.88}{46.39}\% of importance (note that many solutions utilize different techniques). In addition, DL-based techniques are gaining more importance, especially for time series processing, due to their performance handling raw data without pre-calculated features. \rmvtxtRev{Besides, in }\tablename~\ref{tab:model_identification_table}, \ref{tab:identical_device_table}, \ref{tab:attack_anomaly_table}, and \ref{tab:malfunction_table} \changeRev{it can be appreciated that in both behavior fingerprinting scenarios considered, identification and misbehavior detection,}{show that in both behavior fingerprinting scenarios (identification and misbehavior detection),} ML and DL approaches are \changeRev{the most common processing and evaluation techniques.}{gaining importance in the last years. Overall,} ML and DL algorithms are applied in the \change{69.56}{71.87}\% of identification and in the \change{64.44}{62.5}\% of misbehavior detection solutions.
   
\textbf{Statistical and knowledge-based algorithms relegation}. As \figurename~\ref{fig:technique_year} shows, processing and evaluation \addtxtRev{techniques }based on statistical and knowledge-based algorithms are losing importance as evaluation approaches, in favor of ML and DL \addtxt{trend}.

\addtxtRev{\textbf{IoT and ML/DL convergence.} In modern IoT scenarios where devices are massively deployed, behavior fingerprinting is critical management solution, grouping similar devices and detecting faults. ML and DL techniques are the best alternative when it comes to leverage the vast amount of data generated with the required performance and adaptability. This fact can be observed in the solution comparisons located in \tablename~\ref{tab:model_identification_table}, \ref{tab:identical_device_table}, \ref{tab:attack_anomaly_table}, and \ref{tab:malfunction_table}.}

\textbf{Dataset publication}. As it can be appreciated in \figurename~\ref{fig:dataset_chart} and in \tablename~\ref{tab:dataset_identification} and \tablename~\ref{tab:dataset_anomalies}, a good number of datasets have been published for the last years. In the last five years (2016-2020), \change{20}{23} public datasets were released, while in the previous five years (2011-2015) were only 7. \addtxtRev{This trend is influenced by the AI explosion, as ML and DL are powered by datasets.}
    
\textbf{Attack detection \addtxt{and model identification} \change{is gaining importance}{are the prominent application scenarios}}. \figurename~\ref{fig:solution_distribution} shows how attack detection \addtxt{and type or model identification} solutions have been gaining prominence in the last years, increasing from \change{33.3}{50}\% in 2017 to \change{71.4}{85.71}\% in 2020. \change{In contrast, the focus on type or model identification and fault detection has decreased for the last years}{This trend is a direct consequence of the explosion in IoT deployments, as new requirements rise associated with the heterogeneous variety of devices and the new security issues generated by them.}

\subsection{Future Challenges}

Based on the current state-of-the-art, the following points represent the main challenges that future behavior fingerprinting solutions might consider to enhance current solutions.

\textbf{Usage of public datasets for behavior-based solution performance comparison.} Many solutions are based on private datasets, which makes it difficult, if not impossible, to compare performance between different solutions. Among the solutions analyzed, only \change{41.66}{45}\% of device model/type identification used public datasets. The same goes for the \change{18.18}{16.66}\% about individual device identification, \change{43.75}{42.85}\% tackling attacks, and 7.69\% concerning malfunction detection, by using public datasets. Thus, a right direction for future approaches is to evaluate and compare their performance through public datasets.

\textbf{Diverse and quality behavior dataset publication.} Regarding device identification, the main publicly available datasets are focused on the network communications source. However, there is a lack of modern and variate datasets based on other sources. Then, \rmvtxtRev{to build a comprehensive enough dataset background, }it would be interesting for novel proposals addressing behavioral fingerprinting to publish the collected datasets, if any. Besides, datasets should have enough quality to ensure that research results are not influenced or damaged by low-quality data.


\textbf{Solution scalability regarding the number of monitored devices and deployment architecture.} Scalability is an issue that affects various aspects of behavior monitoring solutions. Many solutions covering individual device identification have \changeRev{noticed the number of devices to be identified as an issue}{detected that the number of devices is a challenge} \cite{lanze2012skew,radhakrishnan2014gtid,polcak2015clock}\changeRev{, as with the increase of devices, the classification results got worse.}{. The more devices in the scenario, the worse classification results.} Furthermore, centralized deployments may suffer if too many devices send behavioral data, or blockchain-based solutions may suffer block validation issues. Finally, during data evaluation, solutions based on statistical approaches that require one to one evaluation \cite{sanchez2018clock} may not scale at all when the number of devices increases.
    
\textbf{Define anomaly countermeasures to apply when an attack or fault is detected.} Many solutions solve the misbehavior detection problem, both when \addtxtRev{it is }caused by a cyberattack or a system fault. However, most solutions do not propose any countermeasure \cite{Nespoli2018Countermeasures} to mitigate the detected misbehavior. Only a few works propose some remedies for misbehavior, such as \cite{Haefner2019CompleIoT,Golomb2018CIoTACI}. 

\textbf{Secure the behavior monitoring and analysis process against attacks.}  The fingerprinting solutions can suffer attacks or modifications performed by malicious entities. This fact can jeopardize the entire fingerprinting mechanism, and in the case of centralized processing solutions, even affect other device behavior evaluation. However, few works took behavior monitoring security into account \cite{wang2015confirm}. To solve this issue, additional security mechanisms, such as encryption, should be added to current solutions. \addtxtRev{Besides, there is an emerging area on adversarial attacks to ML/DL models that should also be considered in future solutions \cite{wang2019security}}. \changeRev{Also}{Finally}, trust frameworks \cite{braga2018survey} can be included in behavior monitoring deployments to guarantee system safety.

\addtxtRev{\textbf{Private device model and type to guarantee security.} In some circumstances, like when there are known vulnerabilities, the model and type of devices should be private to avoid targeted attacks \cite{bastos2018internet}. It has been demonstrated how privacy leakage attacks can be used to identify device model and type \cite{acar2020peek,bugeja2016privacy} and further countermeasures, such as dummy traffic generation are required. In this context, there is a growing research area on device privacy enhancement \cite{apthorpe2017closing} working in different solutions such as blocking traffic, concealing DNS, tunneling traffic, and shaping and injecting traffic.}

\addtxtRev{\textbf{User's privacy impact and awareness.} Device  behavior analysis can be leveraged to perform users’ activity tracking and behavior monitoring \cite{ren2019information}. The inference of users’ activity has been demonstrated possible by behavior analysis \cite{apthorpe2019keeping} in health care and smart home IoT environments, even with encrypted traffic \cite{apthorpe2017spying}. As an example, the TV channels watched by a given subject have been inferred in \cite{xu2014watching}. Therefore, manufacturers and service providers should include  solutions  to  improve  users  privacy  and  defend  them against activity inference attacks. These solutions are aligned with the ones commented in the previous challenge, as they can cover both user and device privacy at the same time.}

\textbf{Guarantee behavioral data \addtxtRev{and model} privacy}. As in user behavior\rmvtxtRev{[203]}, \addtxtRev{data and model} privacy is a crucial aspect to consider when performing data analysis. From an ethical perspective, behavior analysis solutions should be employed to fingerprint devices in a non-intrusive way. However, privacy laws, such as GDPR \cite{CELDRAN2020107448} in Europe, are mainly focused on user perspective, leaving some device behavior fingerprinting methods out of their scope. To solve this problem, privacy-preserving solutions, such as federated learning \cite{yang2019federated} combined with differential privacy \cite{dwork2014algorithmic}, allow training \addtxtRev{ML/DL} models \changeRev{to}{that} ensure data privacy.
    

\textbf{Apply novel ML/DL approaches for behavior processing and evaluation.} As ML and DL are fast-evolving fields, some recent techniques have not been applied yet. For example, UMAP \cite{mcinnes2018umap} for dimensionality reduction, or XGBoost \cite{chen2016xgboost} for classification, could improve current solution performance. Besides, DL architectures may combine convolutional and recurrent neuron layers for DL-based time series processing \cite{lai2018modeling,karim2017lstm}. Finally, any of the analyzed solutions addressed an approach based on \textit{Reinforcement learning} \cite{henderson2018deep}, which has gained notable relevance in communications and networking areas \cite{luong2019applications}, and human behavior analysis \cite{lockwood2019computational}.

\textbf{Consider ML/DL models behavior in the device analysis.} Nowadays, devices usually include embedded ML and DL models that perform specific tasks with the data the device manipulates. However, the ML and DL models deployed on the devices have their own behavior \cite{rahwan2019machine}, which influences the general device behavior. Then, understanding AI-powered applications and services is critical to identify the device behavior and its anomalies.

\section{Conclusions}
\label{sec:conclusions}

Device behavior fingerprinting has been determined in recent years as a promising solution to identify devices with different granularity levels, as well as to detect misbehavior originated by cyberattacks or faulty components. The article at hand studies the evolution of the device behavior research field, performing a comprehensive review of the devices, behavioral sources, datasets, and techniques used in both application scenarios. In this context, the present work has been performed with the goal of answering the following research questions. 

\textit{Q1. Which scenarios, device types, and sources are present in behavior-based solutions?} Section~\ref{sec:sources} reviews how these three aspects are used in the most recent and representative works of the literature. The performed analysis shows a relevant heterogeneity of device types and behavioral sources in the existing solutions, and highlights the usage of network communications in the majority of the solutions.

\textit{Q2. What and how behavior processing and evaluation tasks are used in each scenario?} Section~\ref{sec:processing} analyzes the main techniques and algorithms --rule-based, statistical, knowledge-based, ML and DL, and time-series approaches-- used by works dealing with device and misbehavior identification. The analysis results show how ML and DL-based approaches are gaining importanc due to their versatility and excellent performance when enough training data is available, and to the detriment of statistical and knowledge-based solutions.

\textit{Q3. What characteristics do the most recent and representative solutions of each application scenario have?} In the core section of this article, Section~\ref{sec:solutions}, the reviewed solutions are described, analyzed, and compared according to their application scenario, device types, sources, techniques, and results. Regarding sources, this section shows that in device type or model identification solutions, network source is the dominant approach. In individual device identification, clock skew and electromagnetic signals are the main data sources. Attack detection is also mainly tackled using network communications. In contrast, for fault detection, the main approach is to utilize resource usage data. In terms of processing and evaluation techniques, ML and DL techniques are dominant in all the considered scenarios. 

\textit{Q4. Which behavior datasets are available and which are their characteristics?} In Section~\ref{sec:datasets}, the main public datasets containing device behavioral data are analyzed according to their application scenario. It also details the characteristics of the data they contain and how they were collected. This section shows the prominence of network source in the current public datasets, and the lack of other sources such as resource usage or hardware events.

\textit{Q5. How have application scenarios evolved for the last years?} Lessons learned, current trends, and future challenges have been drawn in Section~\ref{sec:lessons}, which details how network source and ML/DL algorithms are gaining prominence. Furthermore, it is also remarkable that novel ML/DL approaches, such as recurrent and convolutional neuron layer combination or Reinforcement learning, have not yet been applied in the area, which opens up pathways for future research. It also depicts how dataset publication is gaining importance during the last years; however, more relevant datasets are still required for sources and devices that are not covered in recent ones, e.g., resource usage or system logs in IoT devices or ICSs.

Aligned with the current trend and challenges drawn in this work, we will focus our next efforts on designing and implementing scalable behavior-based solutions to identify individual devices and detect cyberattacks affecting IoT devices. In both scenarios, we plan to utilize privacy-preserving ML and DL techniques, such as distributed and federated learning, to protect behavioral data while guaranteeing performance capabilities. Finally, we plan to build datasets for both scenarios, which will be publicly accessible to improve current dataset diversity and quality.

\section*{Acknowledgment}

This work has been partially supported by the Swiss Federal Office for Defence Procurement (armasuisse) (project codes Aramis R-3210/047-31 and CYD-C-2020003).


\bibliographystyle{unsrt}  
\bibliography{references}

\end{document}